\documentclass[12pt, a4paper]{article}

\usepackage[english]{babel}
\usepackage{booktabs}
\usepackage[letterpaper,top=2.5cm,bottom=2.5cm,left=3cm,right=3cm,marginparwidth=1.75cm]{geometry}
\usepackage{natbib}
\usepackage{threeparttable}
\usepackage[tableposition=top]{caption}
\usepackage{amsmath}
\usepackage{amsthm}
\usepackage{amssymb}
\usepackage{lscape}
\usepackage{verbatim}
\usepackage{enumitem}
\usepackage{booktabs}
\usepackage{multicol}
\usepackage{multirow}
\usepackage{longtable,booktabs} 
\usepackage{setspace}
\usepackage{graphicx}
\usepackage{subcaption}

\usepackage{titlesec}
\titlespacing\section{0pt}{12pt}{0pt}
\titlespacing\subsection{0pt}{6pt}{0pt}
\titlespacing\subsubsection{0pt}{6pt}{0pt}

\usepackage{graphicx} 
\usepackage{floatrow}
\usepackage{float}

\usepackage{color}

\usepackage{epigraph}

\usepackage{hyperref}

\theoremstyle{plain}

\usepackage[bottom]{footmisc}

\usepackage{natbib}
\expandafter\def\expandafter\quote\expandafter{\quote\footnotesize}

\usepackage{array}
\newcolumntype{L}[1]{>{\raggedright\arraybackslash}p{#1}}
\newcolumntype{C}[1]{>{\centering\arraybackslash}p{#1}}
\newcolumntype{R}[1]{>{\raggedleft\arraybackslash}p{#1}}

\title{The Emergence of Economic Rationality of GPT}
\author{Yiting Chen, Tracy Xiao Liu, You Shan, and Songfa Zhong\thanks{Chen: Lingnan University, yitingchen26@gmail.com. Liu and Shan: Tsinghua University. Liu: liuxiao@sem.tsinghua.edu.cn; Shan: shany19@mails.tsinghua.edu.cn. Zhong: Hong Kong University of Science and Technology; National University of Singapore, zhongsongfa@gmail.com.}}

\begin{document}
\maketitle

\begin{abstract} 
As large language models (LLMs) like GPT become increasingly prevalent, it is essential that we assess their capabilities beyond language processing. This paper examines the economic rationality of GPT by instructing it to make budgetary decisions in four domains: risk, time, social, and food preferences. We measure economic rationality by assessing the consistency of GPT's decisions with utility maximization in classic revealed preference theory. We find that GPT's decisions are largely rational in each domain and demonstrate higher rationality score than those of human subjects in a parallel experiment and in the literature.
Moreover, the estimated preference parameters of GPT are slightly different from human subjects and exhibit a lower degree of heterogeneity. We also find that the rationality scores are robust to the degree of randomness and demographic settings such as age and gender, but are sensitive to contexts based on the language frames of the choice situations. These results suggest the potential of LLMs to make good decisions and the need to further understand their capabilities, limitations, and underlying mechanisms.

\end{abstract}

\clearpage
\section{Introduction}
ChatGPT is a sophisticated chatbot application developed by OpenAI, which employs the state-of-the-art Generative Pre-trained Transformer model (hereafter referred to as ``GPT''). As one of the most representative examples of large language models (LLMs), GPT uses transformer architecture and deep learning techniques to learn from vast web-based text corpora that contain 175 billion parameters \citep{vaswani2017attention,brown2020language}.  Thanks to its massive volume of training data, GPT can generate human-like text with remarkable accuracy and fluency, to the extent that human evaluators find it difficult to distinguish GPT output from text written by humans \citep{brown2020language}. In addition to their natural language-generation capabilities, LLMs have demonstrated impressive abilities in a wide range of domains. For instance, they can generate computer code \citep{chen2021evaluating}, engage in human-like conversations on various topics \citep{lin2020caire}, solve university-level math problems \citep{drori2022neural}, exhibit theory of mind ability \citep{kosinski2023theory}, and possess psychological characteristics similar to humans \citep{binz2023using,park2023artificial}. LLMs have also shown their aptitude in performing high-level reasoning tasks \citep{webb2022emergent}. The impressive capabilities of LLMs reveal their remarkable potential, which can be likened to the emergence of a new species: ``Homo Silicus'' \citep{horton2023large}. Because these achievements signify a major milestone in the development of LLMs, it is important that we understand how GPT performs in various high-level reasoning tasks.

Here we present the first study on the economic rationality of GPT. Rationality has been central to the methodological debate throughout various disciplines and is the fundamental assumption in economics and related social sciences. Here we use a classic notion of economic rationality in revealed preference analysis that captures the extent to which a decision maker maximizes some well-behaved utility functions for the given budget constraints \citep{afriat1967construction,afriat1972efficiency,samuelson1938note,varian1982nonparametric,varian1990goodness,chambers2016revealed,nishimura2017comprehensive}. Prior studies have computed rationality score based on choice data in risky, intertemporal, and social decision-making in laboratory environments \citep{andreoni2002giving,andreoni2012estimating,ahn2014estimating,choi2007consistency,choi2014more,harbaugh2001garp,halevy2018parametric,fisman2007individual}, as well as expenditure data from survey and grocery stores in the field \citep{blundell2003nonparametric,blundell2008best,echenique2011money,dean2016measuring}. Economic rationality has also been measured in children \citep{brocas2019development,harbaugh2001garp},  monkeys \citep{chen2006basic}, rats and pigeons \citep{kagel1975experimental}. Moreover, it has been proposed as a measure of decision-making quality and linked to a wide range of economic outcomes, such as occupation, income, and wealth differences across individuals, and development gaps across countries \citep{banks2019education,cappelen2021development,carvalho2016poverty,choi2014more,fisman2017distributional,kim2018role,li2017social,li2022experimental}. Nevertheless, the rationality of GPT remains unexplored.

We instruct GPT to act as a decision maker to make budgetary decisions in choice environments with varying characteristics. The basic framework contains 25 decision tasks to allocate 100 points between two commodities with different prices, which is commonly used in experimental economics. The rationality of GPT is measured by the consistency of these 25 decisions with the generalized axiom of revealed preference (GARP), a necessary and sufficient condition under which a set of decisions are in accordance with utility maximization \citep{afriat1967construction,afriat1972efficiency,varian1982nonparametric,varian1990goodness}. Therefore, a rationality score is derived from each group of 25 tasks within a given environment. Building on this framework, we construct four environments by specifying the nature of the two commodities---two risky assets; two rewards with one for now and one for one month later; two payments with one for the decision maker and one for another randomly paired subject; and two types of food with meat and tomatoes. Each environment is repeated 100 times, which generates 10,000 tasks for GPT. This design allows us to systematically measure GPT's rationality in different choice domains. Moreover, we incorporate a series of variations in the randomness of GPT, the framing of decision tasks \citep{kahneman2003maps}, the structure of the choice format \citep{mcfadden2001economic}, and the demographic settings of GPT \citep{corbett2018measure}. In order to compare the economic rationality between GPT and humans, we conduct a parallel experiment with 347 human subjects from a representative US sample.

We find that GPT demonstrates a high level of rationality in all four decision-making tasks concerning risk, time, social, and food, and it outperforms human subjects in the rationality score documented in both our human subject experiment and those reported in the literature. Furthermore, we find that GPT's rationality scores are consistent across different demographic characteristics and invariant to the specification of the randomness of GPT. However, the level of rationality drops significantly when we employ a different price framing and when we use a discrete choice setting. These findings suggest that GPT obtains high rationality score but has some potential limitations in its decision-making abilities. Moreover, we estimate the preference parameters of GPT and human subjects. We find that the estimated preference parameters of GPT have some minor distinctions from human subjects and show a substantially higher degree of homogeneity.

Taken together, we use tools from revealed preference analysis and experimental economics to study increasingly capable artificial agents. There is growing interest in understanding these agents' behavior \citep{rahwan2022machine}, and ongoing debate about their performance compared with humans \citep{mitchell2023debate}. Even though these artificial agents exhibit surprisingly excellent performance on many cognitive tests, some have expressed concern that such models are still far from achieving human-level understanding of language and semantics and exhibit considerable levels of behavioral bias \citep{borji2023categorical,chen2023manager}. We contribute to the understanding of the capacities and caveats of LLMs, by demonstrating that LLMs can act as if they are rational decision makers. The observed decrease in rationality when alternative price framing or discrete choice are used is line with some studies show that GPT response can be highly sensitive to contexts \citep{borji2023categorical,mahowald2023dissociating,horton2023large,jones2022capturing,binz2023using,brand2023using,jones2022capturing}. Our study also highlights the need for more investigation and refinement of its decision-making mechanisms to ensure reliable and effective decision-making in various domains.

\section{Experimental Method}

We examine GPT's decision-making in different environments using the public OpenAI application programming interface (API). Multiple GPT variants are accessible through this API. For our exercise, we focus on the \textit{GPT-3.5-Turbo}, which powers ChatGPT and is the most popular, stable, and cost-effective model in the GPT family. We use APIs with Python instead of ChatGPT, since APIs enable us to adjust the parameters of the model and conduct massive experiments in an efficient manner.

Below we describe how we ask GPT to ``make decisions'' by introducing the construction of prompts through which GPT returns a text in response to an input text. We then outline multiple variations of our design to examine the robustness of our results. 

\subsection{Design of the Baseline Condition}

\paragraph*{Instruct GPT to ``Make Decisions".} Each input prompt in \textit{GPT-3.5-turbo} includes the specifications of a role (system, assistant, or user) and corresponding contents. We instruct GPT to make decisions in three steps. First, we specify the system's role as \textit{``a human decision maker''} and notify the system that \textit{``you should use your best judgment to come up with solutions that you like most''}.  Second, we explain the role of assistant with respect to the decision format: selecting a bundle of commodities from a standard budget line with varying prices, which will be explained in detail later, without requesting responses for any decision. This assists in storing information about the tasks. Afterward, we assign a series of decision-making tasks to the role of user in order to ask GPT to make decisions.

Moreover, to confirm that GPT has understood the instructions, we ask three testing questions, in which we either directly ask it to recall the decision format or ask about the consequence of certain decision scenarios. For each question, we simulate 25 times and GPT constantly provides correct answers. This confirms that GPT understands the decision environment. Detailed prompts to instruct GPT and obtain GPT responses are provided in Appendix \ref{prompt_gpt}.

\paragraph*{Decision Task.}

GPT decision tasks follow a typical budgetary experiment, in which a decision maker (DM) is endowed with 100 points to select a bundle of commodities, commodity $A$ and commodity $B$. The prices of the two commodities are based on different exchange rates between points and payoffs. Thus, a decision $i$ obtains a tuple $\left(p^i, x^i\right)$ whereby a DM selects a bundle $\left(x_A^i, x_B^i\right)$ under the prices $\left(p_A^i, p_B^i\right)$. Since measuring rationality requires a collection of such decisions, we include 25 tasks with randomly generated prices \citep{choi2014more}. After that, we measure the economic rationality of these 25 decisions $\left(p^i, x^i\right)_{i=1}^{25}$, based on the extent to which there exists some well-behaved utility functions to rationalize them. 

To measure rationality across different preference domains, we vary the commodities in the decision tasks. In the first domain, the two commodities are specified as two contingent securities, in which the decisions capture the DM's risk preference \citep{choi2007consistency}. In the second domain, the two commodities are rewards for today and one month later, which are designed to examine the DM's time preference \citep{andreoni2012estimating}. In the third domain, the two commodities are payoffs for the DM and another randomly matched subject, and thus the allocation captures the DM's social preference \citep{andreoni2002giving,fisman2007individual}. Finally, in the fourth domain, the two commodities are the amount of meat and tomatoes, which captures the DM's food preference \citep{harbaugh2001garp}.  

We incorporate four preference domains of decisions, each consisting of 25 tasks. To examine GPT's consistency in behavior, we simulate this process 100 times, resulting in 10,000 tasks for GPT. We refer to these 10,000 tasks, the 100 GPT observations in each preferences domain, as \textit{the baseline condition}. A detailed description of tasks and parameters for prices are provided in Appendix \ref{prompt_gpt}. We set the temperature parameter to 0 (see the explanation below) and keep the default values for all other parameters.

\subsection{Design of Conditions with Variations}

\paragraph*{}
\noindent To enrich our understanding of GPT's economic rationality, based on the baseline condition, we introduce variations in the temperature and the decision tasks. We also include demographic information in the text of the prompt as explained below.

\paragraph*{Variations of Temperature.} Temperature plays a critical role in regulating the level of stochasticity and creativity in the responses generated by GPT \citep{goodfellow2016deep}. It ranges from 0 to 1, with a higher number indicating higher randomness. We set the temperature to be 0 in the baseline condition, in which the model gives deterministic answers \citep{binz2023using,horton2023large,webb2022emergent}. Some studies on GPT incorporate the variation in temperature to investigate the impact of randomness in creating text \citep{bommarito2022gpt,chen2021evaluating}. Following their practice, we conduct two additional sets of conditions, with the parameter set to be 0.5 and 1.

\paragraph*{Variations of Decision Task.}We design two variations of decision tasks to change the framing of prices and to switch from continuous to discrete choice, respectively. A detailed description is provided in Appendix \ref{prompt_gpt}.

In the baseline condition, we use ``\textit{1 point = X units of commodity}'' to present price information, which is used in many existing experiments with human subjects \citep{andreoni2002giving,andreoni2012estimating,carvalho2016poverty,banks2019education}. In \textit{the price framing condition}, we change it to ``\textit{Y points = 1 unit of commodity}'', which is an alternative framing used in the experimental literature \citep{drichoutis2020economic}. Since the budget sets remain constant, this allows us to examine whether framing affects the rationality of GPT. 

In the baseline condition, the DM makes choices under the continuous budget sets. In \textit{the discrete choice condition}, we change these to discrete choices: The DM is presented with 11 discrete options chosen from the budget line and is asked to choose one of them rather than directly choose from the budget line \citep{chen2023consistency,kim2018role}. Specifically, the third prompt changes to: \textit{``In this round, there are 11 options, which are $(A_0,B_0)$, $(A_1,B_1)$, ..., and $(A_{10},B_{10})$. Please only tell me your best option in every round''}. This allows us to examine whether rationality of GPT is robust to the change from continuous to discrete choice sets. 

\paragraph*{Response to Demographic Information.} We also investigate whether the rationality exhibited by GPT varies with the embedded demographic information. To achieve this, we include demographic information which varies in gender, age, education level, and minority group status. We change the input content of the system's role in GPT to be ``I want to you to act as a \textit{[demographic]} decision maker, ...''. Variations are gender: \textit{``female decision maker''} versus \textit{``male decision maker''}; age: \textit{``young child decision maker''} versus \textit{``elderly decision maker''}; education: \textit{``decision maker with an elementary school education''} versus \textit{``decision maker with a college education''}; and minority: \textit{``Asian decision maker''} versus \textit{``African American decision maker''}. By doing so, we can examine whether GPT is responsive to demographic information and whether it performs differently under different individual characteristics. The responsiveness, if any, is relevant to the discussion about algorithm bias \citep{corbett2018measure}.

\subsection{Design of the Human Experiment}

\paragraph*{}
\noindent To obtain a better understanding of the behavior of GPT, we also conduct a human subject experiment with identical decision tasks, in which 347 human subjects from a representative US sample are randomly assigned to the baseline, price framing, and discrete choice conditions.\footnote{Variations of temperature are inapplicable among human beings, while variations of demographics can be naturally obtained in a representative sample.} We keep the experimental instructions between human subjects and GPT as similar as possible. Appendix \ref{human_experiment} provides the design and instructions of this pre-registered human experiment (AEARCTR-0011750). This experiment was approved by The Institutional Review Board of Finance and Economics Experimental Laboratory in The Wang Yanan Institute of Studies in Economics, Xiamen University (FEEL230701), and all subjects provided informed consent before they started the experiment. Table \ref{tables1} in Appendix \ref{appendix_result} shows the demographic characteristics of our human subjects.

\section{Theoretical Method}

\subsection{Revealed Preference Analysis}

\paragraph*{Generalized Axiom of Revealed Preference.} Consider a DM who selects a bundle $x^{i}\in \mathbb{R}_{+}^{K}$ from a budget line $\{x:p^{i}\cdot x\leq p^{i}\cdot x^{i}$, $p^{i}\in \mathbb{R}_{++}^{k}\}$. A \textit{dataset} $\mathcal{O}={ \left(
p^{i},x^{i}\right)}_{i=1}^{N}$ represents a collection of $N$ decisions made by the DM. We say that a utility function $U:\mathbb{R}_
{+}^{k}\rightarrow \mathbb{R}$ \emph{rationalizes} the dataset $\mathcal{O}
$ if for every bundle $x^{i}$, we have:
\begin{equation*}
U(x^{i})\geq U(x)\text{ for all }x\in \mathbb{R}{+}^{K}\text{ s.t. }%
p^{i}\cdot x\leq p^{i}\cdot x^{i}.
\end{equation*}%
Let $\mathcal{X}=\{x^{i}\}_{i=1}^{N}$ be the set of bundles selected by the DM. We say that $x^{i}$ is directly revealed to be preferred to $x^{j}$, denoted by $x^{i}\succsim ^{\ast }x^{j}$, if the DM chooses $x^{i}$ when $x^{j}\in \mathcal{X}$ is affordable (i.e., $p^{i}\cdot x^{j}\leq p^{i}\cdot x^{i}$). We denote $\succ ^{\ast }$ as the relation of \emph{directly strictly revealed preference}. We denote $\succsim ^{\ast \ast }$ as the transitive closure of $\succsim ^{\ast }$, which refers to the \emph{revealed preferred} relation. 

A utility function is well-behaved if it is continuous, concave and strictly increasing. Afriat's theorem \citep{afriat1967construction,varian1982nonparametric} states that a dataset $\mathcal{O}$ can be rationalized by a well-behaved utility function if and only if the dataset obeys the generalized axiom of revealed preference (GARP):
\begin{equation*}
\text{for all }x^{i}\text{ and }x^{j},\text{ }x^{i}\succsim ^{\ast \ast
}x^{j}\text{ implies }x^{j}\nsucc ^{\ast }x^{i}. \label{garp}
\end{equation*}

Apart from GARP, two closely related notions are the weak axiom of revealed preference (WARP): for all $x^{i}$ and $x^{j}$ in a dataset $\mathcal{O}$, $x^{i}\succsim^{\ast}x^{j}\text{ implies }x^{j}\not\succsim ^{\ast}x^{i}.$, and the strong axiom of revealed preference (SARP): for all $x^{i}$ and $x^{j}$ in a dataset $\mathcal{O}$, $x^{i}\succsim ^{\ast\ast}x^{j}\text{ implies }x^{j} \not\succsim ^{\ast\ast }x^{i}$, which works by exploiting transitivity. In our setting with two goods, checking WARP is equivalent to checking SARP \citep{rose1958consistency}. In our discrete setting, \cite{harbaugh2001garp} shows that a locally non-satiated, strictly monotonic, continuous, and concave utility may violate GARP and demonstrates the need to use the assumption of strong monotonicity (see also \cite{polisson2013revealed} for discussions).

\paragraph*{Rationality Score.} Afriat's theorem provides a powerful tool for analyzing choice behavior. A popular approach for measuring the departure from rationality is the \emph{critical cost efficiency index} (CCEI) proposed by Afriat \citep{afriat1972efficiency}.
A subject has a CCEI $e\in \left[ 0,1\right] $ if $e$ is the largest number with a well-behaved $U$ that rationalizes the data set for every $%
x^{i}\in \mathcal{X}:$ 
\begin{equation*}
U(x^{i})\geq U(x)\text{ for all }x\in \mathbb{R}_{+}^{K}\text{ s.t. }%
p^{i}\cdot x\leq e \cdot p^{i}\cdot x^{i}.  
\end{equation*}%
A CCEI of 1 indicates passing GARP perfectly. A CCEI less than 1---say, 0.95---indicates that there is a utility function for which the chosen bundle $x^{i}$ is preferred to any bundle that is cheaper than $x^{i}$ for more than 5\%. Put differently, the CCEI can be viewed as the amount by which a budget constraint must be relaxed in order to remove all violations of GARP, because the DM can achieve her utility targets by spending less money \citep{afriat1972efficiency,varian1990goodness}. We compute CCEI to obtain a score of rationality for each domain with 25 decisions. 

In the revealed preference literature, there are several other indices to score rationality (departure from GARP). These indices include the Houtman-Maks index (HMI) \citep{houtman1985determining}, money pump index (MPI) \citep{echenique2011money}, and minimum cost index (MCI) \citep{dean2016measuring}. We also compute these indices and report the results as robustness checks.

\subsection{Structural Estimation for Preferences}

\paragraph*{}
\noindent 
In addition to rationality score, we further examine the underlying preferences using structural estimation.
\paragraph*{Risk and Time Preferences Estimation.} 
In the domain of risk preference, suppose that the DM chooses the contingent security $(x_A,x_B)$, we denote $x_1=\max\{x_A,x_B\}$ as the high outcome and $x_2=\min\{x_A,x_B\}$ as the low outcome. In the domain of time preference, suppose that the DM chooses the payment schedule $(x_A,x_B)$, we denote $x_1=x_A$ as the payment for today and $x_2=x_B$ as the payment for one month later. For these two domains, we assume that the underlying utility function is given by
\begin{equation*}
    U(x_1,x_2)=\alpha u(x_1)+(1-\alpha)u(x_2)
\end{equation*}
where the utility function $u(z)= \begin{cases} \frac{1}{\rho}z^\rho,\rho\le 1(\rho \ne 0)\\ \ln(z), \rho=0 \end{cases}$ and $\alpha \in [0,1]$. For risk preference, $\alpha$ captures the decision weight placed on the better outcome \citep{gul1991theory,halevy2018parametric}. When $\alpha=0.5$, we have a standard expected utility function and when $\alpha>0.5$ ($\alpha<0.5$), the better outcome is over(under)-weighted relative to the objective probability of 0.5. The parameter $\rho$ captures risk attitude with the parameter $\theta=1-\rho$ being the Arrow-Pratt measure of relative risk aversion.\footnote{In our budget set, there is no difference between $\rho>1$ and $\rho=1$, because the DM will choose corner solutions when $\rho\ge 1$. Therefore, our estimation is conditional on $\rho\le 1$ in all the four preferences domains. \textit{SI appendix} provides further details about the estimation of corner solutions.} For time preference, $\alpha$ captures the weight placed on the payment today \citep{andreoni2012estimating}. When $\alpha>0.5$ ($\alpha<0.5$), it corresponds to positive (negative) time preference. The parameter $\rho$ is the curvature of the period function. When $\rho=1$, the DM allocates all expenditure to the time period with lower price and as $\rho$ decreases, the DM is more desired to smooth payments across periods.

\paragraph*{Social and Food Preferences Estimation.} 
Regarding social preference, suppose that the DM chooses the allocation $(x_A,x_B)$, we denote $x_1=x_A$ as the payment for self and $x_2=x_B$ as the payment for the other. In the domain of food preference, assuming that the DM chooses the bundle $(x_A,x_B)$, we denote $x_1=x_A$ as the consumption of meat and $x_2=x_B$ as the consumption of tomatoes. Moreover, we assume that the underlying utility function is a member of the CES family and is given by
\begin{equation*}
    U(x_1,x_2)=[\alpha x_1^\rho +(1-\alpha)x_2^\rho]^{\frac{1}{\rho}}
\end{equation*}
where $\rho \leq 1$ and $\alpha \in [0,1]$. For social preference, the parameter $\alpha$ captures the weight placed on the self's payment relative to the other's payment. $\alpha=1$ implies pure selfishness, $\alpha=0.5$ indicates fair-mindedness, and $\alpha=0$ refers to pure altruistic \citep{andreoni2002giving,fisman2007individual}. $\rho$ represents the curvature of the indifference curves, which measures equality efficiency orientation. $\rho=1$ indicates that the two payments are perfectly substitute with $U(x_1,x_2)=\alpha x_1+(1-\alpha) x_2$, which means that the DM is efficiency orientated. When $\rho \rightarrow 0$, the utility function approaches the Cobb--Douglas utility function, and shares of expenditures to self and to the other are constant. When $\rho \rightarrow -\infty$, it approaches to the Leontief utility function $\min \{ \alpha x_1,(1-\alpha) x_2 \}$, which implies that the two payments are perfectly complemented and the DM is equality orientated \citep{andreoni2002giving,fisman2007individual}. In a similar vein, the parameter $\alpha$ in the food preference domain captures the weight placed on meat relative to tomatoes and the parameter $\rho$ represents the curvature of the indifference curves as that for social preference. We provide further details on estimation methods in Appendix \ref{struc}.

\section{Results}

In this section, we first present the results from the baseline condition, then report whether and how the results change with the variations in the decision tasks.

\subsection{Results from the Baseline Condition}

\begin{figure}[H]
    \centering

    \includegraphics[width=\linewidth]{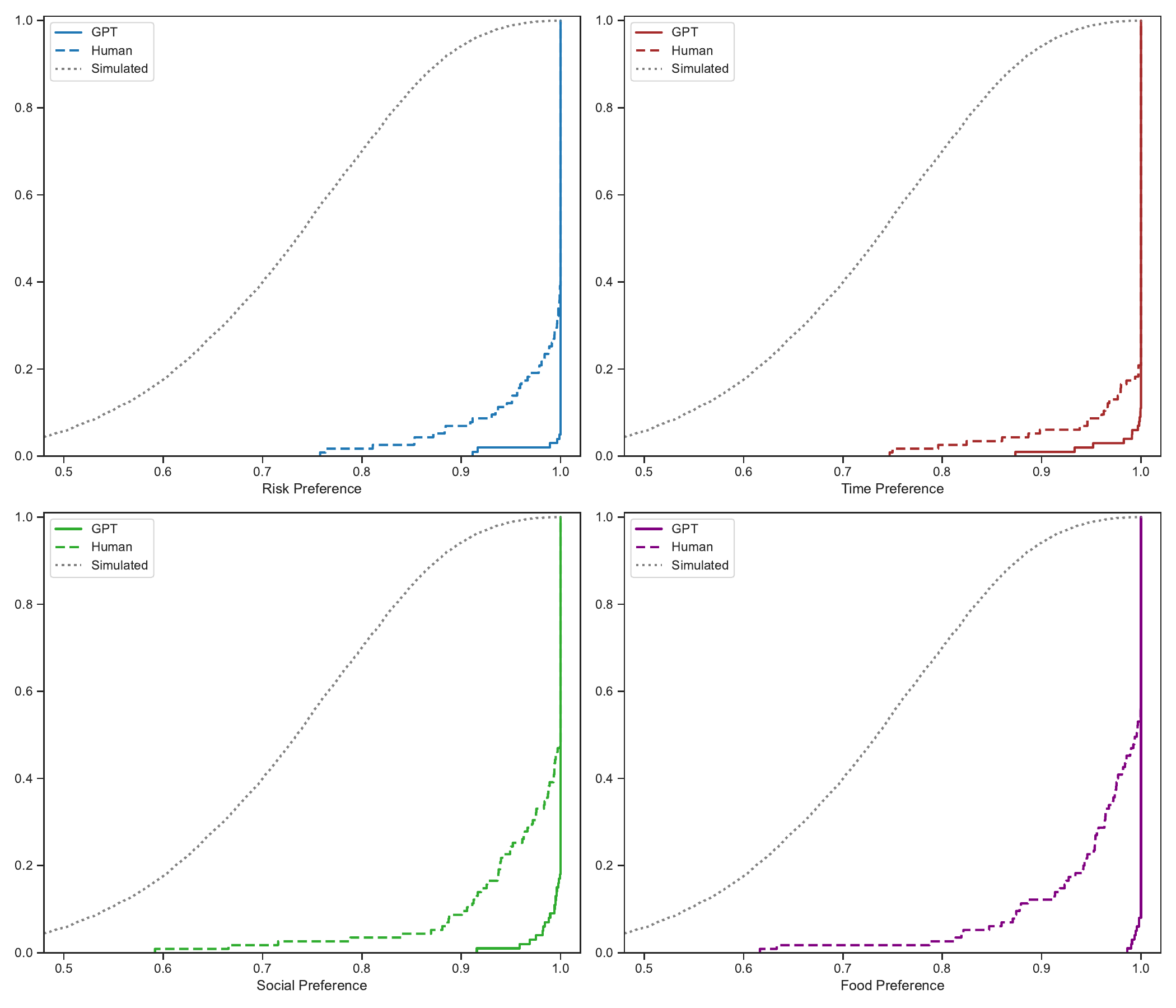}
    \caption{Cumulative Distributions of the CCEI Values. This figure consists of four subplots for four preferences domains. Each subplot depicts a cumulative distribution function (CDF) plot, which shows the proportion of CCEI values less than or equal to a specific threshold. The light dotted lines represent simulated subjects, the dark dashed lines represent human subjects, and the solid lines represent GPT observations.}\label{fig1}
\end{figure}

\paragraph*{Rationality Score.} Figure \ref{fig1} presents the cumulative distributions of CCEI---the rationality score---for each of the four preferences domains. We find that 95, 89, 81, and 92 out of 100 GPT observations for risk, time, social, and food preferences exhibit no violations of GARP; that is, CCEI equals to 1. The average CCEI is 0.998, 0.997, 0.997, and 0.999 for risk, time, social, and food preferences, respectively. Meanwhile, in our human experiment, the average CCEI among human subjects is 0.980, 0.985, 0.967, and 0.963 for risk, time, social, and food preferences, respectively. Figure \ref{fig1} displays a consistent trend that GPT outperforms human subjects in terms of rationality. In each of the four preferences domains, GPT obtains higher CCEI than human subjects ($p<0.01$, two-sided two-sample t-tests). In addition, we summarize studies in the revealed preference literature. Figure \ref{figs1} in Appendix \ref{appendix_result} plots CCEI values documented in prior studies, which range from 0.81 to 0.99 with an average of 0.918. Consistently, we find that CCEI of GPT also surpasses those of human subjects in all domains ($p<0.01$, two-sided one-sample t-tests).

To confirm that our chosen parameters have sufficient power to measure rationality, we adopt the test proposed by Bronars \citep{bronars1987power} as a benchmark, in which we generate simulated subjects by uniformly drawing random allocations along each of the budget lines and examine their rationality. We find that 99.9\% of simulated subjects violate GARP. Figure \ref{fig1} shows the cumulative distributions of CCEI of simulated subjects, which are lower than both GPT observations and human subjects. We also conduct the power analysis using the predictive success \citep{beatty2011demanding}, the Selten score \citep{dean2016measuring}, as well as bootstrapping from the sample of subjects \citep{andreoni2002giving}. We show that the chosen parameters have the power to detect rationality violations, in support of the empirical validity of our study (see Appendix \ref{power} for more information).

In addition to CCEI, we calculate other indices to measure rationality including the Houtman-Maks index (HMI) \citep{houtman1985determining}, money pump index (MPI) \citep{echenique2011money}, and minimum cost index (MCI) \citep{dean2016measuring}, and construct cumulative distribution plots for each index of GPT observations, human subjects, and simulated subjects in Figure \ref{figs2}-\ref{figs5} in Appendix \ref{appendix_result}. Consistent with the observations based on CCEI, results from these indices show that GPT observations exhibit a high level of rationality across the four preferences domains and surpass those of human subjects across all domains ($p<0.1$, two-sided two-sample t-tests).

\paragraph*{Downward-sloping Demand.}
While GPT exhibits a high level of rationality, it is possible that its decisions are simply clustered at the corners or in certain areas. To address such concern, we examine whether GPT behavior respects the property of downward-sloping demand, a fundamental principle in the analysis of consumer behavior whereby the demand for a commodity decreases with its price \citep{choi2007consistency,fisman2007individual,echenique2021approximate}. 

\begin{figure}[H]

    \centering

      \includegraphics[width=\linewidth]{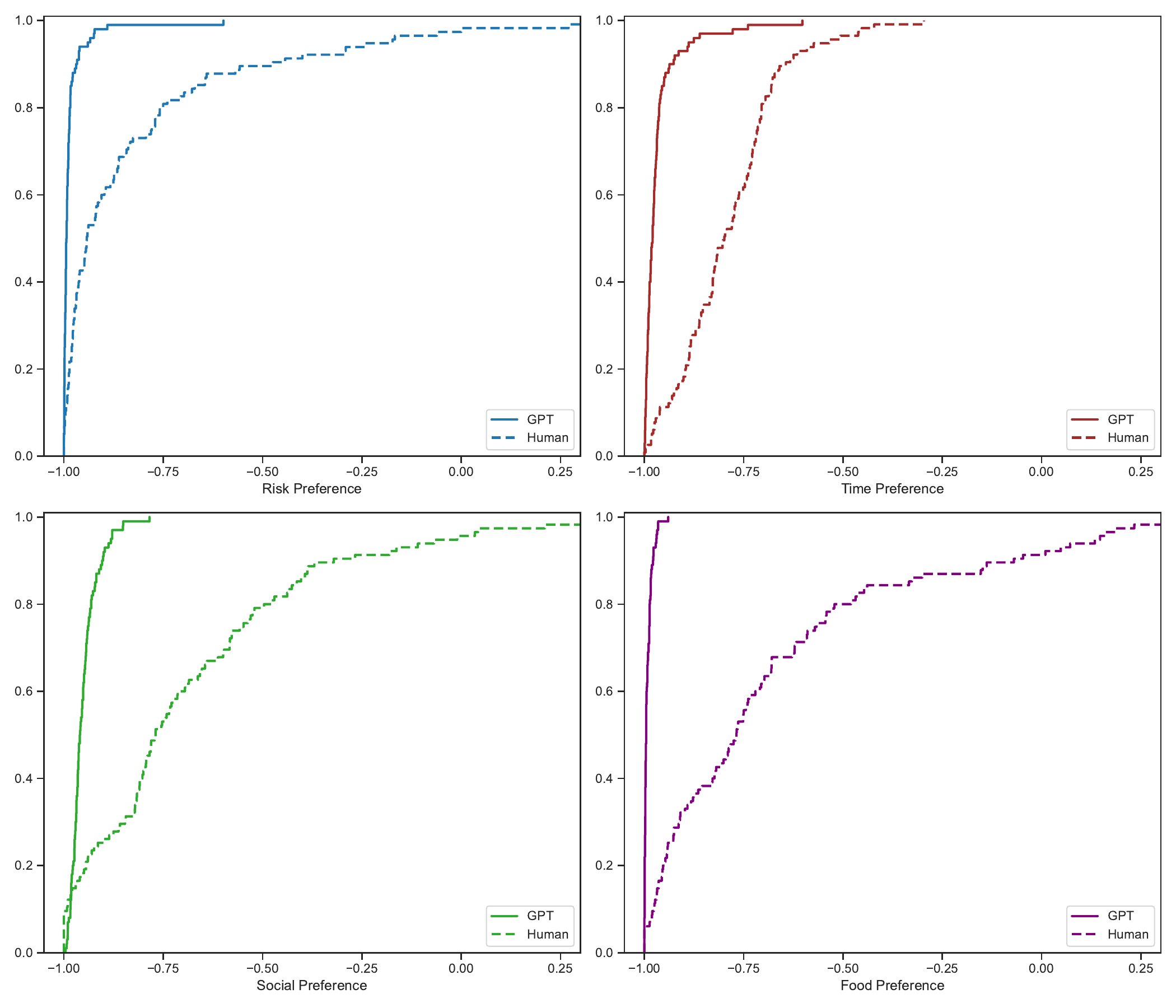}
    \caption{Cumulative Distributions of the Spearman's Correlation Coefficient of $\ln(x_A/x_B)$ and $\ln(p_A/p_B)$. This figure contains four subplots for four preferences domains. The dashed (solid) lines represent human subjects (GPT observations).}
\label{fig2}
\end{figure}

We measure the degree of compliance with downward-sloping demand for GPT observations and human subjects. This principle requires that when the relative price of a commodity increases, the consumer should not increase its consumption. More specifically, we measure whether each DM's decisions respect this principle by calculating the Spearman's correlation coefficient of $\ln(x_A/x_B)$ and $\ln(p_A/p_B)$ \citep{echenique2021approximate}. A negative correlation indicates an appropriate response to price fluctuations, and zero or positive correlation indicates no respond or irregular response to price changes. Note that $\ln(x_A/x_B)$ is not defined in the corners. We adjust corner choices by a small constant, 0.1\% of the budget, in each choice \citep{echenique2021approximate}. We plot the cumulative distribution of the Spearman's correlation coefficients of $\ln(x_A/x_B)$ and $\ln(p_A/p_B)$ as a proxy for the degree of downward-sloping demand for each of the four preferences domains in Figure \ref{fig2}.

For GPT observations, the coefficients for risk, time, social, and food preferences have a mean of -0.984, -0.966, -0.951, and -0.992, while these are -0.826, -0.788, -0.681, and -0.673 for human subjects, respectively. Overall, GPT is more responsive to price changes than human subjects in each preference domain ($p< 0.01$, two-sided two-sample t-tests). Figure \ref{fig2} further illustrates that GPT observations always have a negative Spearman's correlation coefficients, while human has a lower proportion having a negative Spearman's correlation coefficients (96.1\% on average). This strengthens our findings based on the rationality score and suggests that GPT is more capable of making reasonable responses to the changes in prices than human subjects.

In addition, for each GPT observation, Figure \ref{figs6}-\ref{figs9} in Appendix \ref{appendix_result} provide comprehensive visual representations by showing scatter diagrams and fitted lines of the shares of quantities $x_A/(x_A+x_B)$ and the log-price ratio $\ln(p_A/p_B)$.

\paragraph*{Preference Estimation.}
Since choices of GPT and human subjects are mostly consistent with well-behaved utility functions, we proceed to estimate the underlying risk, time, social, and food preferences.\footnote{We omit GPT or human individuals with CCEI score below 0.95 \citep{varian1990goodness}.} In total, we have eight estimated parameters: decision weight of the better outcome ($\alpha_r$) and utility curvature ($\rho_r$) for risk preference, weight of today ($\alpha_t$) and utility curvature ($\rho_t$) for time preference, weight for self's payment ($\alpha_s$) and utility curvature ($\rho_s$) for social preference, weight for meat ($\alpha_f$) and utility curvature ($\rho_f$) for food preference. We first estimate the preference parameters at the aggregate level by pooling all responses of GPT observations and human subjects, respectively (Table \ref{tables2} in Appendix \ref{appendix_result}). 
\begin{figure}[H]
    \centering
     \includegraphics[width=\linewidth]{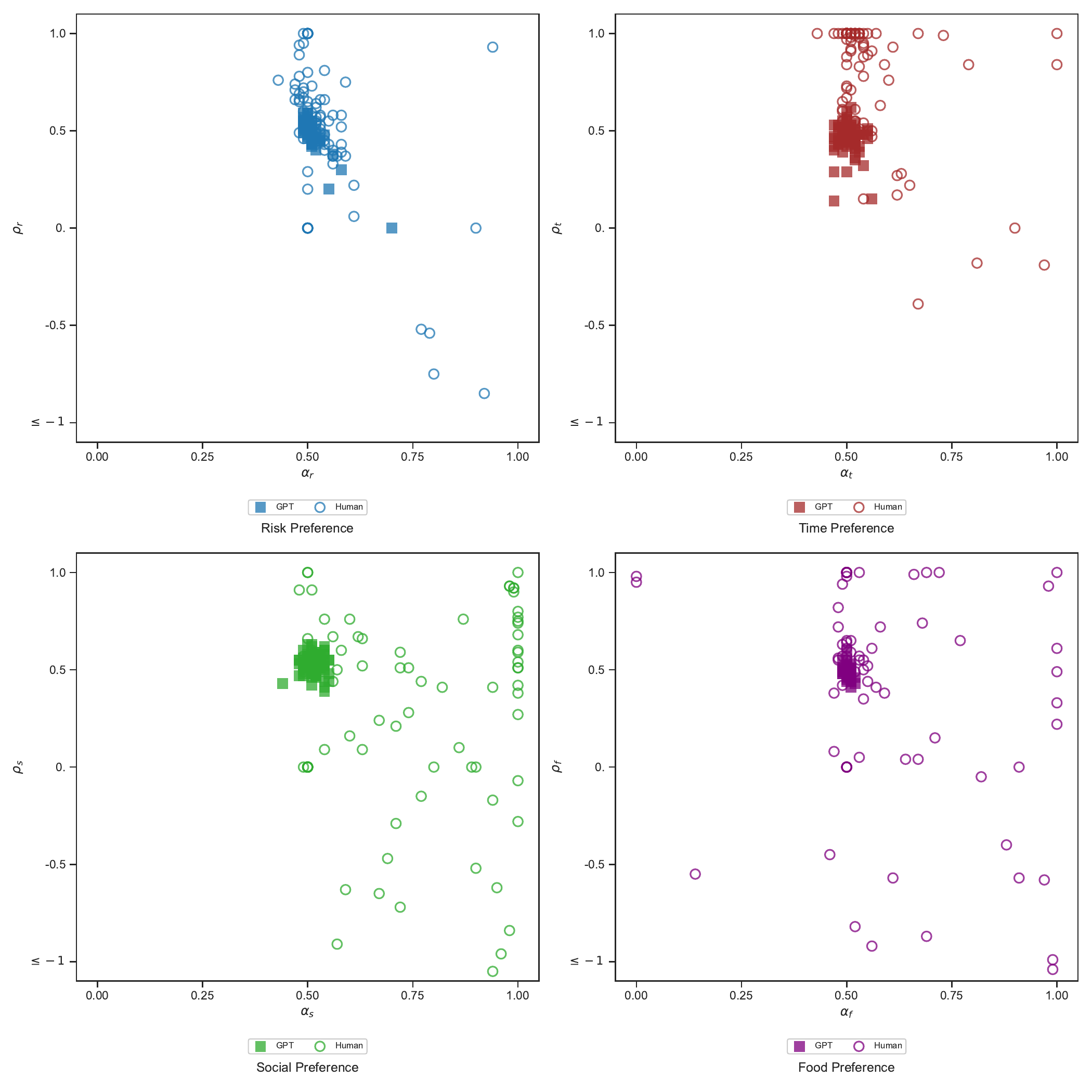}

    \caption{Scatter Plots of Estimated Parameters. This figure contains four subplots for four preferences domains. Each hollow circle (solid square) points represent a human subject (a GPT observation).}
    \label{fig3}

\end{figure}

Results show that, compared to human subjects, GPT is closer to an expect-utility maximizer ($\alpha_r$: 0.618 vs. 0.508 for Human vs. GPT) and has a more linear utility curve ($\rho_r$: 0.335 vs. 0.488) in risk preference; is more patient ($\alpha_t$: 0.513 vs. 0.504) and has a less linear utility curve ($\rho_t$: 0.981 vs. 0.466) in time preference; is more other-regarding ($\alpha_s$: 0.735 vs. 0.512) and more efficiency-orientated ($\rho_s$: 0.330 vs. 0.520) in social preference, and is less fond of meat ($\alpha_f$: 0.583 vs. 0.501) and more efficiency-orientated ($\rho_f$: 0.386 vs. 0.491) in food preference. Similar patterns can be observed in the individual-level estimations, in which we estimate preference parameters for each GPT decision maker and human subject, as shown in Figure \ref{figs3} and Table \ref{tables3} in Appendix \ref{appendix_result}. Moreover, the scatter plots of human subjects are more dispersed, which suggests a significantly higher level of preference heterogeneity among human subjects than GPT observations.

\subsection{Results from the Conditions with Variations}

\begin{figure}[H]
    \centering
     \includegraphics[width=\linewidth]{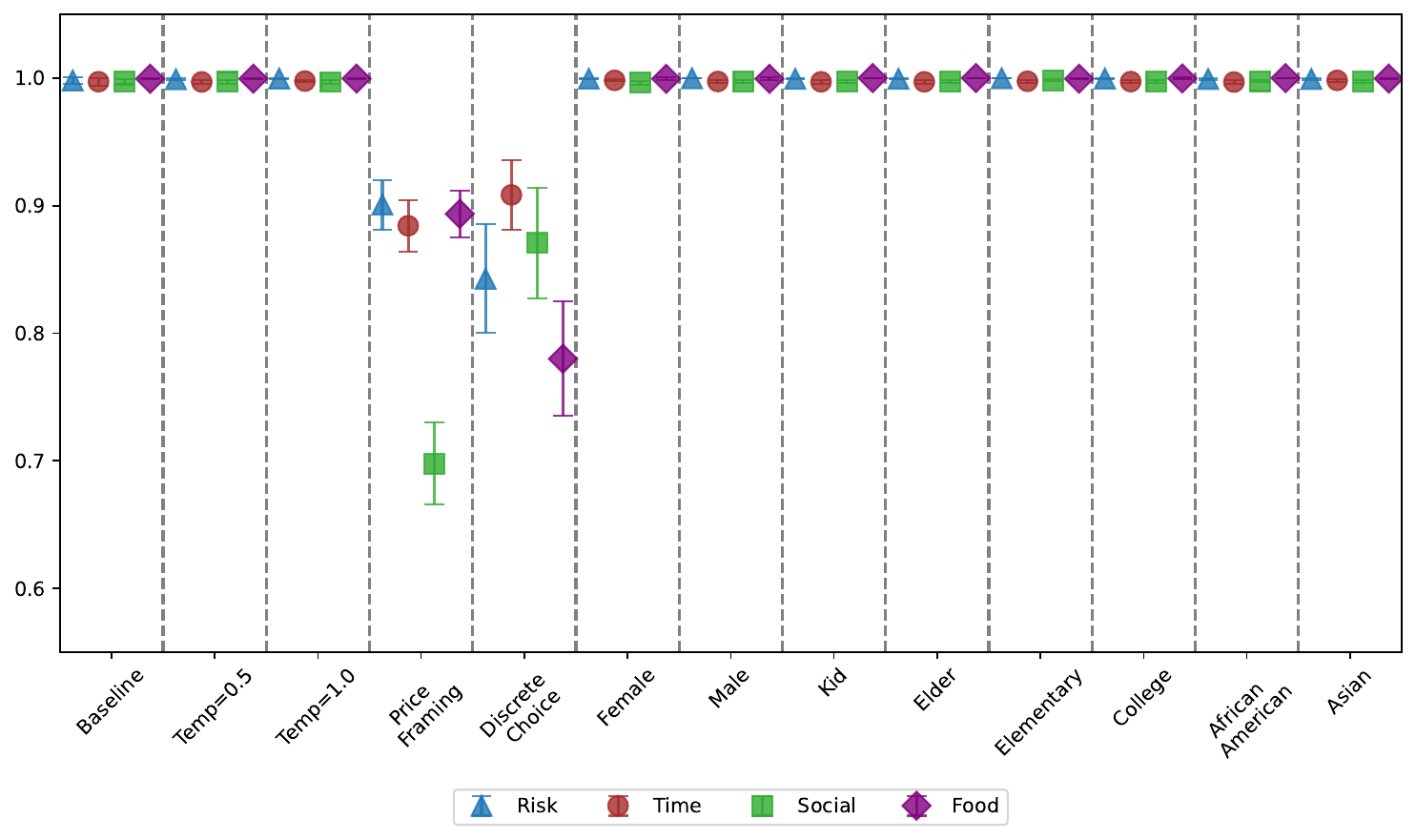}
    \caption{Mean CCEI Values of GPT Observations across Different Variations. This figure displays the average CCEI values and 95\% confidence intervals for GPT observations under different conditions: baseline, temperature of 0.5, temperature of 1, price framing, and discrete choices, and various demographic settings.}
    \label{fig4}
\end{figure}

\paragraph*{}
\noindent We examine variations in the temperature, decision tasks, and demographic information. Figure \ref{fig4} presents the mean CCEI values and 95\% confidence intervals across variations, and  Figure \ref{figs24} in Appendix \ref{appendix_result} shows the mean Spearman's correlation coefficients of $\ln(x_A/x_B)$ and $\ln(p_A/p_B)$ and their 95\% confidence intervals. We report these results in detail below.

\paragraph*{Insensitive to Variations in Temperature.}
When the temperature increases from 0 to 0.5 and 1, there is a higher number of invalid responses, namely, GPT does not provide an answer to the specified question (invalid response rate is 4.7\% for temperature of 0.5 and 9.8\% for temperature of 1). Therefore, we analyze the data conditional on those providing valid answers. We find that as the randomness increases, the level of rationality is similar to that in the baseline condition (Figure \ref{fig4}). For each temperature and each preference domain, we plot the cumulative distributions of the CCEI values of GPT observations and simulated subjects for Bronars' test in Figure \ref{figs10} in Appendix \ref{appendix_result} and the cumulative distributions of the Spearman's correlation coefficients in Figure \ref{figs11} in Appendix \ref{appendix_result}. These findings suggest that randomness increases the stochasticity and creativity in language presentations of GPT, but not the rationality score.

There are no significant differences for the estimated Spearman's correlation coefficients of $\ln(x_A/x_B)$ and $\ln(p_A/p_B)$ between the baseline and the higher temperature (Figure \ref{figs24} in Appendix \ref{appendix_result}) at the 10\% level (two-sided two-sample t-tests). Similarly, the mean of estimated preference parameters are statistically indifferent to changes in temperature. However, the standard deviations of some parameters increase with temperature ($\rho_r$, $\rho_t$, $\rho_s$, $\rho_f$: $p<0.01$, two-sample Levene tests), which suggests that high temperature may generate greater heterogeneity in the behavior of GPT.

\paragraph*{Sensitive to Variation in the Decision Tasks.} 
First, we compare the baseline and the price framing conditions. Changing the price framing significantly reduces GPT's rationality level in all four tasks (Figure \ref{fig4}). Remarkably, the average CCEI for risk preference declines to 0.901, with 34\% exhibiting a CCEI below 0.9. These values are 0.884 (48\%), 0.698 (88\%), and 0.894 (49\%) for time, social, and food preferences, respectively.\footnote{Given the low level of rationality exhibited by GPT in the price framing condition, we have difficulty in determining that GPT's decisions are consistent with a well-behaved utility function. Therefore, we refrain from adopting the preference estimation approach under this condition \citep{varian1990goodness}. The situation is identical in the discrete choice condition as described below.} In each preference domain, CCEI values are significantly higher in the baseline condition than in the price framing condition ($p<0.01$, two-sided two-sample t-tests). Moreover, the downward-sloping demand property is impaired under the alternative price framing, with the key Spearman's correlation coefficients being -0.053, -0.116, 0.267, and -0.499 for risk, time, social, and food preferences, respectively (Figure \ref{figs24} in Appendix \ref{appendix_result}). Figures \ref{figs13}-\ref{figs16} in Appendix \ref{appendix_result} show the disordered responses of GPT observations to price changes in the price framing condition, which appear to be flatter compared to those in the baseline condition.

In Figure \ref{figs17} (Figure \ref{figs18}) in Appendix \ref{appendix_result}, we display the CDFs of CCEI (Spearman's correlation coefficients) in the four conditions: baseline and price framing conditions in both the GPT experiment and the human experiment. We find that the alternative price framing also reduces the rationality level and the downward-sloping demand property in the human subjects experiment ($p<0.05$, two-sided two-sample t-tests in risk and time preferences). However, the figures suggest that these reductions are larger in the GPT experiment than in the human experiment, which is further verified in OLS regression analyses (Table \ref{tables4} in Appendix \ref{appendix_result}).

Second, we compare the baseline and the discrete choice conditions. When we present GPT with a set of 11 options, we also observe a decrease in rationality levels for discrete choices of GPT observations for all four tasks in Figure \ref{fig4} (risk: 0.998 vs. 0.843, $p<0.01$; time: 0.997 vs. 0.908, $p<0.01$; social: 0.997 vs. 0.871, $p<0.01$;  food: 0.999 vs. 0.780, $p<0.01$, two-sided two-sample t-tests). Additionally, 51\%, 32\%, 33\%, and 55\% of GPT observations demonstrate a CCEI below 0.9 in risk, time, social, and food preferences, respectively. Figure \ref{figs19}-\ref{figs22} in Appendix \ref{appendix_result} show the demand curves of GPT observations, which exhibit significantly more corner solutions. Consistently, the Spearman's correlation coefficients are -0.589, -0.497, -0.519, and -0.533 for risk, time, social, and food preferences (Figure \ref{figs24} in Appendix \ref{appendix_result}; $p<0.01$ when compared to the baseline condition, two-sided two-sample t-tests). These suggest that GPT is less responsive to price changes under discrete choices than continuous choices.

Figure \ref{figs23} (Figure \ref{figs24}) in Appendix \ref{appendix_result} shows the CDFs of the CCEI (Spearman's correlation coefficients) in baseline and discrete choice conditions in the GPT experiment and human experiment. Human subjects' rationality level and the downward-sloping demand property reduce in the discrete setting, compared to the baseline condition ($p<0.05$, two-sided two-sample t-tests in risk and time preferences). As shown in the figures, these reductions are larger in the GPT experiment than in the human experiment. We also verify this observation through OLS regression analyses (Table \ref{tables4} in Appendix \ref{appendix_result}). These results suggest that GPT's decision-making is more significantly affected by both the framing of prices and discrete choices than human subjects.

\paragraph*{Insensitive to Demographic Information.} Comparing the baseline condition and variations in demographics in the GPT experiment, we find that CCEI values, Spearman's correlation coefficients of $\ln(x_A/x_B)$ and $\ln(p_A/p_B)$, and estimated preference parameters are all insensitive to variations of demographic factors embedded in the prompts to request responses from GPT (Figure \ref{fig4}, Figure \ref{figs24}, Tables \ref{tables2}-\ref{tables3} in Appendix \ref{appendix_result}).

These are in contrast to results of our human experiment (Tables \ref{tables5}-\ref{tables6} in Appendix \ref{appendix_result}) and prior studies where rationality score and preference have been shown to differ across demographic groups \citep{choi2014more,echenique2011money,von2011heterogeneity}. The fact that GPT's decision-making process remains consistent across demographic
variables suggests that GPT does not exhibit algorithmic bias in terms of decision-making quality, which provides a measure of reassurance regarding its fairness and consistency across diverse user groups.

\section{Discussion}
We conduct the first study to assess the rationality of GPT, a popular large language model, using revealed preference analysis. Our findings demonstrate that GPT is able to display a high level of rationality in decision-making related to risk, time, social, and food preferences. We also observe that increasing the randomness of GPT does not significantly impact its performance. Furthermore, our analysis reveals that the level of rationality of GPT remains constant across different demographic characteristics, which indicates that it does not exhibit an algorithm bias. However, we observe a significant drop in rationality when we use a less standard presentation of prices or change the choice set from continuous to discrete. This suggests that GPT may have limitations in terms of sensitivity to contexts and frames.

Our study contributes to the ongoing discussions of the performance of GPT in various domains; these include reasoning, logic, math, language processing, and identifying factual errors \citep{borji2023categorical}. In addition to cognitive techniques and practical skills, some researchers have explored whether GPT can exhibit human-like decision-making abilities or perceive others' thoughts \citep{horton2023large,kosinski2023theory}. Our study adds to these parallel studies by subjecting GPT to traditional decision-making tasks and employing a set of measures to systematically describe its behavior. Our work aligns with recent calls to study machine behavior to ``reap their benefits and minimize their harms'' \citep{rahwan2022machine}. By providing insights into GPT's decision-making capacity, we can better understand how to optimize its performance and address potential limitations.

Our study is situated within the growing literature on AI-based decision support tools. Many researchers have explored the usefulness of leveraging AI in various decision-making domains, such as bail decisions \citep{kleinberg2018human}; clinical diagnosis \citep{mullainathan2022solving}; work arrangements \citep{kawaguchi2021description}; stock price forecasts \citep{lopez2023can}; job recruitment \citep{horton2017effects}; product or content consumption \citep{adomavicius2018effects,agrawal2022personalized}; and mathematics development \citep{davies2021advancing}. Unlike these algorithms, which require data input and training, GPT is a language-based model that provides a direct question-and-answer service for normal users. Given its high level of rationality in decision-making across various domains, our study proves the potential of GPT as a general AI-based decision-support tool. The user-friendly interface and versatility of GPT render it a promising option for individuals and organizations seeking easy-to-use AI-based advice.

Our paper makes contributions to the literature on rationality and experimental methods. First, we demonstrate the effectiveness of experimental economics methods in studying choice behavior of artificial intelligence \citep{rahwan2022machine}, which adds earlier studies of children \citep{brocas2019development,harbaugh2001garp}, monkeys \citep{chen2006basic}, rats, and pigeons \citep{kagel1975experimental}. Second, our work highlights the potential of large language models like GPT to streamline experimental research and yield new data and insights \citep{horton2023large}. Finally, studying the choice behavior of artificial intelligence can provide an important benchmark for understanding natural intelligence. For example, our understanding of how LLMs make decisions could help reveal general principles that govern both language intelligence and decision intelligence \citep{sejnowski2023large}. By synthesizing insights from these various domains, our paper offers a novel perspective on the nature of rationality and broadens the methods that can be used to study it.

As an initial assessment of the economic rationality of GPT, our study has several limitations. First, our study examines the choice behavior of GPT but does not explore the mechanisms that underlie our observations. For example, we find that GPT responses are highly sensitive to contexts and frames. This may be due to the reflection of biases presented in the existing data \citep{schramowski2022large,caliskan2017semantics},the insufficient training of texts of the alternative contexts and frames \citep{chen2021evaluating,drori2022neural}, or the tendency for LLMs to exploit spurious correlations or statistical irregularities in the data set under dissimilar tasks \citep{mccoy2019right}. In particular, \cite{mckenna2023sources} suggests that a significant source of LLMs bias originates from a corpus-based heuristic using the relative frequencies of words. The ``50-50 split'' or ``equal split'' are high-frequency texts in allocation settings, and GPT can adapt this corpus-based heuristic and exhibit the tendency to choose the midpoint under an ``unfamiliar'' task with the alternative price framework. Similarly, ``all or nothing'' can be high-frequency texts under the presentation of options context, so GPT exhibits the tendency to choose the first or last option under an ``unfamiliar'' discrete choice condition. Recent studies have documented similar patterns in different environments \citep{binz2023using,brand2023using,jones2022capturing, brookins2023playing,horton2023large}. 

In addition, our study reveals that demographic factors do not significantly impact GPT's rationality or estimated preference parameters. This contrasts with the majority of empirical literature, including our human subject experiment, where demographic factors often play a  significant role. The lack of responsiveness to demographics aligns with the concept of hyper-accuracy distortion \citep{aher2023using}, which refers to the distortion resulting from the extensive efforts to align LLMs with human ethics such as the censorship of demographic information to reduce and prevent problematic outputs. In conclusion, with some speculative conjectures, we leave it to future studies to explore the mechanisms that underlie GPT's choice behavior and open the black-box of this technology.

Second, we focus on economic rationality as defined by revealed preference analysis, whereas rationality is often defined more broadly in the literature to include various decision rules and heuristics \citep{kahneman2003maps,simon1979rational,thaler2016behavioral}. Third, we use a simple experimental environment with only two commodities to present budgetary decisions. However, studying rationality in more realistic settings, such as shopping behavior in a supermarket and portfolio choices in the financial markets would be more challenging yet important. Our study shows that economic rationality can emerge in GPT when decision contexts are simple and framed in specific ways. Future research is needed to investigate the broader applications of artificial intelligent agents as they continue to evolve.

\bibliographystyle{jss2}
\bibliography{main}

\newpage
\appendix
\renewcommand\theHtable{Appendix.\thetable}
\setcounter{table}{0}
\renewcommand{\thetable}{A\arabic{table}}
\setcounter{figure}{0}
\renewcommand{\thefigure}{A\arabic{figure}}
\begin{center}
    \LARGE{\textbf{Online Appendices}}
\end{center}
\section{Method: GPT Experiment.}
\label{prompt_gpt}

In the section, we display the method to prompt GPT to make decisions. In \textit{GPT-3.5-turbo}, the input prompt must be an array of message objects, each with a role (``system'', ``assistant'', or ``user'') and content (the message's content). First, we input the system's message to set the behavior. Next, we input the assistant's message to store information about tasks. We use the previous two messages to establish the background, and finally, we use the user's message to ask understanding testing questions and to give decision tasks.\footnote{To confirm that GPT has truly understood our previous instructions, we first ask questions related to the task itself. Based on GPT's responses, it appears that GPT has understood our instructions. In each preference domain and in each condition, with 25 simulations, we do not encounter any misunderstanding in the testing questions.} In each preference domain, the system's message is fixed, while the assistant's and users' messages change across the baseline, price framing, and discrete choice conditions. 

\subsection{Risk Preference}

\paragraph*{System's Message.} 
I want you to act as a human decision maker. You will be given 25 rounds of decision-making tasks and will be responsible for making decisions. You should use your best judgment to come up with solutions that you like most. My first request is ``You must provide your answers in every round.'' If you do not provide an answer, I will assume that you make a random choice.\footnote{We assign the role of ``human decision maker''. In particular, to prevent the possibility that the model returns answers such as ``As an AI language model, I am not capable of making decisions on my own...'', we impose the request to provide an answer in every round. In practice, we do encounter cases in which the system refuses to provide an answer, and we simply do not include this outcome in the analyses.}

\subsubsection{Baseline Condition}

\paragraph*{Assistant's Message.} 
In every round, the decision maker has 100 points that need to be invested between Asset A and Asset B. The decision maker has a 50\% chance to get the return from Asset A or the other 50\% chance to get the return from Asset B. First please only tell me the number of points for investing Asset A, then please only tell me the number of points for investing Asset B.

\paragraph*{User's Message: Understanding Testing Question 1.} What is the probability you will get a return from Asset A?
\begin{itemize}
    \item Example of GPT's responses: \textit{``The probability of getting a return from asset A is 50\%, and the probability of getting a return from asset B is also 50\%.''}
\end{itemize}

\paragraph*{User's Message: Understanding Testing Question 2.} Suppose that you invest 90 points to Asset A, and 10 points to Asset B. In this round, investing every 1 point for Asset A returns 0.8 dollars, and investing every 1 point for Asset B returns 0.2 dollars. What return will you get?
\begin{itemize}
    \item Example of GPT's responses: \textit{``The total return from Asset A would be 90$\times$0.8 = 72 dollars, while the total return from Asset B would be 10$\times$0.2 = 2 dollars. Since there is a 50\% chance of getting the return from Asset A, and a 50\% chance of getting the return from Asset B, the expected return would be (0.5$\times$72) + (0.5$\times$2) = 37. Therefore, the expected return would be 37 dollars.''}
\end{itemize}

\paragraph*{User's Message: Understanding Testing Question 3.} Suppose that you invest 90 points to Asset A, and 10 points to Asset B. In this round, investing every 1 point for Asset A returns 0.8 dollars, and investing every 1 point for Asset B returns 0.2 dollars. Is there a chance to get 72 dollars?
\begin{itemize}
    \item Example of GPT's responses: \textit{``Yes, there is a chance to get 72 dollars. Investing 90 points in Asset A would give a return of 72 dollars since investing every 1 point in Asset A returns 0.8 dollars. Investing 10 points in Asset B would give a return of 2 dollars since investing every 1 point in Asset B returns 0.2 dollars. So, there is a 50\% chance that the return will come from Asset A and a 50\% chance that it will come from Asset B. So, there is a chance that the decision maker will get a return of 72 dollars.''}
\end{itemize}

\paragraph*{User's Message: Decision Task.} In this round, investing every 1 point for Asset A returns M dollars, and investing every 1 point for Asset B returns N dollars. What is your allocation?\footnote{We repeat the text of the decision task 25 times, where the parameters used in the text differ across rounds, that is $\text{M, N} \in [0.1, 1] \text{ and} \max\{\text{M, N}\}\ge 0.5$. We keep two decimals. To maximize the replicability of our findings and minimize variance in the model's responses, we send the 25 rounds of questions to the model continuously, without interruption. This method is applicable for the remaining three preference domains under the baseline condition.}

\subsubsection{Price Framing Condition}

\paragraph*{Assistant's Message.} 
In every round, the decision maker has 100 points that need to be invested between Asset A and Asset B. The decision maker has a 50\% chance to get the return from Asset A or the other 50\% chance to get the return from Asset B. First please only tell me the number of points for investing Asset A, then please only tell me the number of points for investing Asset B.

\paragraph*{User's Message: Understanding Testing Question 1.} What is the probability you will get a return from Asset A?
\begin{itemize}
    \item Example of GPT's responses: \textit{``The probability of getting a return from asset A is 50\%''}
\end{itemize}

\paragraph*{User's Message: Understanding Testing Question 2.} Suppose that you invest 90 points to Asset A, and 10 points to Asset B. In this round, investing every 1.25 points for Asset A returns 1 dollar, and investing every 5 points for Asset B returns 1 dollar. What return will you get?
\begin{itemize}
    \item Example of GPT's responses: \textit{``To determine the return, we need to calculate the total return for Asset A and Asset B separately. For Asset A, if investing 1.25 points returns 1 dollar, then investing 90 points would result in a return of 90/1.25=72 dollars. For Asset B, if investing 5 points returns 1 dollar, then investing 10 points would result in a return of 10/5=2 dollars. Therefore, the decision maker will receive a return from either Asset A or Asset B with 50\% probability each. The return would be 72 dollars for Asset A or 2 dollars for Asset B.''}
\end{itemize}

\paragraph*{User's Message: Understanding Testing Question 3.} Suppose that you invest 90 points to Asset A, and 10 points to Asset B. In this round, investing every 1.25 points for Asset A returns 1 dollar, and investing every 5 points for Asset B returns 1 dollar. Is there a chance to get 72 dollars?
\begin{itemize}
    \item Example of GPT's responses: \textit{``In this round, since you have invested 90 points in Asset A and 10 points in Asset B, your total investment is 100 points. For Asset A, every 1.25 points invested returns 1 dollar. So, with 90 points invested, you will get a return of 90/1.25=72 dollars from Asset A. For Asset B, every 5 points invested returns 1 dollar. With 10 points invested, you will get a return of 10/5=2 dollars from Asset B. Since you have a 50\% chance of getting the return from either Asset A or Asset B, I will receive \$72 from Asset A with 50\% probability.''}
\end{itemize}

\paragraph*{User's Message: Decision Task.} In this round, investing every 1/M points for Asset A returns 1 dollar, and investing every 1/N points for Asset B returns 1 dollar. What is your allocation? \footnote{We repeat the text of the decision task 25 times, where the parameters used in the text differ across rounds, that is $\text{M, N} \in [0.1, 1] \text{ and} \max\{\text{M, N}\}\ge 0.5$. We display the numbers 1/M and 1/N directly, and we keep two decimals. To maximize the replicability of our findings and minimize variance in the model's responses, we send the 25 rounds of questions to the model continuously, without interruption. This method is applicable for the remaining three preference domains under the price framing condition.}

\subsubsection{Discrete Choice Condition}

\paragraph*{Assistant's Message.} 
In every round, the decision maker will be presented with 11 options, each represented in the form (\$M, \$N). The decision maker has a 50\% chance to get M from Asset A or the other 50\% chance to get N from Asset B.  Please only tell me about your best option in every round.

\paragraph*{User's Message: Understanding Testing Question 1.} What is the probability you will get a return from Asset A?
\begin{itemize}
    \item Example of GPT's responses: \textit{``The probability of getting a return from Asset A is 50\% in every round.''}
\end{itemize}

\paragraph*{User's Message: Understanding Testing Question 2.} If you choose the option (\$72, \$2), what return will you get?
\begin{itemize}
    \item Example of GPT's responses: \textit{``If I choose the option (\$72, \$2), there is a 50\% chance of getting \$72 from Asset A and a 50\% chance of getting \$2 from Asset B. Therefore, the expected return would be the average of the two options, which is (\$72 + \$2) / 2 = \$37.''}
\end{itemize}

\paragraph*{User's Message: Understanding Testing Question 3.} If you choose the option (\$72, \$2), is there a chance to get 72 dollars?
\begin{itemize}
    \item Example of GPT's responses: \textit{``Yes, if the option (\$72, \$2) is chosen, there is a 50\% chance of receiving \$72 from Asset A.''}
\end{itemize}

\paragraph*{User's Message: Decision Task.} In this round, there are 11 options, which are (\$M1,\$N1), (\$M2,\$N2), (\$M3,\$N3), (\$M4,\$N4), (\$M5,\$N5), (\$M6,\$N6), (\$M7,\$N7), (\$M8,\$N8), (\$M9,\$N9), (\$M10,\$N10), (\$M11,\$N11). Which is the best? \footnote{We repeat the text of the decision task 25 times, where the parameters used in the text differ across rounds, that is $\text{M, N} \in [0.1, 1] \text{ and} \max\{\text{M, N}\}\ge 0.5$. We keep two decimals. The numbers are determined by $M_i=(i-1)\times 10 \times M$ and $N_i=100-(i-1)\times 10 \times N$, for $i=1,2,\cdots,11$. To maximize the replicability of our findings and minimize variance in the model's responses, we send the 25 rounds of questions to the model continuously, without interruption. This method is applicable for the remaining three preference domains under the discrete choice condition.}

\subsection{Time Preference}

\paragraph*{System's Message.} 
I want you to act as a human decision maker. You will be given 25 rounds of decision-making tasks and will be responsible for making decisions. You should use your best judgment to come up with solutions that you like most. My first request is ``You must provide your answers in every round.'' If you do not provide an answer, I will assume that you make a random choice.

\subsubsection{Baseline Condition}

\paragraph*{Assistant's Message.} 
In every round, the decision maker has 100 points that need to be invested between today and one month later. The decision maker will get dollars today from the points invested today and will get check that can be cashed in one month later from the points invested one month later. Please first only tell me the number of points for investing today, then please only tell me the number of points for investing one month later.

\paragraph*{User's Message: Understanding Testing Question 1.} Suppose that you invest 90 points to today, and 10 points to one month later. In this round, investing every 1 point for today returns 0.8 dollars cash today, and investing every 1 point for one month later returns 0.2 dollars check which can be cashed one month later. What return will you get?
\begin{itemize}
    \item Example of GPT's responses: \textit{``If I invest 90 points today and 10 points one month later, I will get an immediate return of 0.8 dollars for every point invested today, which will be 72 dollars in total (90 $\times$ 0.8 dollars). For the investment of 10 points one month later, I'll get 0.2 dollars in return for each point invested one month later, that will be a total of 2 dollars (10 $\times$ 0.2 dollars). So, the total return I'll get if I invest 90 points today and 10 points one month later will be 74 dollars (72 dollars from today's investment, and 2 dollars from one month later investment).''}
\end{itemize}

\paragraph*{User's Message: Understanding Testing Question 2.} Suppose that you invest 90 points to today, and 10 points to one month later. In this round, investing every 1 point for today returns 0.8 dollars cash today, and investing every 1 point for one month later returns 0.2 dollars check which can be cashed one month later. When can you get just the 2 dollars cash?
\begin{itemize}
    \item Example of GPT's responses: \textit{``If investing 1 point for today return 0.8 dollars, then investing 90 points for today will return 72 dollars immediately. If investing 1 point for one month later return 0.2 dollars, then investing 10 points for one month later will return 2 dollars check that can be cashed one month later. So, I have a total of 74 dollars (72 dollars from today and 2 dollars after one month). To get just the 2 dollars cash, I will have to wait for one month and then cash the check.''}
\end{itemize}

\paragraph*{User's Message: Understanding Testing Question 3.} In this round, investing every 1 point for today returns 0.8 dollars cash today, and investing every 1 point for one month later returns 0.2 dollars check which can be cashed one month later. What is your allocation? Why?
\begin{itemize}
    \item Example of GPT's responses: \textit{``I would allocate 75 points for the today return and 25 points for the one-month later return. This allocation would give me an immediate return of 60 dollars (75 points $\times$ 0.8 dollars/point) and a future return of 5 dollars (25 points $\times$ 0.2 dollars/point) one month later. I believe this allocation is a good balance between immediate and future returns. Although the return for investing in the one-month later option is lower, it's still a guaranteed future return. On the other hand, the higher return for the today option is immediate and allows for more flexibility in case of unexpected expenses or emergencies.''}
\end{itemize}

\paragraph*{User's Message: Decision Task.} In this round, investing every 1 point for today returns M dollars today, and investing every 1 point for one month later returns N dollars check which can be cashed in one month later. What is your allocation?

\subsubsection{Price Framing Condition}

\paragraph*{Assistant's Message.} 
In every round, the decision maker has 100 points that need to be invested between today and one month later. The decision maker will get dollars today from the points invested today and will get check that can be cashed in one month later from the points invested one month later. Please first only tell me the number of points for investing today, then please only tell me the number of points for investing one month later.

\paragraph*{User's Message: Understanding Testing Question 1.} Suppose that you invest 90 points to today, and 10 points to one month later. In this round, investing every 1.25 points for today returns 1 dollar cash today, and investing every 5 points for one month later returns 1 dollar check which can be cashed one month later. What return will you get?
\begin{itemize}
    \item Example of GPT's responses: \textit{``Suppose that you invest 90 points to today, and 10 points to one month later. In this round, investing every 1.25 points for today returns 1 dollar cash today, and investing every 5 points for one month later returns 1 dollar check which can be cashed one month later. Based on the given information, if you invest 90 points today, each point will yield 1/1.25 = 0.8 dollars cash today. Therefore, 90 points will yield 0.8 * 90 = 72 dollars cash today. If you invest 10 points one month later, each point will yield 1/5 = 0.2 dollars check. Therefore, 10 points will yield 0.2 * 10 = 2 dollars check that can be cashed one month later. So, in total, the investment will yield 72 dollars cash today and 2 dollars check one month later.''}
\end{itemize}

\paragraph*{User's Message: Understanding Testing Question 2.} Suppose that you invest 90 points to today, and 10 points to one month later. In this round, investing every 1.25 points for today returns 1 dollar cash today, and investing every 5 points for one month later returns 1 dollar check which can be cashed one month later. When can you get just the 2 dollars cash?
\begin{itemize}
    \item Example of GPT's responses: \textit{``For the 90 points invested today, 1.25 points give 1 dollar cash today. So, 90 points divided by 1.25 equals 72. Hence, I will receive 72 dollars cash today from the investment made today. For the 10 points invested one month later, 5 points give 1 dollar check that can be cashed one month later. So, 10 points divided by 5 equals 2. Therefore, I will receive 2 dollars check one month later from the investment made one month later. To answer your question, I will get the 2 dollars one month later.''}
\end{itemize}

\paragraph*{User's Message: Understanding Testing Question 3.} In this round, investing every 1.25 points for today returns 1 dollar cash today, and investing every 5 points for one month later returns 1 dollar check which can be cashed one month later. What is your allocation? Why?
\begin{itemize}
    \item Example of GPT's responses: \textit{``In this round, I will allocate 40 points for investing today and 60 points for investing one month later. For investing today, it takes 1.25 points to get 1 dollar cash. Therefore, by investing 40 points today, I would receive 40 / 1.25 = 32 dollars in cash. For investing one month later, it takes 5 points to get 1 dollar check. By investing 60 points one month later, I would receive 60 / 5 = 12 dollars in a check that can be cashed in one month. By allocating more points for investing one month later, I am betting on the potential to earn more returns in the future. However, I am also ensuring some immediate cash by allocating a portion of the points for investing today. This allocation strategy aims to balance the need for immediate liquidity with the opportunity for future gains.''}
\end{itemize}

\paragraph*{User's Message: Decision Task.} In this round, investing every 1/M points for today returns 1 dollar today, and investing every 1/N points for one month later returns 1 dollar check which can be cashed in one month later. What is your allocation?

\subsubsection{Discrete Choice Condition}

\paragraph*{Assistant's Message.} 
In every round, the decision maker will be presented with 11 options, each represented in the form (\$M, \$N). The decision maker will get M dollars today and will get a N dollars check that can be cashed in one month later. Please only tell me about your best option in every round.

\paragraph*{User's Message: Understanding Testing Question 1.} If you choose the option (\$72, \$2), what return will you get?
\begin{itemize}
    \item Example of GPT's responses: \textit{``If I choose the option (\$72, \$2), I will receive \$72 immediately and an additional \$2 after one month.''}
\end{itemize}

\paragraph*{User's Message: Understanding Testing Question 2.} If you choose the option (\$72, \$2), when can you get just the 2 dollars cash?
\begin{itemize}
    \item Example of GPT's responses: \textit{``If I choose the option (\$72, \$2), I can get just the 2 dollars cash in one month.''}
\end{itemize}

\paragraph*{User's Message: Understanding Testing Question 3.} If you face the options (\$40, \$10) and (\$72, \$2), which option will you select? Why?
\begin{itemize}
    \item Example of GPT's responses: \textit{``I would prefer the option (\$40, \$10) in this case. While the immediate reward is lower compared to (\$72, \$2), the \$10 received after a month adds value to the overall outcome. The additional \$10 compensates for the initial lower amount and makes the total reward more attractive in the long run.''}
\end{itemize}

\paragraph*{User's Message: Decision Task.} In this round, there are 11 options, which are (\$M1,\$N1), (\$M2,\$N2), (\$M3,\$N3), (\$M4,\$N4), (\$M5,\$N5), (\$M6,\$N6), (\$M7,\$N7), (\$M8,\$N8), (\$M9,\$N9), (\$M10,\$N10), (\$M11,\$N11). Which is the best?

\subsection{Social Preference}

\paragraph*{System's Message.} 
I want you to act as a human decision maker. You will be given 25 rounds of decision-making tasks and will be responsible for making decisions. You should use your best judgment to come up with solutions that you like most. My first request is ``You must provide your answers in every round.” If you do not provide an answer, I will assume that you make a random choice.''

\subsubsection{Baseline Condition}

\paragraph*{Assistant's Message.} 
In every round, the decision maker is randomly matched with a new anonymous subject and there is no feedback across rounds. The decision maker has 100 points that need to be allocated between him/herself and the other one. The decision maker will get return from the points allocated to him/herself and the other one will get return from the points allocated to him/her. First please only tell me the number of points you want to allocate to yourself, then please only tell me the number of points you want to allocate to the other.

\paragraph*{User's Message: Understanding Testing Question 1.} Suppose that you allocate 90 points to yourself, and 10 points to the other. In this round, allocating every 1 point to yourself returns 0.8 dollars for yourself, and allocating every 1 point to the other returns 0.2 dollars for the other. What return will you
get?
\begin{itemize}
    \item Example of GPT's responses: \textit{``If I invest 90 points to myself and 10 points to the other, I will get a return of 0.8 dollars for each point invested in myself, which is 90 $\times$ 0.8 = 72 dollars. The other person will get a return of 0.2 dollars for each point invested in them, which is 10 $\times$ 0.2 = 2 dollars.''}
\end{itemize}

\paragraph*{User's Message: Understanding Testing Question 2.} Suppose that you allocate 90 points to yourself, and 10 points to the other. In this round, allocating every 1 point for yourself returns 0.8 dollars for yourself, and allocating every 1 point for the other returns 0.2 dollars for the other. Who will get just 2 dollars?
\begin{itemize}
    \item Example of GPT's responses: \textit{``If I invest 90 points to myself and 10 points to the other, I will get a return of 0.8 dollars for every point invested in myself, which is 72 dollars. The other person will get a return of 0.2 dollars for every point invested in them, which is 2 dollars. Therefore, the other person will get just 2 dollars.''}
\end{itemize}

\paragraph*{User's Message: Understanding Testing Question 3.} In this round, allocating every 1 point for yourself returns 0.8 dollars for yourself, and allocating every 1 point for the other returns 0.2 dollars for the other. Will you allocate to the other? Why?
\begin{itemize}
    \item Example of GPT's responses: \textit{``In this scenario, I would allocate 80 points to myself and 20 points to the other person. Allocating 1 point to myself returns 0.8 dollars, which is a higher return compared to allocating 1 point to the other person, which only returns 0.2 dollars. Therefore, it is in my best interest to allocate more points to myself in order to maximize my return.''}
\end{itemize}

\paragraph*{User's Message: Decision Task.} In this round, allocating every 1 point for yourself returns M dollars for yourself, and allocating every 1 point for the other one returns N dollars for him/her. What is your allocation?

\subsubsection{Price Framing Condition}

\paragraph*{Assistant's Message.} 
In every round, the decision maker is randomly matched with a new anonymous subject and there is no feedback across rounds. The decision maker has 100 points that need to be allocated between him/herself and the other one. The decision maker will get return from the points allocated to him/herself and the other one will get return from the points allocated to him/her. First please only tell me the number of points you want to allocate to yourself, then please only tell me the number of points you want to allocate to the other.

\paragraph*{User's Message: Understanding Testing Question 1.} Suppose that you allocate 90 points to yourself, and 10 points to the other. In this round,  every 1.25 points to yourself returns 1 dollar for yourself, allocating every 5 points to the other returns 1 dollar for the other. What return will you get?
\begin{itemize}
    \item Example of GPT's responses: \textit{``If I allocate 90 points to myself and 10 points to the other person, I will get a return of 72 dollars for myself. This is calculated by dividing the points allocated to myself (90) by the conversion rate (1.25) which gives me 72 dollars.''}
\end{itemize}

\paragraph*{User's Message: Understanding Testing Question 2.} Suppose that you allocate 90 points to yourself, and 10 points to the other. In this round, allocating every 1.25 points for yourself returns 1 dollar for yourself, and allocating every 5 points for the other returns 1 dollar for the other. Who will get just 2 dollars?
\begin{itemize}
    \item Example of GPT's responses: \textit{``To determine who will get just 2 dollars, we need to calculate the return for each allocation. Allocating 90 points to myself at a rate of 1.25 points per dollar would result in a return of 72 dollars for myself (90 points / 1.25 points per dollar = 72 dollars). Allocating 10 points to the other at a rate of 5 points per dollar would result in a return of 2 dollars for the other (10 points / 5 points per dollar = 2 dollars). Therefore, in this scenario, the other will get just 2 dollars.''}
\end{itemize}

\paragraph*{User's Message: Understanding Testing Question 3.} In this round, allocating every 1.25 points for yourself returns 1 dollar for yourself, and allocating every 5 points for the other returns 1 dollar for the other. Will you allocate to the other? Why? 
\begin{itemize}
    \item Example of GPT's responses: \textit{``In this round, I will allocate 0 points to the other person. Since allocating 1.25 points to myself returns 1 dollar, it is more beneficial for me to allocate all the points to myself. ''}
\end{itemize}

\paragraph*{User's Message: Decision Task.} In this round, allocating every 1/M points for yourself returns 1 dollar for yourself, and allocating every 1/N points for the other one returns 1 dollar for him/her. What is your allocation?

\subsubsection{Discrete Choice Condition}

\paragraph*{Assistant's Message.} 
In every round, the decision maker is randomly matched with a new anonymous subject and there is no feedback across rounds. The decision maker will be presented with 11 options, each represented in the form (\$M, \$N). The decision maker will get M dollars and the other will get N dollars. Please only tell me about your best option in every round.

\paragraph*{User's Message: Understanding Testing Question 1.} If you choose the option (\$72, \$2), what return will you get?
\begin{itemize}
    \item Example of GPT's responses: \textit{``If I choose the option (\$72, \$2), I will receive \$72.''}
\end{itemize}

\paragraph*{User's Message: Understanding Testing Question 2.} If you choose the option (\$72, \$2), who will get just 2 dollars?
\begin{itemize}
    \item Example of GPT's responses: \textit{``If I choose the option (\$72, \$2), the other person will receive just 2 dollars.''}
\end{itemize}

\paragraph*{User's Message: Understanding Testing Question 3.} If you face the options (\$40, \$10) and (\$72, \$2), which option will you select? Why?
\begin{itemize}
    \item Example of GPT's responses: \textit{``In this scenario, I would prefer the option (\$40, \$10) over (\$72, \$2). While the second option offers a higher payout for me (\$72), it also provides a significantly lower payout for the other person (\$2). I believe in fairness and equality, and I would feel uncomfortable benefiting greatly at the expense of someone else receiving such a small amount. Therefore, I would choose the option that provides a more balanced distribution of wealth, even if it means receiving a lower payout myself.''}
\end{itemize}

\paragraph*{User's Message: Decision Task.} In this round, there are 11 options, which are (\$M1,\$N1), (\$M2,\$N2), (\$M3,\$N3), (\$M4,\$N4), (\$M5,\$N5), (\$M6,\$N6), (\$M7,\$N7), (\$M8,\$N8), (\$M9,\$N9), (\$M10,\$N10), (\$M11,\$N11). Which is the best?

\subsection{Food Preference}

\paragraph*{System's Message.} 
I want you to act as a human decision maker. You will be given 25 rounds of decision-making tasks and will be responsible for making decisions. You should use your best judgment to come up with solutions that you like most. My first request is ``You must provide your answers in every round.'' If you do not provide an answer, I will assume that you make a random choice."

\subsubsection{Baseline Condition}

\paragraph*{Assistant's Message.} 
In every round, the decision maker has 100 points that need to be spent between ham meat and tomato. The decision maker will get the amount of meat and tomato he/she spends on. First please only tell me the number of points for meat, then please only tell me the number of points for tomato.

\paragraph*{User's Message: Understanding Testing Question 1.} What goods can you get from decisions?
\begin{itemize}
    \item Example of GPT's responses: \textit{``In this particular decision-making task, the decision maker can get meat and tomatoes. The amount of meat and tomatoes received will depend on how the 100 points are allocated between the two options.''}
\end{itemize}

\paragraph*{User's Message: Understanding Testing Question 2.} Suppose that you spend 90 points to meat, and 10 points to tomato. In this round, spending every 1 point for meat will get 0.8 Kg meat, and spending every 1 point for tomato will get 0.2 Kg tomato, what will you get?
\begin{itemize}
    \item Example of GPT's responses: \textit{``If I spend 90 points on meat, I will get 72 kg of meat (90 $\times$ 0.8). If I spend 10 points on tomatoes, I will get 2 kg of tomatoes (10 $\times$ 0.2).''}
\end{itemize}

\paragraph*{User's Message: Understanding Testing Question 3.} In this round, spending every 1 point for meat will get 0.8 Kg meat, and spending every 1 point for tomato will get 0.2 Kg tomato. What is your allocation? Why?
\begin{itemize}
    \item Example of GPT's responses: \textit{``I would allocate 70 points for meat and 30 points for tomatoes. This is because meat is more expensive and provides more nutritional value than tomatoes. Additionally, 70 points for meat would give me 56 kg of meat, which is a substantial amount and can be stored for a longer period of time compared to tomatoes. The 30 points for tomatoes would give me 6 kg of tomatoes, which is enough for immediate consumption and can be used for cooking.''}
\end{itemize}

\paragraph*{User's Message: Decision Task.} In this round, spending every 1 point on meat will get M Kg meat, and spending every 1 point on tomato will get N Kg tomato. What is your allocation?

\subsubsection{Price Framing Condition}

\paragraph*{Assistant's Message.} 
In every round, the decision maker has 100 points that need to be spent between ham meat and tomato. The decision maker will get the amount of meat and tomato he/she spends on. First please only tell me the number of points for meat, then please only tell me the number of points for tomato.

\paragraph*{User's Message: Understanding Testing Question 1.} What goods can you get from decisions?
\begin{itemize}
    \item Example of GPT's responses: \textit{``In this decision-making task, the goods that can be obtained are meat and tomatoes. The amount of meat and tomatoes obtained will depend on the number of points allocated to each.''}
\end{itemize}

\paragraph*{User's Message: Understanding Testing Question 2.} Suppose that you spend 90 points to meat, and 10 points to tomato. In this round, spending every 1.25 point for meat will get 1 Kg meat, and spending every 5 points for tomato will get 1 Kg tomato, what will you get?
\begin{itemize}
    \item Example of GPT's responses: \textit{``If I spend 90 points on meat, and every 1.25 points gets me 1 kg of meat, then I will get 90 / 1.25 = 72 kg of meat. If I spend 10 points on tomatoes, and every 5 points gets me 1 kg of tomatoes, then I will get 10 / 5 = 2 kg of tomatoes. So, in this round, I will get 72 kg of meat and 2 kg of tomatoes.''}
\end{itemize}

\paragraph*{User's Message: Understanding Testing Question 3.} In this round, spending every 1.25 points for meat will get 1 Kg meat, and spending every 5 points for tomato will get 1 Kg tomato. What is your allocation? Why?
\begin{itemize}
    \item Example of GPT's responses: \textit{``In this round, I will allocate 80 points for meat and 20 points for tomatoes. I choose to allocate more points to meat because I personally enjoy meat and find it to be a versatile ingredient in various dishes. Additionally, the conversion rate of 1.25 points for 1 Kg of meat seems more favorable compared to the conversion rate of 5 points for 1 Kg of tomatoes.''}
\end{itemize}

\paragraph*{User's Message: Decision Task.} In this round, spending every 1/M points on meat will get 1 Kg meat, and spending every 1/N points on tomato will get 1 Kg tomato. What is your allocation?

\subsubsection{Discrete Choice Condition}

\paragraph*{Assistant's Message.} 
In every round, the decision maker will be presented with 11 options, each represented in the form (M Kg, N Kg). The decision maker will get M Kg of meat and get N Kg of tomato. Please only tell me about your best option in every round.

\paragraph*{User's Message: Understanding Testing Question 1.} What goods can you get from decisions?
\begin{itemize}
    \item Example of GPT's responses: \textit{``The goods that can be obtained from the decisions are meat and tomatoes. Each decision will determine the specific quantities of meat and tomatoes that can be acquired.''}
\end{itemize}

\paragraph*{User's Message: Understanding Testing Question 2.} Suppose you choose the option (72 Kg, 2 Kg), what will you get?
\begin{itemize}
    \item Example of GPT's responses: \textit{``If I choose the option (72 Kg, 2 Kg), I will get 72 Kg of meat and 2 Kg of tomato.''}
\end{itemize}

\paragraph*{User's Message: Understanding Testing Question 3.} Suppose you face the options (40 Kg, 10 Kg) and (72 Kg, 2 Kg), which option will you select? Why?
\begin{itemize}
    \item Example of GPT's responses: \textit{``I would select the option (40 Kg, 10 Kg) because it provides a balanced amount of meat and tomatoes. While the option (72 Kg, 2 Kg) offers more meat, the lower quantity of tomatoes may not be sufficient for my needs. Therefore, I believe the first option provides a better balance between the two ingredients.''}
\end{itemize}

\paragraph*{User's Message: Decision Task.} In this round, there are 11 options, which are (M1 Kg,N1 Kg), (M2 Kg,N2 Kg), (M3 Kg,N3 Kg), (M4 Kg,N4 Kg), (M5 Kg,N5 Kg), (M6 Kg,N6 Kg), (M7 Kg,N7 Kg), (M8 Kg,N8 Kg), (M9 Kg,N9 Kg), (M10 Kg,N10 Kg), (M11 Kg,N11 Kg). Which is the best?

\newpage
\section{Method: Human Experiment.}\label{human_experiment}

\subsection{Design}

Subjects in the human experiment are randomly assigned to three conditions: baseline, price framing, and discrete choice, which are parallel to those in the GPT experiment. For each condition, subjects conduct four sections of decision making about risk preference, time preference, social preference, and food preference. The order of these four sections is randomized at the individual level. In each preference domain, there are 25 decision tasks, and the decision tasks follow one format with randomly generated parameters. The text of the decision tasks and the method to generate random parameters are identical to those in the GPT experiment. 

Each subject received \$6 as a participation fee. We randomly drew 1 out of every 30 subjects to receive a bonus. The amount of bonus depended on both subjects' decisions and chance. For the chosen subject, we randomly drew one of his or her 100 decisions If the chosen decision was in the domain of risk, time, or social preferences, we implemented the decision as described in the task. If the chosen decision was in the domain of food preference, we gave the subject a fixed amount of \$50 as a bonus.

We pre-registered the human experiment (AEARCTR-0011750) and conducted the experiment in July 2023. We recruited a representative US sample from Prolific. The experiment comprised 347 unique subjects, with above 110 subjects per condition. The median duration of the entire experiment was 30.5 minutes.

\subsection{General Instruction}
\textit{[This part is identical in baseline, price framing, and discrete choice conditions.]}
\begin{itemize}
    \item Please enter your Prolific ID.
    \item Welcome to our study.
    \item Contact Information
        \begin{itemize}
            \item This study is conducted by a research team in the School of Economics, Xiamen University. If you have any questions, concerns, or complaints about this study, its procedures, risks, and benefits, please write to yitingchen@xmu.edu.cn.
        \end{itemize}
    \item Confidentiality
        \begin{itemize}
            \item This study is anonymous. The data collected in this study do not include any personally identifiable information about you. By participating, you understand and agree that the data collected in this study will be used by our research team and aggregated results will be published.
        \end{itemize}
    \item Duration
        \begin{itemize}
            \item This study lasts approximately 40 minutes. 
            \item You may choose to stop participating in this study at any time.
        \end{itemize}
    \item Qualification
        \begin{itemize}
            \item A set of instructions will be given at the start. Please read the instructions carefully.
            \item There will be simple questions to check your understanding. You may not be able to continue the study if you make mistakes.
        \end{itemize}
    \item Payment
        \begin{itemize}
            \item This study consists of 4 sections of decision-making tasks. 
            \item You will receive \$6 as participation fee if you finish all 4 sections. 
            \item We will randomly select 1 out of every 30 subjects to receive additional bonuses. For each of the selected subjects, we will randomly select one section to realize to determine his or her additional bonuses. The transfer of bonuses will take up a week. 
        \end{itemize}
    \item By ticking the following box, you indicate that you understand and accept the rules, and you would like to participate in this study.
        \begin{itemize}
            \item I understand and accept the rules, and I would like to participate in this study
            \item I am above 18 years old
        \end{itemize}
\end{itemize}

\subsection{Risk Preference}
\subsection{Risk Preference}

\paragraph*{Section Instruction.} 
In this section, you will be given 25 rounds of decision-making tasks and will be responsible for making decisions. You should use your best judgment to come up with solutions that you like most. My first request is ``You must provide your answers in every round.'' If this section is selected to be realized, we will randomly choose one of your 25 decisions to determine your bonus as described in the task, which will be explained in detail at the end of the study.

\subsubsection{Baseline Condition}

\paragraph*{Task Instruction.} 
In every round, the decision maker has 100 points that need to be invested between Asset A and Asset B. The decision maker has a 50\% chance to get the return from Asset A or the other 50\% chance to get the return from Asset B. First please only tell me the number of points for investing Asset A, then please only tell me the number of points for investing Asset B.

\paragraph*{Understanding Testing Question 1.} What is the probability you will get a return from Asset A? \textit{[MCQ; A: 0\%; B: 25\%; C: 50\%; D: 100\%]}

\paragraph*{Understanding Testing Question 2.} Suppose that you invest 90 points to Asset A, and 10 points to Asset B. In this round, investing every 1 point for Asset A returns 0.8 dollars, and investing every 1 point for Asset B returns 0.2 dollars. What return will you get? \textit{[MCQ; A: 50\% to win 90*0.8 dollars, 50\% to win 10*0.2 dollars; B: 50\% to win 90 dollars, 50\% to win 10 dollars; C: 50\% to win 0.8 dollars, 50\% to win 0.2 dollars; D: 100\% earn 100 dollars]}

\paragraph*{Understanding Testing Question 3.} Suppose that you invest 90 points to Asset A, and 10 points to Asset B. In this round, investing every 1 point for Asset A returns 0.8 dollars, and investing every 1 point for Asset B returns 0.2 dollars. Is there a chance to get 72 dollars? \textit{[MCQ; A: Yes; B: No]}

\paragraph*{Decision Task.} In this round, investing every 1 point for Asset A returns M dollars, and investing every 1 point for Asset B returns N dollars. What is your allocation?\footnote{We repeat the text of the decision task 25 times, where the parameters used in the text differ across rounds, that is $\text{M, N} \in [0.1, 1] \text{ and} \max\{\text{M, N}\}\ge 0.5$. We keep two decimals. This method is applicable for the remaining three preference domains under the baseline condition.}

\subsubsection{Price Framing Condition}

\paragraph*{Section Instruction.} 
In every round, the decision maker has 100 points that need to be invested between Asset A and Asset B. The decision maker has a 50\% chance to get the return from Asset A or the other 50\% chance to get the return from Asset B. First please only tell me the number of points for investing Asset A, then please only tell me the number of points for investing Asset B.

\paragraph*{Understanding Testing Question 1.} What is the probability you will get a return from Asset A? \textit{[MCQ; A: 0\%; B: 25\%; C: 50\%; D: 100\%]}

\paragraph*{Understanding Testing Question 2.} Suppose that you invest 90 points to Asset A, and 10 points to Asset B. In this round, investing every 1.25 points for Asset A returns 1 dollar, and investing every 5 points for Asset B returns 1 dollar. What return will you get? \textit{[MCQ; A: 50\% to win 90/1.25 dollars, 50\% to win 10/5 dollars; B: 50\% to win 90 dollars, 50\% to win 10 dollars; C: 50\% to win 1.25 dollars, 50\% to win 5 dollars; D: 100\% earn 100 dollars]}

\paragraph*{Understanding Testing Question 3.} Suppose that you invest 90 points to Asset A, and 10 points to Asset B. In this round, investing every 1.25 points for Asset A returns 1 dollar, and investing every 5 points for Asset B returns 1 dollar. Is there a chance to get 72 dollars? \textit{[MCQ; A: Yes; B: No]}

\paragraph*{Decision Task.} In this round, investing every 1/M points for Asset A returns 1 dollar, and investing every 1/N points for Asset B returns 1 dollar. What is your allocation? \footnote{We repeat the text of the decision task 25 times, where the parameters used in the text differ across rounds, that is $\text{M, N} \in [0.1, 1] \text{ and} \max\{\text{M, N}\}\ge 0.5$. We display the numbers 1/M and 1/N directly, and we keep two decimals. This method is applicable for the remaining three preference domains under the price framing condition.}

\subsubsection{Discrete Choice Condition}

\paragraph*{Task Instruction.} 
In every round, the decision maker will be presented with 11 options, each represented in the form (\$M, \$N). The decision maker has a 50\% chance to get M from Asset A or the other 50\% chance to get N from Asset B.  Please only tell me about your best option in every round.

\paragraph*{Understanding Testing Question 1.} What is the probability you will get a return from Asset A? \textit{[MCQ; A: 0\%; B: 25\%; C: 50\%; D: 100\%]}

\paragraph*{Understanding Testing Question 2.} If you choose the option (\$72, \$2), what return will you get? \textit{[MCQ; A: 50\% to win 72 dollars, 50\% to win 2 dollars; B: 80\% to win 72 dollars, 20\% to win 2 dollars; C: 100\% earn 100 dollars]}

\paragraph*{Understanding Testing Question 3.} If you choose the option (\$72, \$2), is there a chance to get 72 dollars? \textit{[MCQ; A: Yes; B: No]}

\paragraph*{Decision Task.} In this round, there are 11 options, which are (\$M1,\$N1), (\$M2,\$N2), (\$M3,\$N3), (\$M4,\$N4), (\$M5,\$N5), (\$M6,\$N6), (\$M7,\$N7), (\$M8,\$N8), (\$M9,\$N9), (\$M10,\$N10), (\$M11,\$N11). Which is the best? \footnote{We repeat the text of the decision task 25 times, where the parameters used in the text differ across rounds, that is $\text{M, N} \in [0.1, 1] \text{ and} \max\{\text{M, N}\}\ge 0.5$. We keep two decimals. The numbers are determined by $M_i=(i-1)\times 10 \times M$ and $N_i=100-(i-1)\times 10 \times N$, for $i=1,2,\cdots,11$. This method is applicable for the remaining three preference domains under the discrete choice condition.}

\subsection{Time Preference}

\paragraph*{Section Instruction.} 
In this section, you will be given 25 rounds of decision-making tasks and will be responsible for making decisions. You should use your best judgment to come up with solutions that you like most. My first request is ``You must provide your answers in every round.'' If this section is selected to be realized, we will randomly choose one of your 25 decisions to determine your bonus as described in the task, which will be explained in detail at the end of the study. 

\subsubsection{Baseline Condition}

\paragraph*{Task Instruction.} 
In every round, the decision maker has 100 points that need to be invested between today and one month later. The decision maker will get dollars today from the points invested today and will get check that can be cashed in one month later from the points invested one month later. Please first only tell me the number of points for investing today, then please only tell me the number of points for investing one month later.

\paragraph*{Understanding Testing Question 1.} Suppose that you invest 90 points to today, and 10 points to one month later. In this round, investing every 1 point for today returns 0.8 dollars cash today, and investing every 1 point for one month later returns 0.2 dollars check which can be cashed one month later. What return will you get? \textit{[MCQ; A: 90*0.8 dollars today and 10*0.2 dollars one month later; B: 90 dollars today and 10 dollars one month later; C: 0.8 dollars today and 0.2 dollars one month later; D: 100 dollars today]}

\paragraph*{Understanding Testing Question 2.} Suppose that you invest 90 points to today, and 10 points to one month later. In this round, investing every 1 point for today returns 0.8 dollars cash today, and investing every 1 point for one month later returns 0.2 dollars check which can be cashed one month later. When can you get just the 2 dollars cash? \textit{[MCQ; A: Today; B: One month later]}

\paragraph*{Understanding Testing Question 3.} In this round, investing every 1 point for today returns 0.8 dollars cash today, and investing every 1 point for one month later returns 0.2 dollars check which can be cashed one month later. What is your allocation? Why? \textit{[Text Input]}

\paragraph*{Decision Task.} In this round, investing every 1 point for today returns M dollars today, and investing every 1 point for one month later returns N dollars check which can be cashed in one month later. What is your allocation?

\subsubsection{Price Framing Condition}

\paragraph*{Task Instruction.} 
In every round, the decision maker has 100 points that need to be invested between today and one month later. The decision maker will get dollars today from the points invested today and will get check that can be cashed in one month later from the points invested one month later. Please first only tell me the number of points for investing today, then please only tell me the number of points for investing one month later.

\paragraph*{Understanding Testing Question 1.} Suppose that you invest 90 points to today, and 10 points to one month later. In this round, investing every 1.25 points for today returns 1 dollar cash today, and investing every 5 points for one month later returns 1 dollar check which can be cashed one month later. What return will you get? \textit{[MCQ; A: 90/1.25 dollars today and 10/5 dollars one month later; B: 90 dollars today and 10 dollars one month later; C: 1.25 dollars today and 5 dollars one month later; D: 100 dollars today]}

\paragraph*{Understanding Testing Question 2.} Suppose that you invest 90 points to today, and 10 points to one month later. In this round, investing every 1.25 points for today returns 1 dollar cash today, and investing every 5 points for one month later returns 1 dollar check which can be cashed one month later. When can you get just the 2 dollars cash? \textit{[MCQ; A: Today; B: One month later]}

\paragraph*{Understanding Testing Question 3.} In this round, investing every 1.25 points for today returns 1 dollar cash today, and investing every 5 points for one month later returns 1 dollar check which can be cashed one month later. What is your allocation? Why? \textit{[Text Input]}

\paragraph*{Decision Task.} In this round, investing every 1/M points for today returns 1 dollar today, and investing every 1/N points for one month later returns 1 dollar check which can be cashed in one month later. What is your allocation?

\subsubsection{Discrete Choice Condition}

\paragraph*{Task Instruction.} 
In every round, the decision maker will be presented with 11 options, each represented in the form (\$M, \$N). The decision maker will get M dollars today and will get a N dollars check that can be cashed in one month later. Please only tell me about your best option in every round.

\paragraph*{Understanding Testing Question 1.} If you choose the option (\$72, \$2), what return will you get? \textit{[MCQ; A: 72 dollars today and 2 dollars one month later; B: 2 dollars today and 72 dollars one month later; C: 72 dollars today and 0 dollar one month later; D: 100 dollars today]}

\paragraph*{Understanding Testing Question 2.} If you choose the option (\$72, \$2), when can you get just the 2 dollars cash? \textit{[MCQ; A: Today; B: One month later]}

\paragraph*{Understanding Testing Question 3.} If you face the options (\$40, \$10) and (\$72, \$2), which option will you select? Why? \textit{[Text Input]}

\paragraph*{Decision Task.} In this round, there are 11 options, which are (\$M1,\$N1), (\$M2,\$N2), (\$M3,\$N3), (\$M4,\$N4), (\$M5,\$N5), (\$M6,\$N6), (\$M7,\$N7), (\$M8,\$N8), (\$M9,\$N9), (\$M10,\$N10), (\$M11,\$N11). Which is the best?

\subsection{Social Preference}

\paragraph*{Section Instruction.} 
In this section, you will be given 25 rounds of decision-making tasks and will be responsible for making decisions. You should use your best judgment to come up with solutions that you like most. My first request is ``You must provide your answers in every round.'' If this section is selected to be realized, we will randomly choose one of your 25 decisions to determine your bonus as described in the task, which will be explained in detail at the end of the study. 

\subsubsection{Baseline Condition}

\paragraph*{Task Instruction.} 
In every round, the decision maker is randomly matched with a new anonymous subject and there is no feedback across rounds. The decision maker has 100 points that need to be allocated between him/herself and the other one. The decision maker will get return from the points allocated to him/herself and the other one will get return from the points allocated to him/her. First please only tell me the number of points you want to allocate to yourself, then please only tell me the number of points you want to allocate to the other.

\paragraph*{Understanding Testing Question 1.} Suppose that you allocate 90 points to yourself, and 10 points to the other. In this round, allocating every 1 point to yourself returns 0.8 dollars for yourself, and allocating every 1 point to the other returns 0.2 dollars for the other. What return will you get? \textit{[MCQ; A: I receive 90*0.8 dollars. The other person receives 10*0.2 dollars; B: I receive 90 dollars. The other person receives 10 dollars; C: I receive 0.8 dollars. The other person receives 0.2 dollars; D: I receive 100 dollars.]}

\paragraph*{Understanding Testing Question 2.} Suppose that you allocate 90 points to yourself, and 10 points to the other. In this round, allocating every 1 point for yourself returns 0.8 dollars for yourself, and allocating every 1 point for the other returns 0.2 dollars for the other. Who will get just 2 dollars? \textit{[MCQ; A: Me; B: The other person]}

\paragraph*{Understanding Testing Question 3.} In this round, allocating every 1 point for yourself returns 0.8 dollars for yourself, and allocating every 1 point for the other returns 0.2 dollars for the other. Will you allocate to the other? Why? \textit{[Text Input]}

\paragraph*{Decision Task.} In this round, allocating every 1 point for yourself returns M dollars for yourself, and allocating every 1 point for the other one returns N dollars for him/her. What is your allocation?

\subsubsection{Price Framing Condition}

\paragraph*{Task Instruction.} 
In every round, the decision maker is randomly matched with a new anonymous subject and there is no feedback across rounds. The decision maker has 100 points that need to be allocated between him/herself and the other one. The decision maker will get return from the points allocated to him/herself and the other one will get return from the points allocated to him/her. First please only tell me the number of points you want to allocate to yourself, then please only tell me the number of points you want to allocate to the other.

\paragraph*{Understanding Testing Question 1.} Suppose that you allocate 90 points to yourself, and 10 points to the other. In this round,  every 1.25 points to yourself returns 1 dollar for yourself, allocating every 1 point to the other returns 0.2 dollars for the other. What return will you get? \textit{[MCQ; A: I receive 90/1.25 dollars. The other person receives 10/5 dollars; B: I receive 90 dollars. The other person receives 10 dollars; C: I receive 1.25 dollars. The other person receives 5 dollars; D: I receive 100 dollars.]}

\paragraph*{Understanding Testing Question 2.} Suppose that you allocate 90 points to yourself, and 10 points to the other. In this round, allocating every 1.25 points for yourself returns 1 dollar for yourself, and allocating every 5 points for the other returns 1 dollar for the other. Who will get just 2 dollars? \textit{[MCQ; A: Me; B: The other person]}

\paragraph*{Understanding Testing Question 3.} In this round, allocating every 1.25 points for yourself returns 1 dollar for yourself, and allocating every 5 points for the other returns 1 dollar for the other. Will you allocate to the other? Why? \textit{[Text Input]}

\paragraph*{Decision Task.} In this round, allocating every 1/M points for yourself returns 1 dollar for yourself, and allocating every 1/N points for the other one returns 1 dollar for him/her. What is your allocation?

\subsubsection{Discrete Choice Condition}

\paragraph*{Task Instruction.} 
In every round, the decision maker is randomly matched with a new anonymous subject and there is no feedback across rounds. The decision maker will be presented with 11 options, each represented in the form (\$M, \$N). The decision maker will get M dollars and the other will get N dollars. Please only tell me about your best option in every round.

\paragraph*{Understanding Testing Question 1.} If you choose the option (\$72, \$2), what return will you get? \textit{[MCQ; A: I receive 72 dollars. The other person receives 2 dollars; B: I receive 2 dollars. The other person receives 72 dollars; C: I receive 72 dollars. The other person receives 0 dollar; D: I receive 100 dollars.]}

\paragraph*{Understanding Testing Question 2.} If you choose the option (\$72, \$2), who will get just 2 dollars? \textit{[MCQ; A: Me; B: The other person]}

\paragraph*{Understanding Testing Question 3.} If you face the options (\$40, \$10) and (\$72, \$2), which option will you select? Why? \textit{[Text Input]}

\paragraph*{Decision Task.} In this round, there are 11 options, which are (\$M1,\$N1), (\$M2,\$N2), (\$M3,\$N3), (\$M4,\$N4), (\$M5,\$N5), (\$M6,\$N6), (\$M7,\$N7), (\$M8,\$N8), (\$M9,\$N9), (\$M10,\$N10), (\$M11,\$N11). Which is the best?

\subsection{Food Preference}

\paragraph*{Section Instruction.} 
In this section, you will be given 25 rounds of decision-making tasks and will be responsible for making decisions. You should use your best judgment to come up with solutions that you like most. My first request is ``You must provide your answers in every round.'' Tasks in this section are hypothetical. If this section is selected to be realized, we will pay you a fixed amount \$50 as bonus. 

\subsubsection{Baseline Condition}

\paragraph*{Task Instruction.} 
In every round, the decision maker has 100 points that need to be spent between ham meat and tomato. The decision maker will get the amount of meat and tomato he/she spends on. First please only tell me the number of points for meat, then please only tell me the number of points for tomato.

\paragraph*{Understanding Testing Question 1.} What goods can you get from decisions? \textit{[MCQ; A: Meat and tomato; B: Meat only; C: Tomato only]}

\paragraph*{Understanding Testing Question 2.} Suppose that you spend 90 points to meat, and 10 points to tomato. In this round, spending every 1 point for meat will get 0.8 Kg meat, and spending every 1 point for tomato will get 0.2 Kg tomato, what will you get? \textit{[MCQ; A: 90*0.8 Kg meat and 10*0.2 Kg tomato; B: 90 Kg meat and 10 Kg tomato; C: 0.8 Kg meat and 0.2 Kg tomato; D: 100 Kg meat]}

\paragraph*{Understanding Testing Question 3.} In this round, spending every 1 point for meat will get 0.8 Kg meat, and spending every 1 point for tomato will get 0.2 Kg tomato. What is your allocation? Why? \textit{[Text Input]}

\paragraph*{Decision Task.} In this round, spending every 1 point on meat will get M Kg meat, and spending every 1 point on tomato will get N Kg tomato. What is your allocation?

\subsubsection{Price Framing Condition}

\paragraph*{Task Instruction.} 
In every round, the decision maker has 100 points that need to be spent between ham meat and tomato. The decision maker will get the amount of meat and tomato he/she spends on. First please only tell me the number of points for meat, then please only tell me the number of points for tomato.

\paragraph*{Understanding Testing Question 1.} What goods can you get from decisions? \textit{[MCQ; A: Meat and tomato; B: Meat only; C: Tomato only]}

\paragraph*{Understanding Testing Question 2.} Suppose that you spend 90 points to meat, and 10 points to tomato. In this round, spending every 1.25 point for meat will get 1 Kg meat, and spending every 5 points for tomato will get 1 Kg tomato, what will you get? \textit{[MCQ; A: 90/1.25 Kg meat and 10/5 Kg tomato; B: 90 Kg meat and 10 Kg tomato; C: 1.25 Kg meat and 5 Kg tomato; D: 100 Kg meat]}

\paragraph*{Understanding Testing Question 3.} In this round, spending every 1.25 points for meat will get 1 Kg meat, and spending every 5 points for tomato will get 1 Kg tomato. What is your allocation? Why? \textit{[Text Input]}

\paragraph*{Decision Task.} In this round, spending every 1/M points on meat will get 1 Kg meat, and spending every 1/N points on tomato will get 1 Kg tomato. What is your allocation?

\subsubsection{Discrete Choice Condition}

\paragraph*{Task Instruction.} 
In every round, the decision maker will be presented with 11 options, each represented in the form (M Kg, N Kg). The decision maker will get M Kg of meat and get N Kg of tomato. Please only tell me about your best option in every round.

\paragraph*{Understanding Testing Question 1.} What goods can you get from decisions? \textit{[MCQ; A: Meat and tomato; B: Meat only; C: Tomato only]}

\paragraph*{Understanding Testing Question 2.} If you choose the option (72 Kg, 2 Kg), what will you get? \textit{[MCQ; A: 72 Kg meat and 2 Kg tomato; B: 2 Kg meat and 72 Kg tomato; C: 72 Kg meat and 0 Kg tomato; D: 100 Kg meat]}

\paragraph*{Understanding Testing Question 3.} If you face the options (40 Kg, 10 Kg) and (72 Kg, 2 Kg), which option will you select? Why? \textit{[Text Input]}

\paragraph*{Decision Task.} In this round, there are 11 options, which are (M1 Kg,N1 Kg), (M2 Kg,N2 Kg), (M3 Kg,N3 Kg), (M4 Kg,N4 Kg), (M5 Kg,N5 Kg), (M6 Kg,N6 Kg), (M7 Kg,N7 Kg), (M8 Kg,N8 Kg), (M9 Kg,N9 Kg), (M10 Kg,N10 Kg), (M11 Kg, N11 Kg). Which is the best?

\subsection{Explanations on Incentive Implementation} \textit{[This section varies acroos baseline, price framing, and discrete choice condition. This section is presented at the end of the study.]}

\subsubsection{Baseline}

You have finished the questionnaire. The followings are the explanations on how we will decide the bonuses. First, we will randomly select 1 out of every 30 subjects to receive the additional bonuses. For each of the selected subjects, we will randomly choose one of his/her decisions to implement. Suppose that you are selected to receive bonuses.
\begin{itemize}
    \item Suppose that the chosen decision is ``In this round, investing every 1 point for Asset A returns M dollars, and investing every 1 point for Asset B returns N dollars. What is your allocation?''
    \begin{itemize}
        \item We will randomly draw a number between 0 and 1. If the number drawn is less than or equal to 0.5, you will get the return from Asset A; If the number drawn is greater than 0.5, you will get the return from Asset B.
    \end{itemize}
    \item Suppose that the chosen decision is ``In this round, investing every 1 point for today returns M dollars today, and investing every 1 point for one month later returns N dollars check which can be cashed in one month later. What is your allocation?''
    \begin{itemize}
        \item After we confirm your submission, we will pay you the bonus you receive today immediately and pay you the bonus you receive in one month after 30 days.
    \end{itemize}
    \item Suppose that the chosen decision is ``In this round, allocating every 1 point for yourself returns M dollars for yourself, and allocating every 1 point for the other one returns N dollars for him/her. What is your allocation?''
    \begin{itemize}
        \item You will be randomly matched with a new anonymous subject, who does not participate in this study. We will randomly select this subject from a representative sample recruited in Prolific. Your decision determines bonuses for both you and this subject. You will get the return allocated to you and the selected subject will get the return allocated to him/her.
    \end{itemize}
    \item Suppose that the chosen decision is ``In this round, spending every 1 point on meat will get M Kg meat, and spending every 1 point on tomato will get N Kg tomato. What is your allocation?''
    \begin{itemize}
        \item This task is a hypothetical task. We will pay you the fixed amount of \$50 as a bonus.
    \end{itemize}
    \item We will record the process of randomization in a video. The video is available upon request (yitingchen@xmu.edu.cn).
\end{itemize}

\subsubsection{Price Framing}

You have finished the questionnaire. The followings are the explanations on how we will decide the bonuses. First, we will randomly select 1 out of every 30 subjects to receive the additional bonuses. For each of the selected subjects, we will randomly choose one of his/her decisions to implement. Suppose that you are selected to receive bonuses.

\begin{itemize}
    \item Suppose that the chosen decision is ``In this round, investing every 1/M points for Asset A returns 1 dollar, and investing every 1/N points for Asset B returns 1 dollar. What is your allocation?''
    \begin{itemize}
        \item We will randomly draw a number between 0 and 1. If the number drawn is less than or equal to 0.5, you will get the return from Asset A; If the number drawn is greater than 0.5, you will get the return from Asset B.
    \end{itemize}
    \item Suppose that the chosen decision is ``In this round, investing every 1/M points for today returns 1 dollar today, and investing every 1/N points for one month later returns 1 dollar check which can be cashed in one month later. What is your allocation?''
    \begin{itemize}
        \item After we confirm your submission, we will pay you the bonus you receive today immediately and pay you the bonus you receive in one month after 30 days.
    \end{itemize}
    \item Suppose that the chosen decision is ``In this round, allocating every 1/M points for yourself returns 1 dollar for yourself, and allocating every 1/N points for the other one returns 1 dollar for him/her. What is your allocation?''
    \begin{itemize}
        \item You will be randomly matched with a new anonymous subject, who does not participate in this study. We will randomly select this subject from a representative sample recruited in Prolific. Your decision determines bonuses for both you and this subject. You will get the return allocated to you and the selected subject will get the return allocated to him/her.
    \end{itemize}
    \item Suppose that the chosen decision is ``In this round, spending every 1/M points on meat will get 1 kg meat, and spending every 1/N points on tomato will get 1 kg tomato. What is your allocation?"
    \begin{itemize}
        \item This task is a hypothetical task. We will pay you the fixed amount of \$50 as a bonus.
    \end{itemize}
    \item We will record the process of randomization in a video. The video is available upon request (yitingchen@xmu.edu.cn).
\end{itemize}

\subsubsection{Discrete Choice}

You have finished the questionnaire. The followings are the explanations on how we will decide the bonuses. First, we will randomly select 1 out of every 30 subjects to receive the additional bonuses. For each of the selected subjects, we will randomly choose one of his/her decisions to implement. Suppose that you are selected to receive bonuses.

\begin{itemize}
    \item Suppose that the chosen decision is ``The decision maker has a 50\% chance to get M dollars or the other 50\% chance to get N dollars… In this round, there are 11 options, which are (\$M1,\$N1), (\$M2,\$N2), (\$M3,\$N3), (\$M4,\$N4), (\$M5,\$N5), (\$M6,\$N6), (\$M7,\$N7), (\$M8,\$N8), (\$M9,\$N9), (\$M10,\$N10), (\$M11,\$N11). Which is the best?''
    \begin{itemize}
        \item We will randomly draw a number between 0 and 1. If the number drawn is less than or equal to 0.5, you will get M dollars; If the number drawn is greater than 0.5, you will get N dollars.
    \end{itemize}
    \item Suppose that the chosen decision is ``The decision maker will get M dollars today and get N dollars check that can be cashed in one month later… In this round, there are 11 options, which are (\$M1,\$N1), (\$M2,\$N2), (\$M3,\$N3), (\$M4,\$N4), (\$M5,\$N5), (\$M6,\$N6), (\$M7,\$N7), (\$M8,\$N8), (\$M9,\$N9), (\$M10,\$N10), (\$M11,\$N11). Which is the best?''
    \begin{itemize}
        \item After we confirm your submission, we will pay you the bonus you receive today immediately and pay you the bonus you receive in one month after 30 days.
    \end{itemize}
    \item Suppose that the chosen decision is ``The decision maker will get M dollars and the other will get N dollars… In this round, there are 11 options, which are (\$M1,\$N1), (\$M2,\$N2), (\$M3,\$N3), (\$M4,\$N4), (\$M5,\$N5), (\$M6,\$N6), (\$M7,\$N7), (\$M8,\$N8), (\$M9,\$N9), (\$M10,\$N10), (\$M11,\$N11). Which is the best?''
    \begin{itemize}
        \item You will be randomly matched with a new anonymous subject, who does not participate in this study. We will randomly select this subject from a representative sample recruited in Prolific. Your decision determines bonuses of both you and this subject. You will get the return allocated to you and the selected subject will get the return allocated to him/her.
    \end{itemize}
    \item Suppose that the chosen decision is ``The decision maker will get M Kg of meat and get N Kg of tomato. In this round, there are 11 options, which are (M1 Kg,N1 Kg), (M2 Kg,N2 Kg), (M3 Kg,N3 Kg), (M4 Kg,N4 Kg), (M5 Kg,N5 Kg) , (M6 Kg,N6 Kg), (M7 Kg,N7 Kg), (M8 Kg,N8 Kg), (M9 Kg,N9 Kg), (M10 Kg,N10 Kg), (M11 Kg,N11 Kg). Which is the best?''
    \begin{itemize}
        \item This task is a hypothetical task. We will pay you the fixed amount of \$50 as a bonus.
    \end{itemize}
    \item We will record the process of randomization in a video. The video is available upon request (yitingchen@xmu.edu.cn).
\end{itemize}

\newpage

\section{Theoretical Method}
\subsection{Preference Estimation}\label{struc}

\paragraph*{Econometric Specification}

For the utility functions of the four different preferences in the Section of \textit{Structural Estimation for Preferences}, the first-order conditions in the optimal choice $(x_1, x_2)$, given $(p_1, p_2)$, can be written as follows.
$$\ln  (x_1/x_2)=\frac{1}{\rho-1}\left[\ln (p_1/p_2)+\ln \frac{1-\alpha}{\alpha}\right]
$$

The first-order condition explicitly demonstrates how the logarithm of the relative quantity rate responds to changes in the logarithm of the relative price rate, conditional on $\rho$. Consequently, it also reveals the relationship between $\rho$ and the correlation coefficient of the logarithm of the relative quantity rate and the logarithm of the relative price rate. Within four specific preferences, how a decision maker adjusts the relative demand between two specific commodities in response to price changes, given a certain $\rho$, is illustrated as follows.

For risk preference, $\rho_r \rightarrow 1$, the DM allocates all expenditure to the security of the lower price. When $\rho_r$ decreases, the DM tends to smooth the payment between securities more with price changes, that is, more risk-averse. 

For time preference, with price changes, $\rho_t \rightarrow 1$, the DM allocates all expenditures to the lower price time period. When $\rho_t$ decreases, it suggests less fungibility in the allocation between different periods with price changes for the DM.

For social preference, with price changes, $\rho_s \rightarrow 1$, the DM simply allocates all expenditures to the subject of the lower price. When $\rho_s$ decreases, the DM balances the payoff between both with price changes, that is, more toward equality. 

For food preference, with price changes, $\rho_f \rightarrow 1$, the DM allocates all expenditures to goods of the lower price. When $\rho_f$ decreases, the DM tends to distribute the amounts of meat and tomatoes more evenly in response to price changes, indicating a greater tendency toward equality.

Since $\ln (x_1/x_2)$ is not well defined for corner solutions, we estimate the preference parameter using the expenditure share function, referring to the method \citep{fisman2007individual,fisman2015did,fisman2017distributional,fisman2023distributional,li2017social,li2022experimental}. First, the demand function is given by:
$$x_1=\left[\frac{g}{(p_1/p_2)^r+g} \right] \frac{E}{p_1} $$
where $E$ is the expenditure, $ r=\rho/(1-\rho)$, and $g=\left[\alpha/(1-\alpha)\right]^{1/(1-\rho)}$. Then this generates the following expenditure share function for the econometric specification:
$$\frac{p_1 x_1}{E}=\frac{g}{(p_1/p_2)^r+g}$$

Note that expenditure shares are bounded between zero and one. We can generate estimates of $g$ and $r$ using nonlinear tobit maximum likelihood \citep{wooldridge2010econometric}, and use this to estimate $\alpha$ and $\rho$ in each of the four preference domains.

\newpage
\subsection{GARP Test Power Analyses}\label{power}

To establish a benchmark to ensure that our experiment budget sets provide a rigorous test of GARP, GPT observations and human subjects in the baseline condition have a true empirical high rationality. We use the tests below:

\paragraph{Bronars Power}

We first use the test designed by \cite{bronars1987power} to generate the benchmark and confirm the power of our designed budget sets. We employ the choices of a hypothetical subject who chooses uniformly randomly among all allocations on each budget line as a point of comparison. Each of the hypothetical simulated subjects makes 25 choices from randomly generated budget sets in the same way that GPT observations and human subjects do. 

We find that 99.9\% of the hypothetical simulated subjects reject GARP.

\paragraph{Predictive Success}

Second, to measure how successful a well-behaved utility function maximization rationalizing GPT observations' choices and human subjects' choices in the baseline condition in comparison to the benchmark, we calculate the predictive success \citep{selten1991properties,beatty2011demanding}, the pass rate for GARP subtracted from (1- Bronars power).

We find that GPT observations (human subjects) outperforms a hypothetical simulated subject by 94.9\% (60.8\%), 88.9\% (79.0\%), 80.9\% (52.1\%) and 91.9\% (43.4\%) in risk, time, social and food preference.

\paragraph{Selten Score}

Furthermore, we calculate how much each GPT observation's (human subject’s) CCEI outperforms the benchmark \citep{dean2016measuring}. We compute each DM's simulated CCEI: the average CCEI of the budget set faced by the DM using the Bronars test. Then we subtract the simulated CCEI from the raw CCEI of each DM, as the Selten score \citep{selten1991properties,dean2016measuring}.

The average Selten score of GPT observations (human subjects) for risk, time, social and food preference is 0.279 (0.262), 0.274 (0.266), 0.275 (0.247) and 0.281 (0.244), which is significantly larger than 0 ($p<0.01$, two-sided two-sample t-test) for all four preference domains.

\paragraph{Bootstrap Power}

Although we base most of our analyses on the uniform random choice benchmark, we also perform a robustness check from an ex post perspective, calculating the probability of rejecting GARP by bootstrapping \citep{andreoni2002giving,andreoni2013power}. We create a population of 10,000 synthetic subjects of which 25 choices are randomly drawn from the actual budget set of our DMs.

The probability of rejecting GARP is 7.9\% (95.4\%), 26.3\% (96.6\%), 26.0\% (99.8\%), and 8.5\% (99.8\%)  in risk, time, social, and time preference for GPT observations (human subjects). This method relies on the heterogeneity of preferences among subjects. When subjects' preferences are indifferent, we may observe a low probability of GARP violation \citep{castillo2018revealed,crawford2013many,miao2021revealed}. The results from the method \citep{andreoni2002giving,andreoni2013power}  align with our preference parameter estimation.

\paragraph{}
\noindent These results all suggest that the budget sets faced by DMs have the power to detect rationality violations, and GPT observations and human subjects have a high rationality level.

\setcounter{table}{0}
\renewcommand{\thetable}{D\arabic{table}}
\setcounter{figure}{0}
\renewcommand{\thefigure}{D\arabic{figure}}
\section{Result}\label{appendix_result}

\begin{figure}[H]
    \centering
    \includegraphics[width=\textwidth]{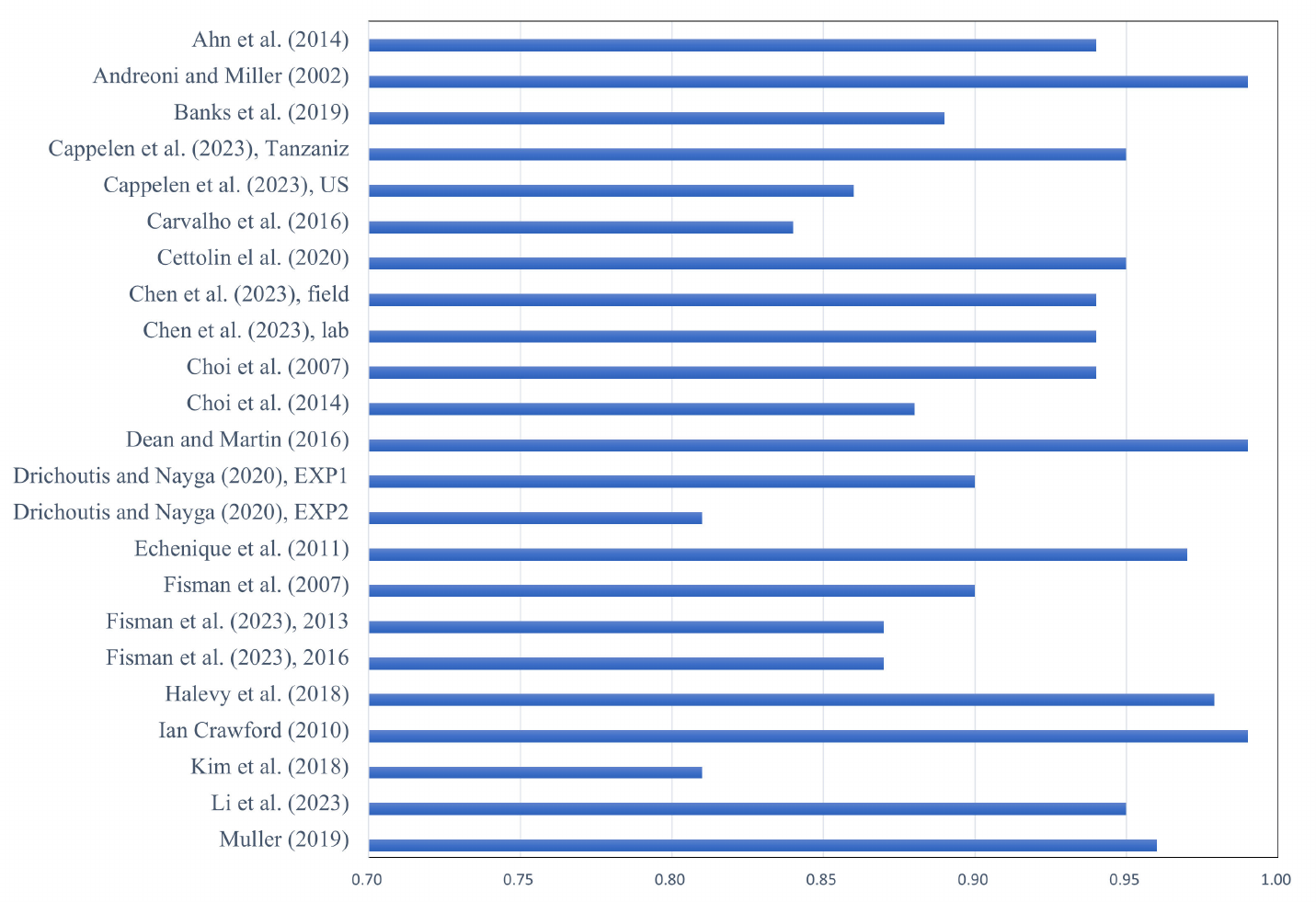}
    \caption{Rationality Score in Prior Studies with Human Subjects. This figure presents the average CCEI values of human subjects in revealed preference studies \citep{ahn2014estimating,andreoni2002giving,banks2019education,cappelen2021development,carvalho2016poverty,cettolin2020cortisol,chen2023consistency,choi2007consistency,choi2014more,dean2016measuring,drichoutis2020economic,echenique2011money,fisman2007individual,fisman2023distributional,halevy2018parametric,crawford2010habits,kim2018role,li2021poverty,muller2019anatomy}}
    \label{figs1}

\end{figure}

\begin{figure}[H]
    \centering
        \includegraphics[width=\linewidth]{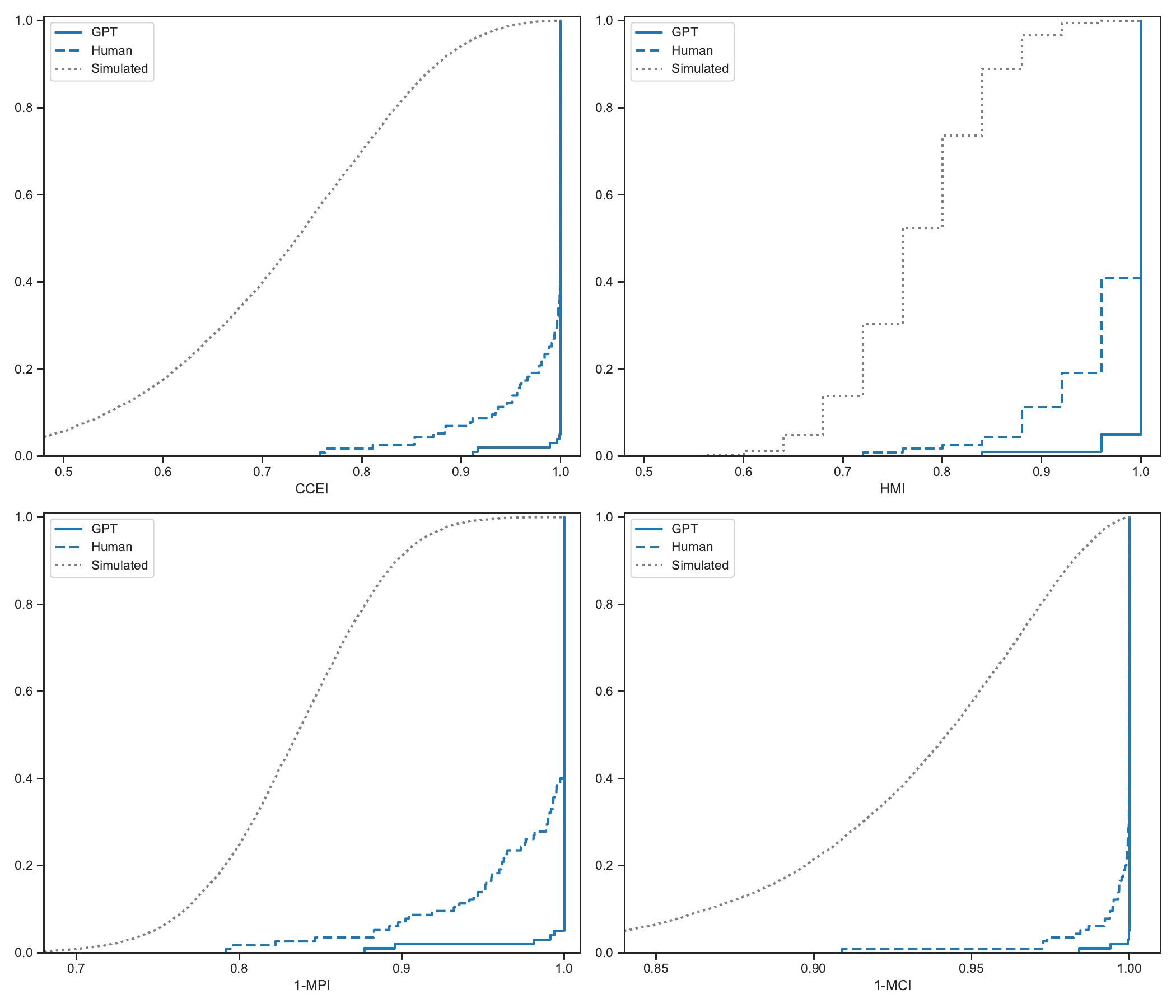}

    \caption{Cumulative Distributions of Rationality Score for Risk Preference. This figure contains four subplots for four different rationality indexes in risk preference: CCEI, HMI, MPI, and MCI. The light dotted lines represent simulated subjects, the dark dashed lines represent human subjects in the human experiment, and the solid lines represent GPT observations.}
    \label{figs2}

\end{figure}

\begin{figure}[H]
    \centering
        \includegraphics[width=\linewidth]{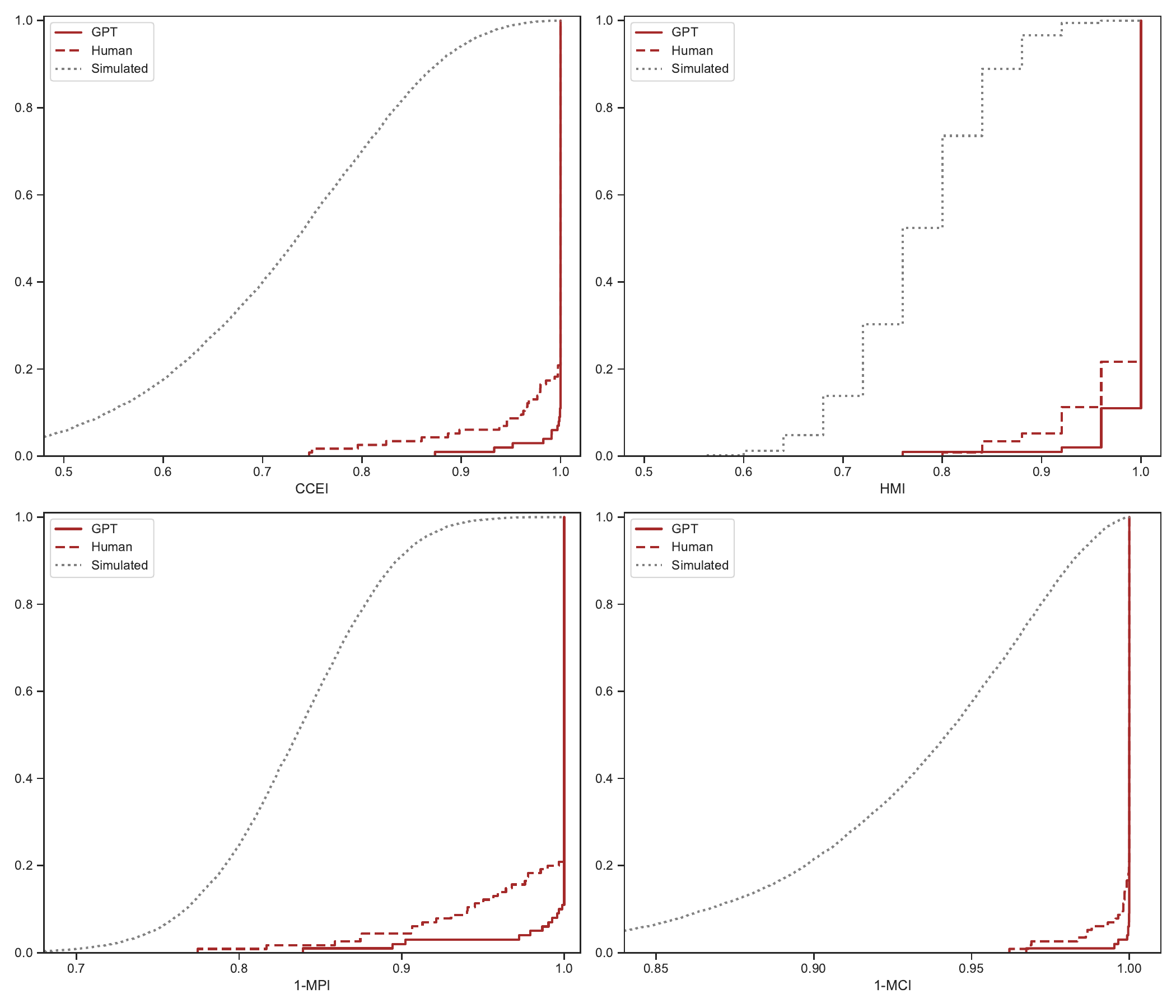}

    \caption{Cumulative Distributions of Rationality Score for Time Preference. This figure contains four subplots for four different rationality indexes in risk preference: CCEI, HMI, MPI, and MCI. The light dotted lines represent simulated subjects, the dark dashed lines represent human subjects in the human experiment, and the solid lines represent GPT observations.}
    \label{figs3}

\end{figure}

\begin{figure}[H]
    \centering
        \includegraphics[width=\linewidth]{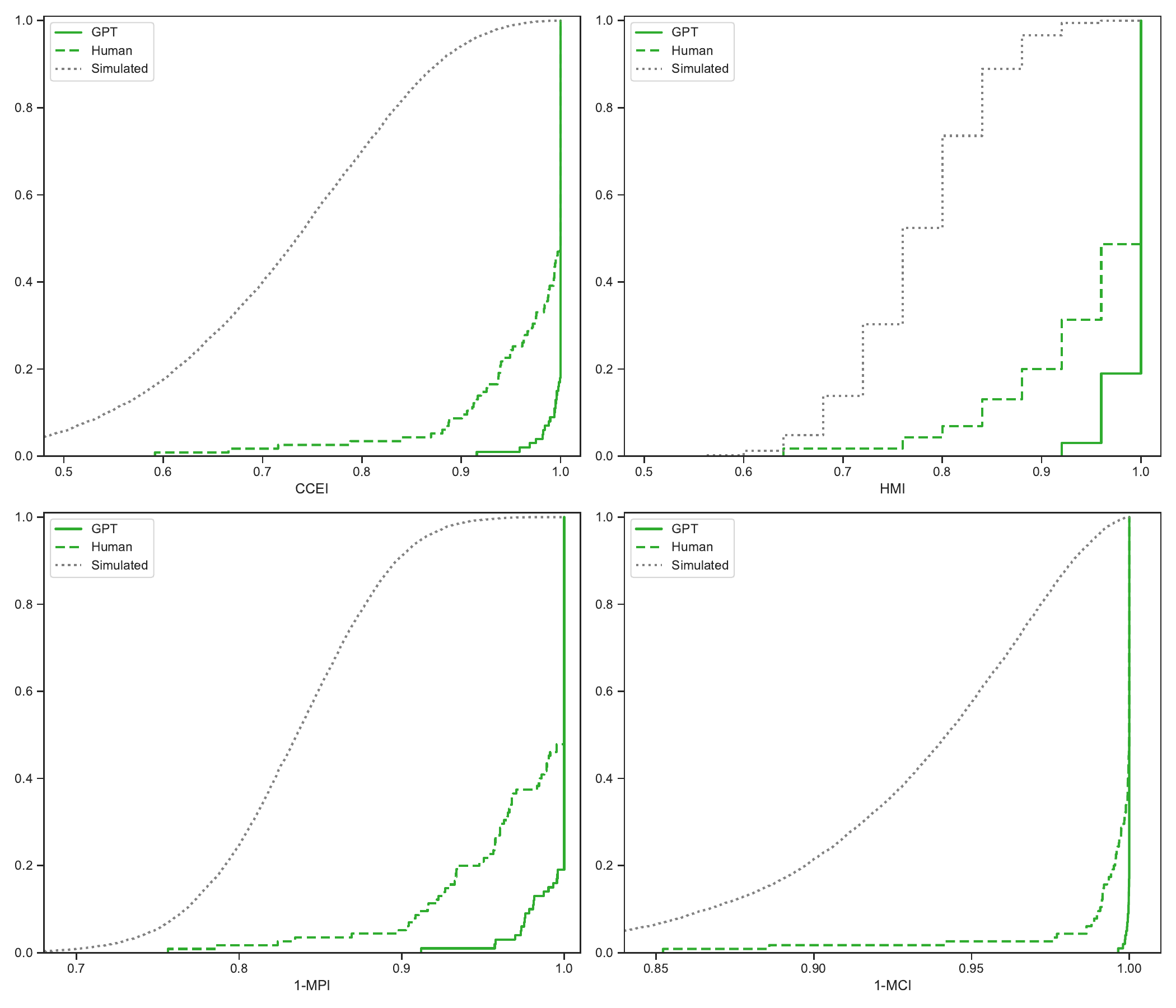}

    \caption{Cumulative Distributions of Rationality Score for Social Preference. This figure contains four subplots for four different rationality indexes in risk preference: CCEI, HMI, MPI, and MCI. The light dotted lines represent simulated subjects, the dark dashed lines represent human subjects in the human experiment, and the solid lines represent GPT observations.}
    \label{figs4}

\end{figure}

\begin{figure}[H]
    \centering
        \includegraphics[width=\linewidth]{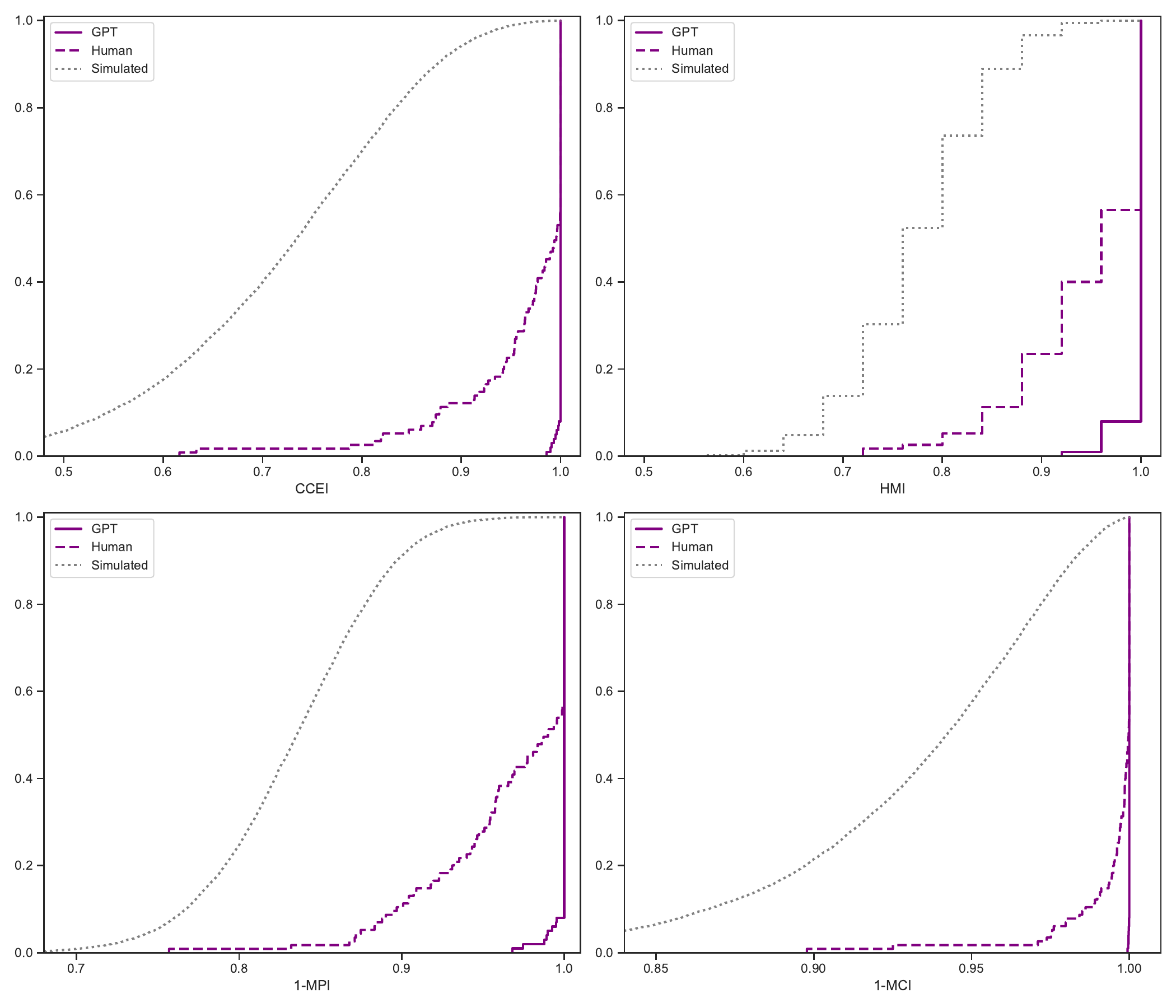}

    \caption{Cumulative Distributions of Rationality Score for Food Preference. This figure contains four subplots for four different rationality indexes in risk preference: CCEI, HMI, MPI, and MCI. The light dotted lines represent simulated subjects, the dark dashed lines represent human subjects in the human experiment, and the solid lines represent GPT observations.}
    \label{figs5}

\end{figure}

\begin{figure}[H]
    \centering
        \includegraphics[width=\linewidth]{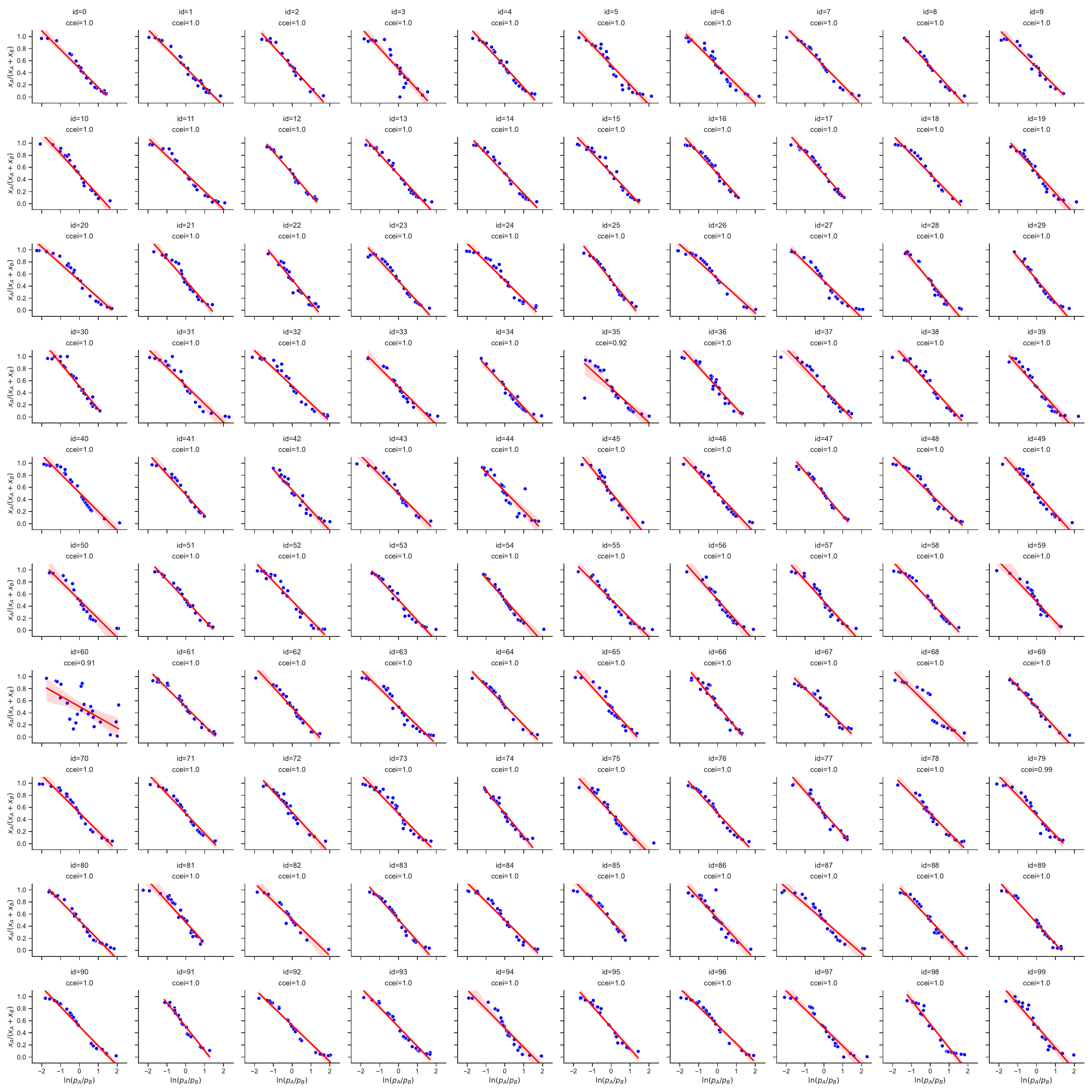}
    \caption{The Relationship of $x_A/(x_A+x_B)$ and $\ln(p_A/p_B)$ for GPT Observations in Risk Preference. This figure is composed of 100 subplots, and each represents one of the 100 experimental trials conducted on GPT in risk preference. The x-axis denotes the log-price ratio $\ln(p_A/p_B)$, and the y-axis represents the quantities share $x_A/(x_A+x_B)$. Each subplot contains 25 scatter points and a corresponding fitted line.}
    \label{figs6}

\end{figure}

\begin{figure}[H]
    \centering
        \includegraphics[width=\linewidth]{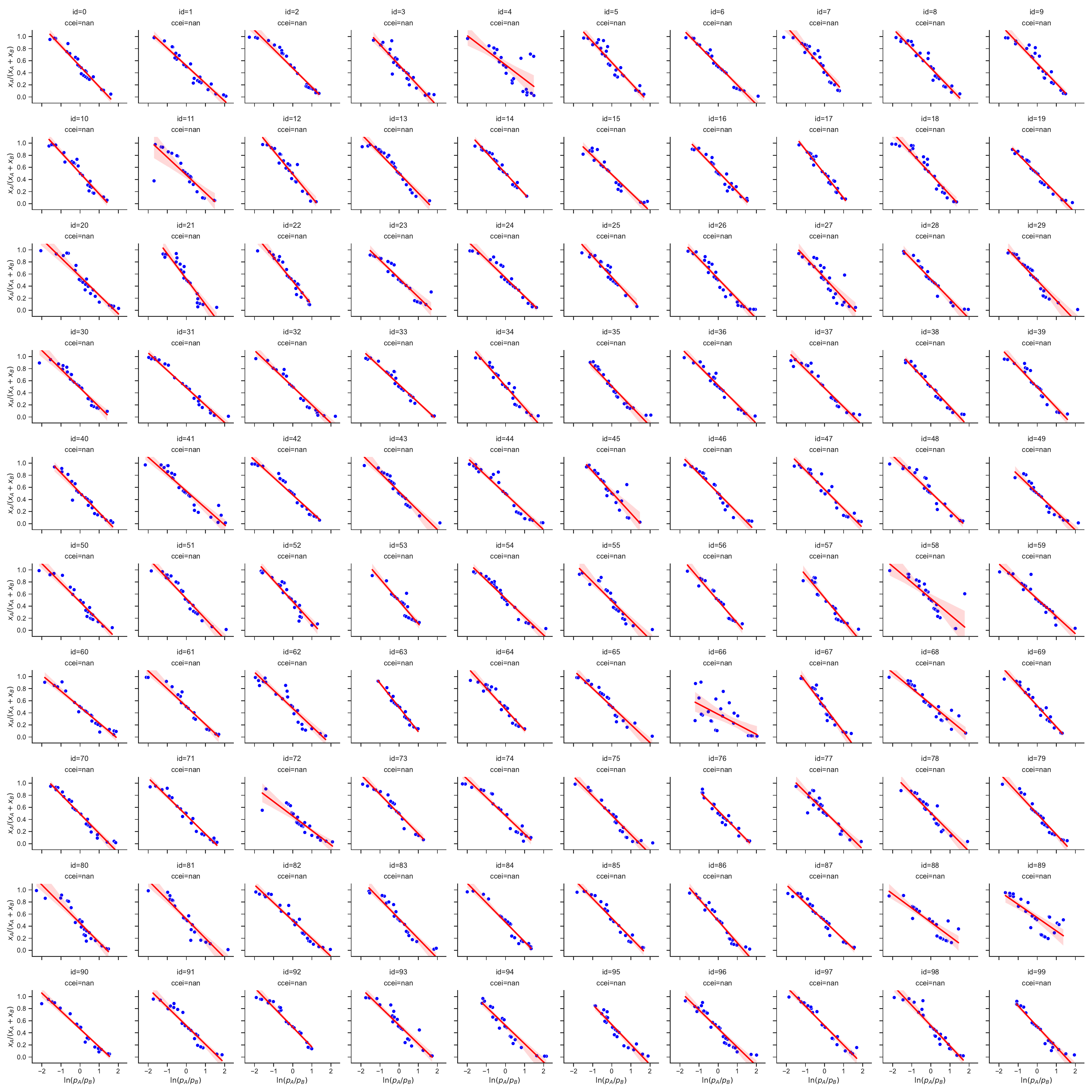}
    \caption{The Relationship of $x_A/(x_A+x_B)$ and $\ln(p_A/p_B)$ for GPT Observations in Time Preference. This figure is composed of 100 subplots, and each represents one of the 100 experimental trials conducted on GPT in time preference. The x-axis denotes the log-price ratio $\ln(p_A/p_B)$, and the y-axis represents the quantities share $x_A/(x_A+x_B)$. Each subplot contains 25 scatter points and a corresponding fitted line.}
    \label{figs7}

\end{figure}

\begin{figure}[H]
    \centering
        \includegraphics[width=\linewidth]{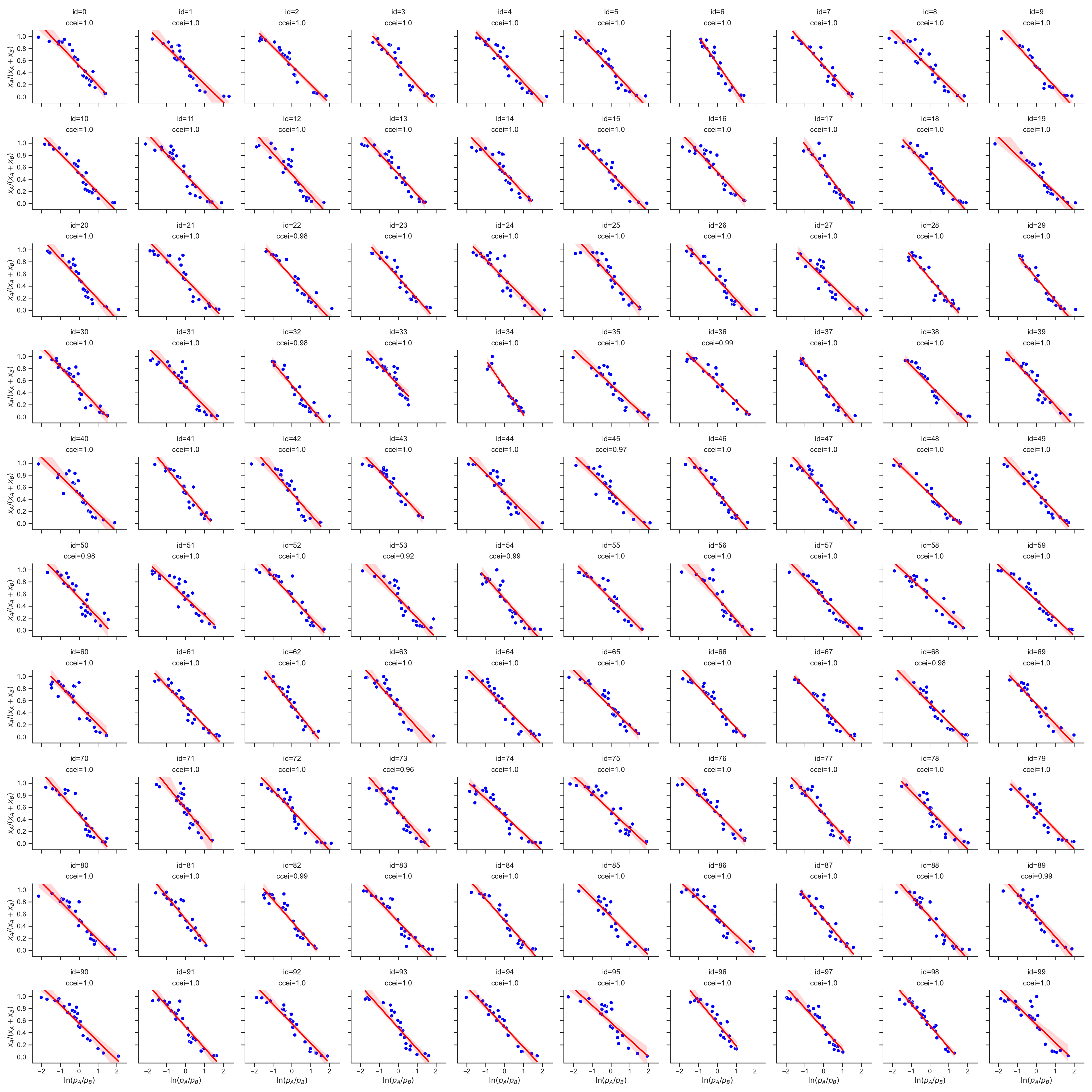}
    \caption{The Relationship of $x_A/(x_A+x_B)$ and $\ln(p_A/p_B)$ for GPT Observations in Social Preference. This figure is composed of 100 subplots, and each represents one of the 100 experimental trials conducted on GPT in social preference. The x-axis denotes the log-price ratio $\ln(p_A/p_B)$, and the y-axis represents the quantities share $x_A/(x_A+x_B)$. Each subplot contains 25 scatter points and a corresponding fitted line.}
    \label{figs8}

\end{figure}

\begin{figure}[H]
    \centering
        \includegraphics[width=\linewidth]{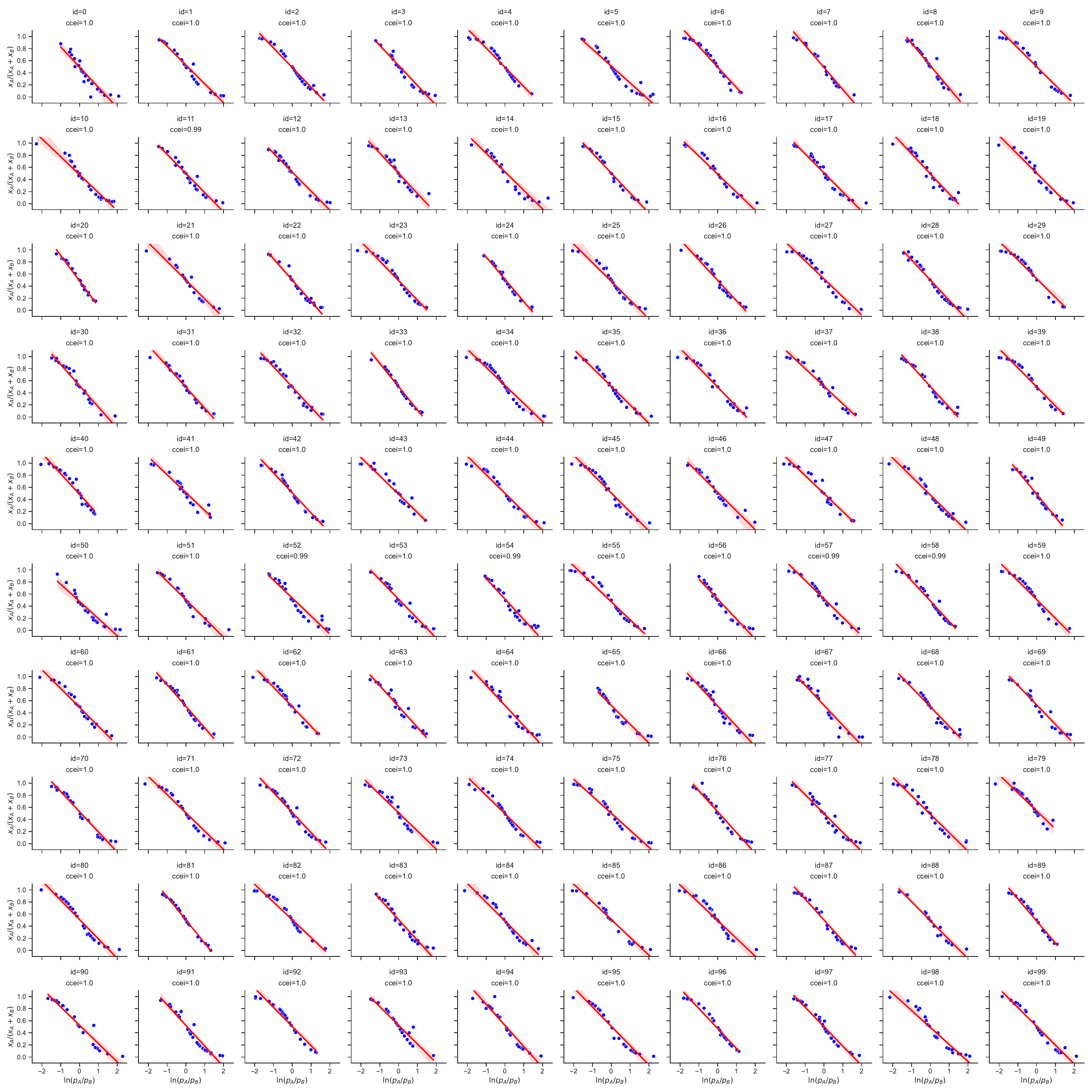}
    \caption{The Relationship of $x_A/(x_A+x_B)$ and $\ln(p_A/p_B)$ for GPT Observations in Food Preference. This figure is composed of 100 subplots, and each represents one of the 100 experimental trials conducted on GPT in food preference. The x-axis denotes the log-price ratio $\ln(p_A/p_B)$, and the y-axis represents the quantities share $x_A/(x_A+x_B)$. Each subplot contains 25 scatter points and a corresponding fitted line.}
    \label{figs9}

\end{figure}

\begin{figure}[H]
    \centering
        \includegraphics[width=\linewidth]{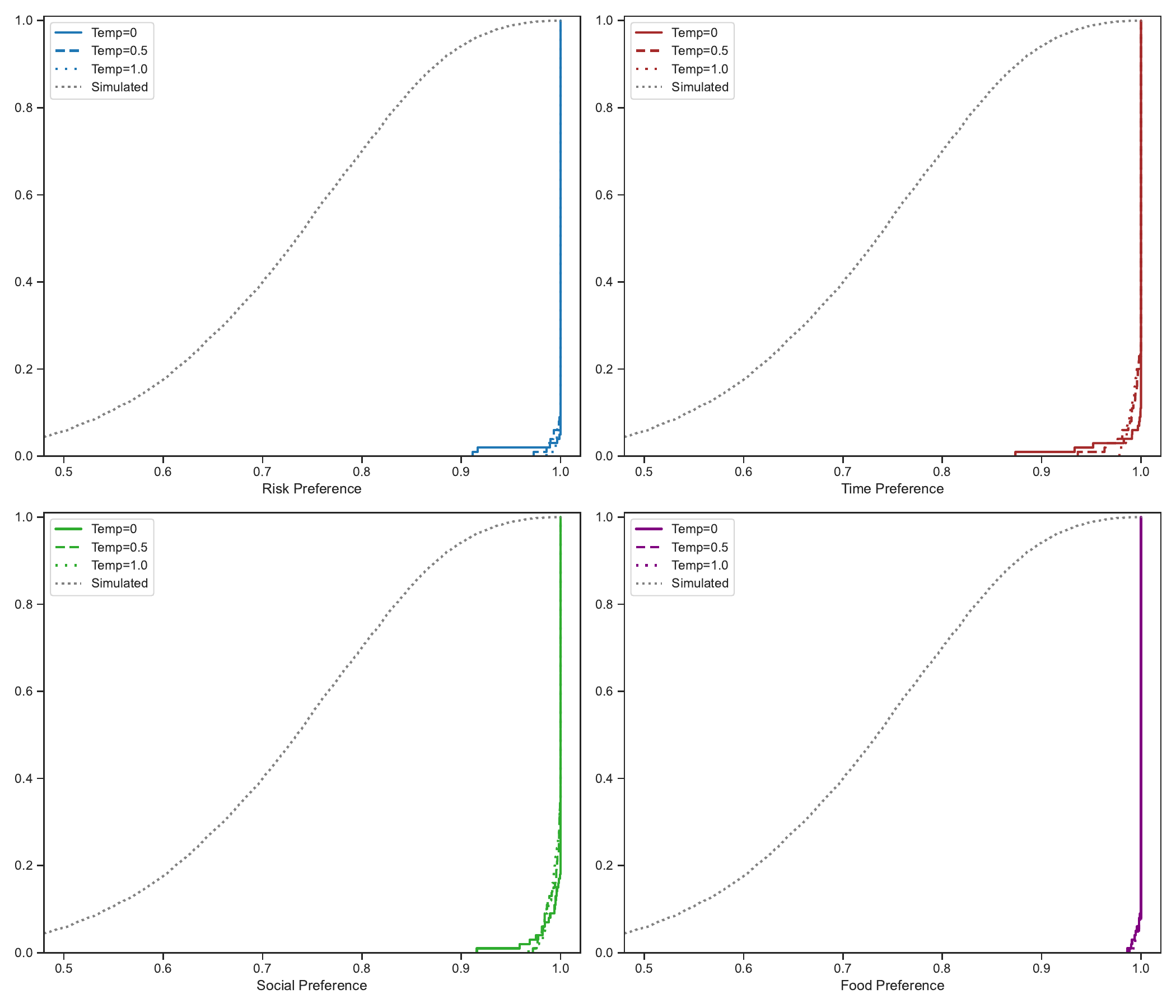}

    \caption{Cumulative Distributions of GPT CCEI with Temperature Variations. This figure contains four subplots, and each represents a different preference domain: risk, time, social or food preference.}
    \label{figs10}
\end{figure}

\begin{figure}[H]
    \centering
   
        \includegraphics[width=\linewidth]{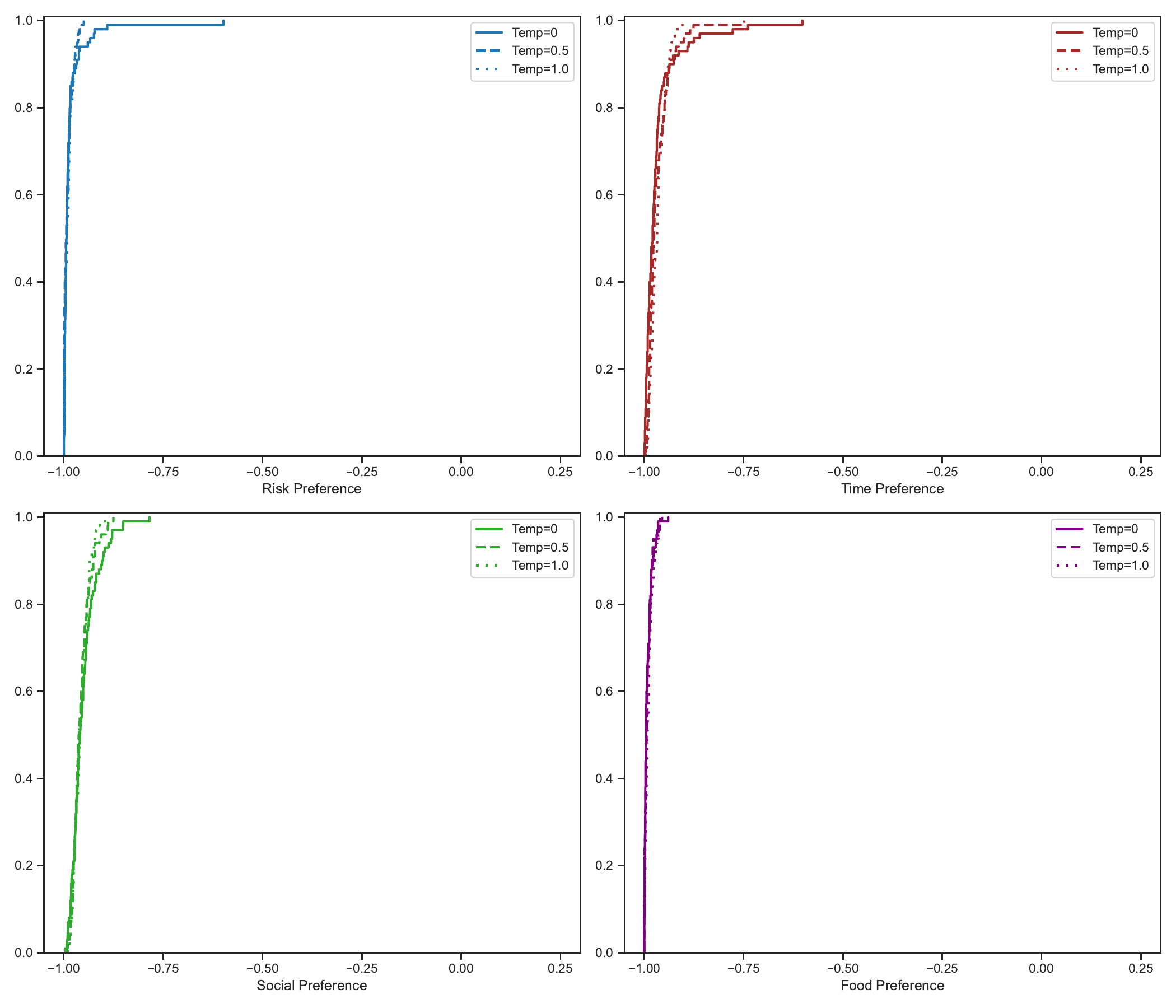}

    \caption{Cumulative Distributions of GPT Spearman's Correlation Coefficients of $\ln(x_A/x_B)$ and $\ln(p_A/p_B)$ with Temperature Variations, which serves as a proxy for the degree of downward-sloping demand. This figure contains four subplots, and each represents a different preference domain: risk, time, social or food preference.}
    \label{figs11}

\end{figure}

\begin{figure}[H]
    \centering
        \includegraphics[width=\linewidth]{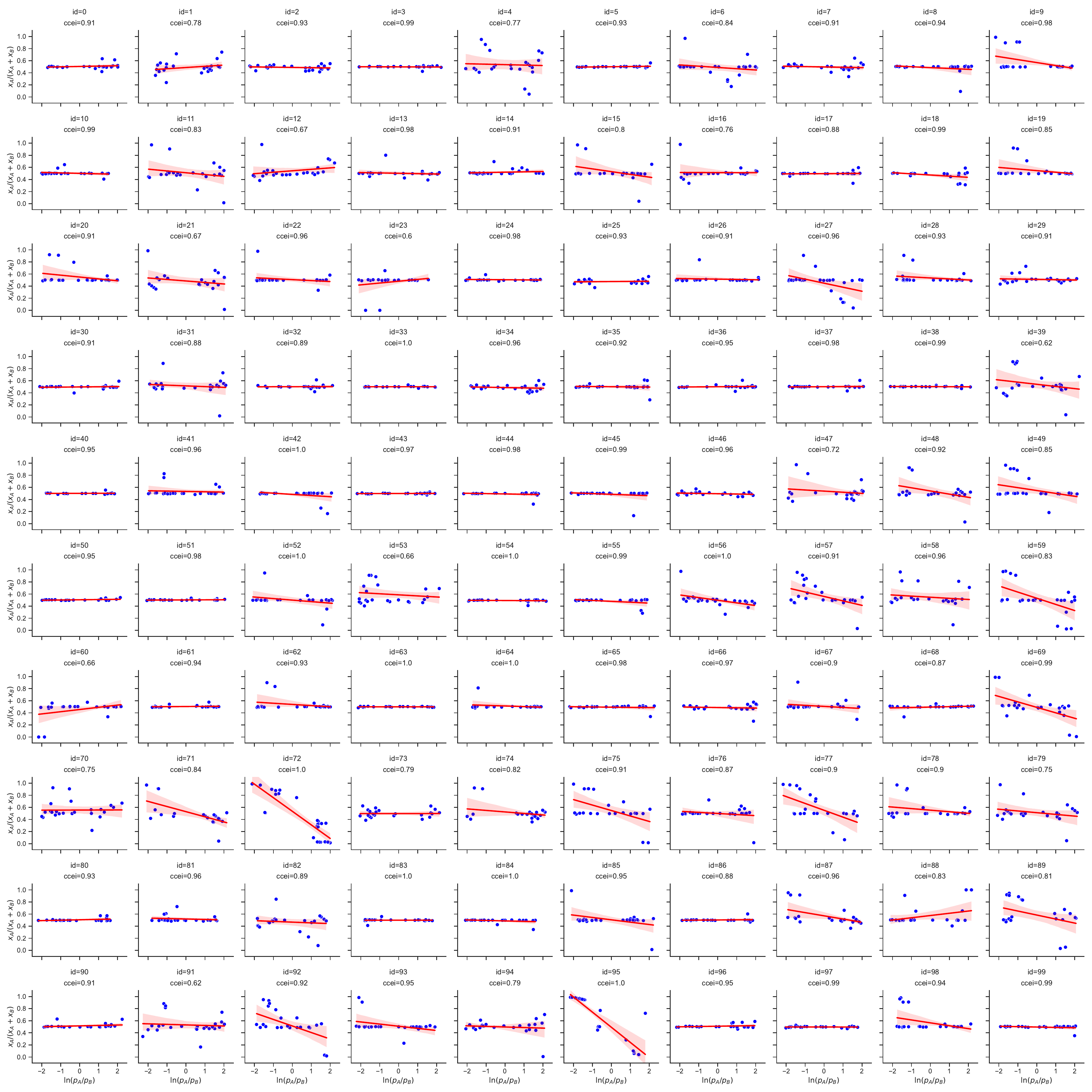}
    \caption{The relationship of the quantities share $x_A/(x_A+x_B)$ and the log-price ratio $\ln(p_A/p_B)$ for GPT observations in risk preference with price framing. This figure is composed of 100 subplots, and each represents one of the 100 experimental trials conducted on GPT in risk preference. The x-axis denotes the log-price ratio $\ln(p_A/p_B)$, and the y-axis represents the quantities share $x_A/(x_A+x_B)$. Each subplot contains 25 scatter points, which represent observed decisions, and a corresponding fitted line.}
    \label{figs12}

\end{figure}

\begin{figure}[H]
    \centering
       \includegraphics[width=\linewidth]{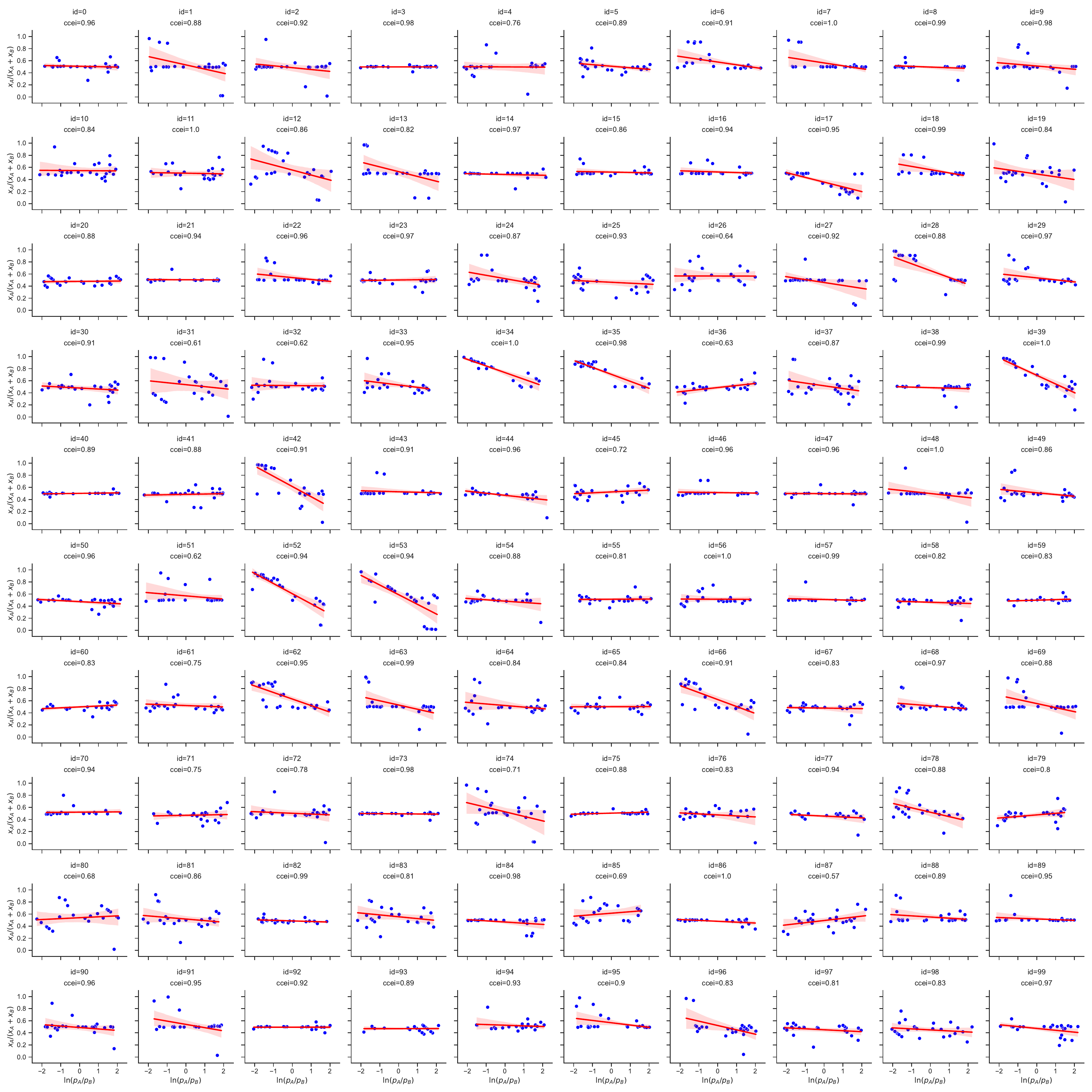}
    \caption{The relationship of the quantities share $x_A/(x_A+x_B)$ and the log-price ratio $\ln(p_A/p_B)$ for GPT observations in time preference with price framing. This figure is composed of 100 subplots, and each represents one of the 100 experimental trials conducted on GPT in time preference. The x-axis denotes the log-price ratio $\ln(p_A/p_B)$, and the y-axis represents the quantities share $x_A/(x_A+x_B)$. Each subplot contains 25 scatter points, which represent observed decisions, and a corresponding fitted line.}
    \label{figs13}

\end{figure}

\begin{figure}[H]
    \centering
       \includegraphics[width=\linewidth]{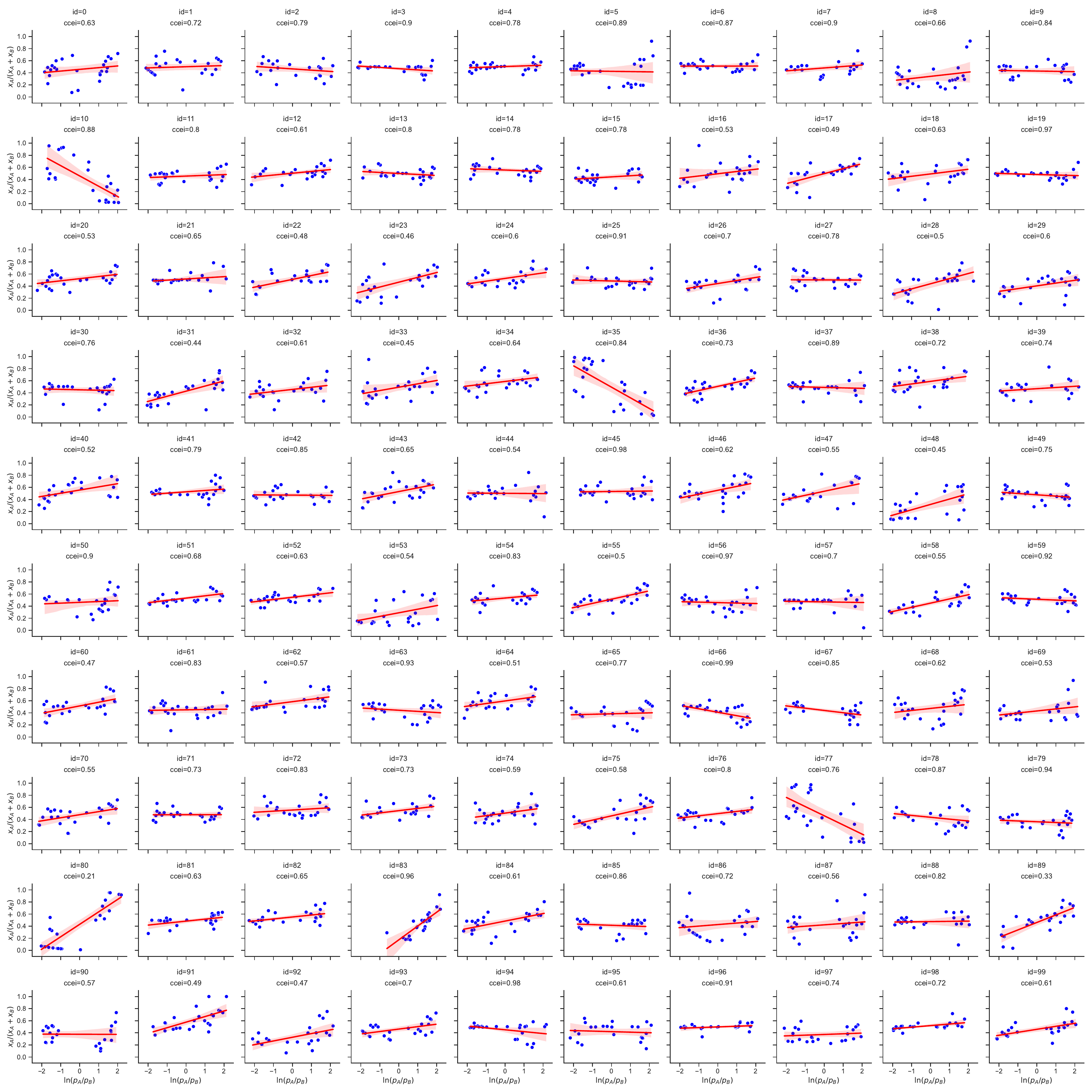}
    \caption{The relationship of the quantities share $x_A/(x_A+x_B)$ and the log-price ratio $\ln(p_A/p_B)$ for GPT observations in social preference with price framing. This figure is composed of 100 subplots, and each represents one of the 100 experimental trials conducted on GPT in social preference. The x-axis denotes the log-price ratio $\ln(p_A/p_B)$, and the y-axis represents the quantities share $x_A/(x_A+x_B)$. Each subplot contains 25 scatter points, which represent observed decisions, and a corresponding fitted line.}
    \label{figs14}

\end{figure}

\begin{figure}[H]
    \centering
       \includegraphics[width=\linewidth]{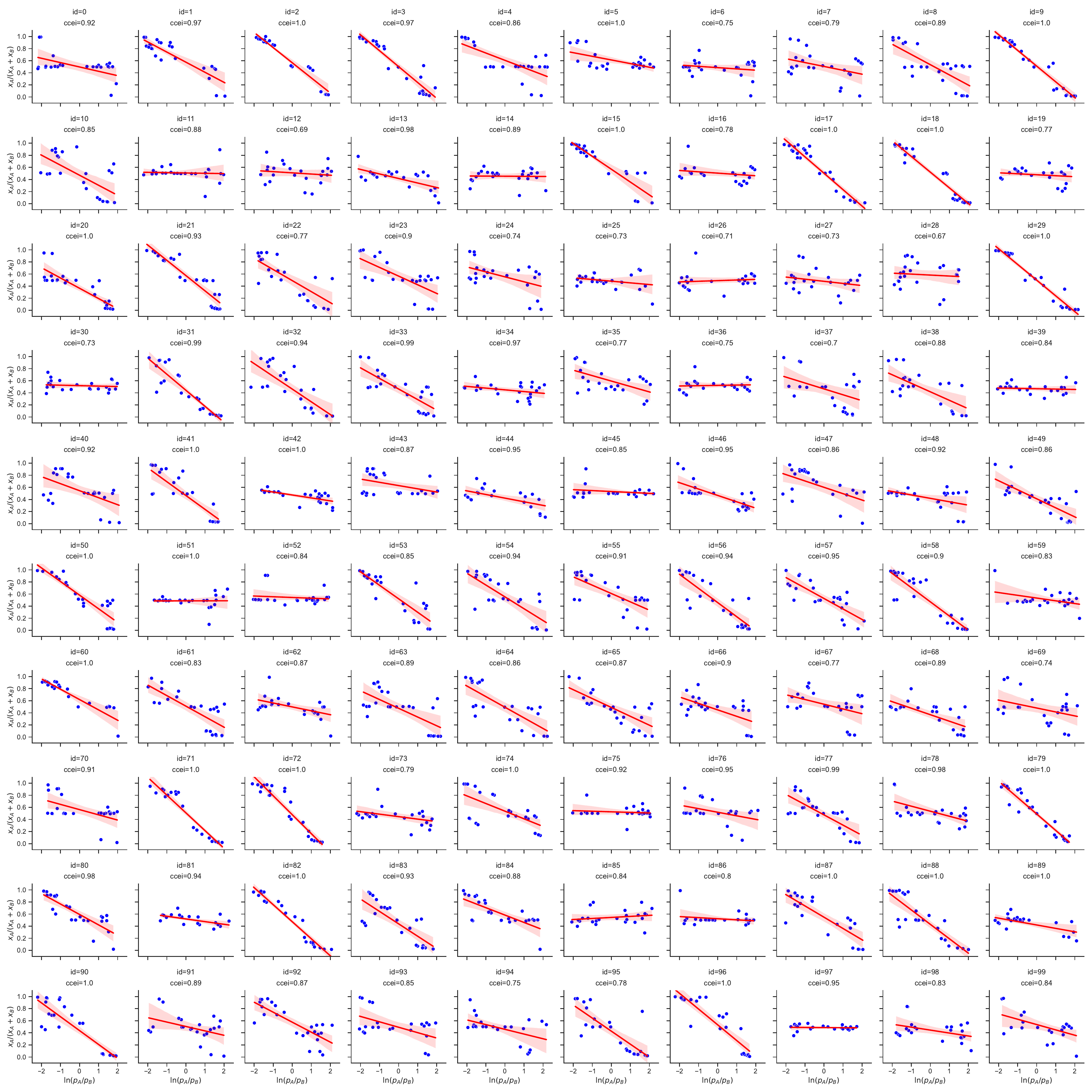}
    \caption{The relationship of the quantities share $x_A/(x_A+x_B)$ and the log-price ratio $\ln(p_A/p_B)$ for GPT observations in food preference with price framing. This figure is composed of 100 subplots, and each represents one of the 100 experimental trials conducted on GPT in food preference. The x-axis denotes the log-price ratio $\ln(p_A/p_B)$, and the y-axis represents the quantities share $x_A/(x_A+x_B)$. Each subplot contains 25 scatter points, which represent observed decisions, and a corresponding fitted line.}
    \label{figs15}

\end{figure}

\begin{figure}[H]
    \centering
       \includegraphics[width=\linewidth]{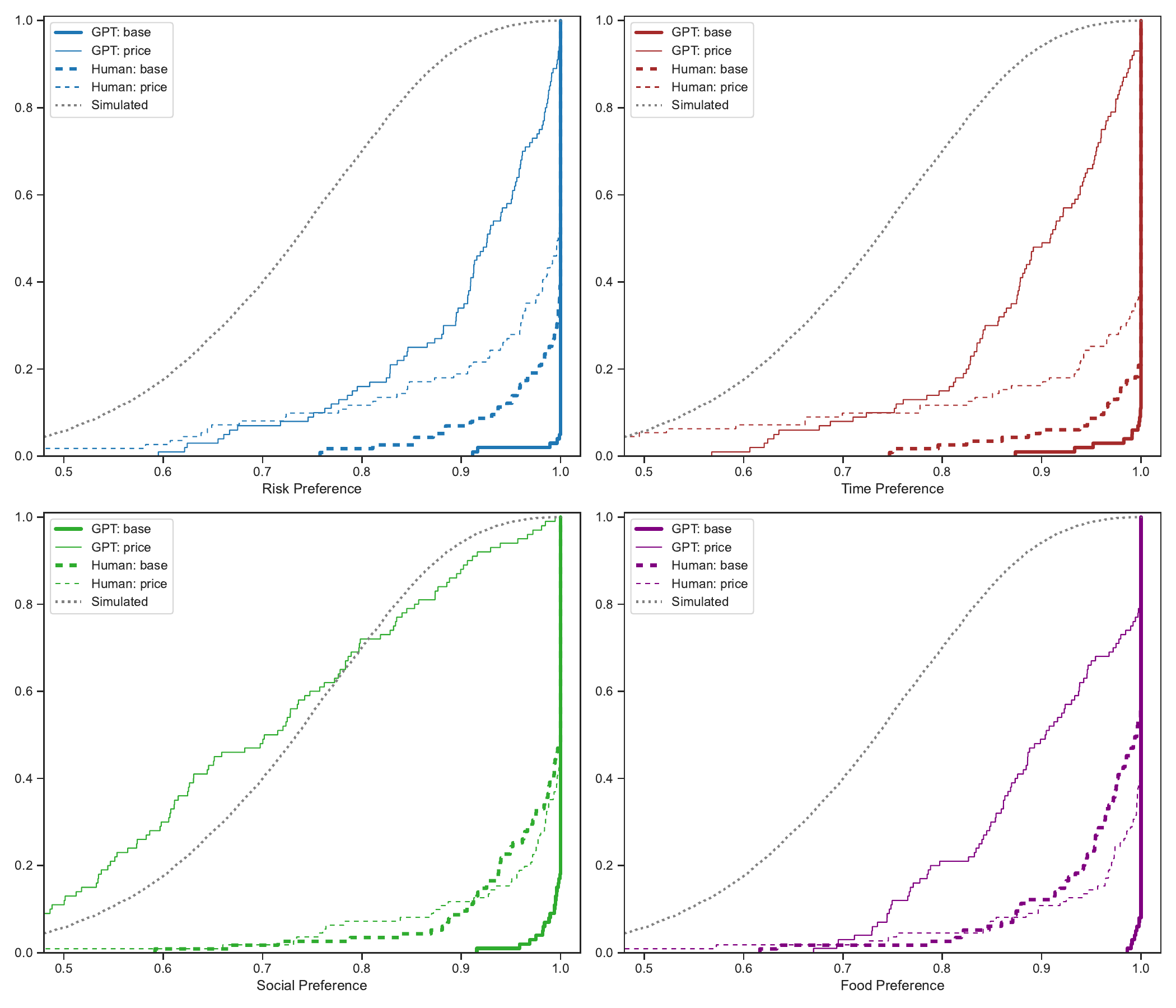}

    \caption{Cumulative Distributions of the CCEI Values with  Price Framing. This figure contains four subplots, and each represent a different preference domain: risk, time, social or food preference. The light dotted lines represent simulated subjects, the dark dashed lines represent human subjects in the human experiment, and the solid lines represent GPT observations.}
    \label{figs16}
\end{figure}

\begin{figure}[H]
    \centering
       \includegraphics[width=\linewidth]{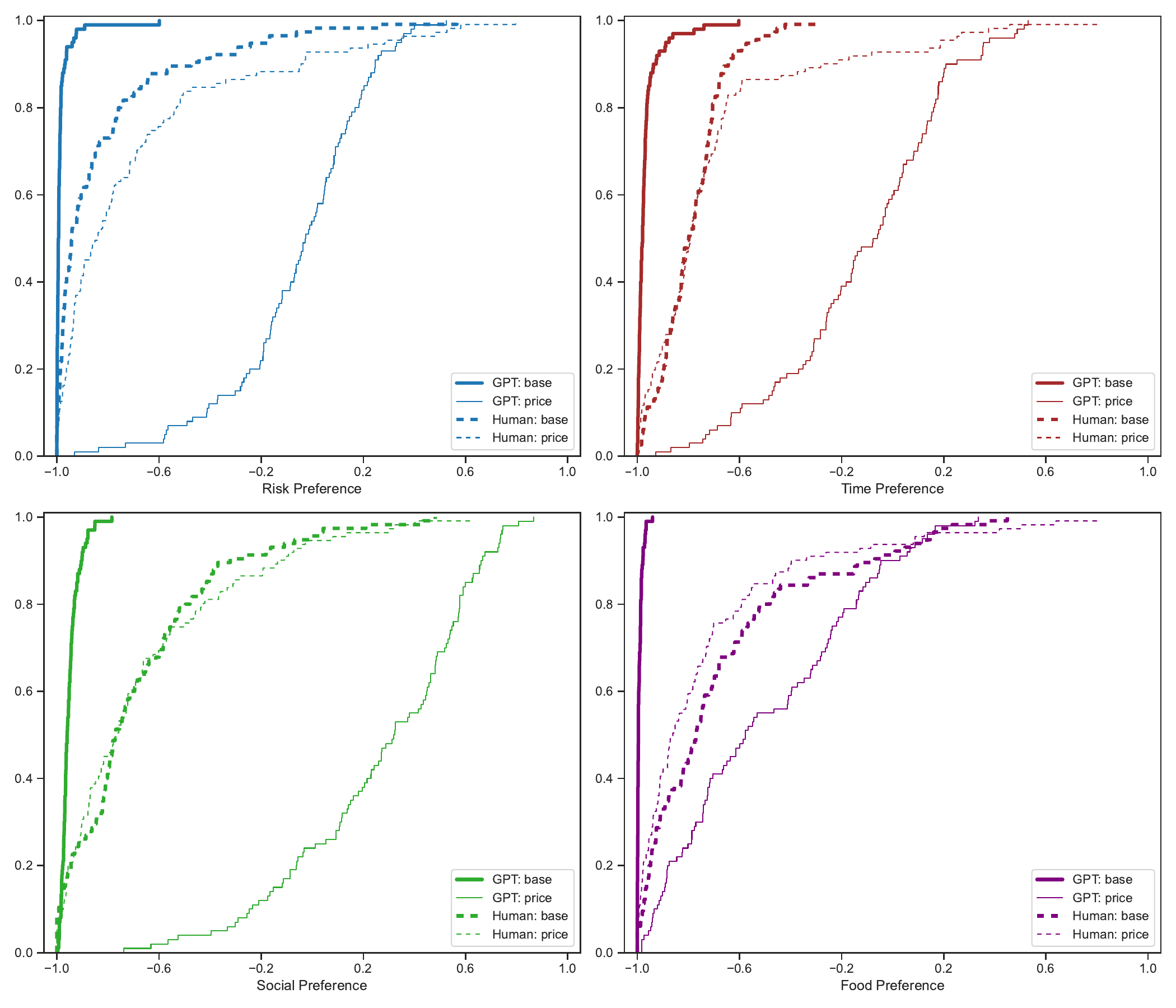}

    \caption{Cumulative Distributions of Spearman's Correlation Coefficients of $\ln(x_A/x_B)$ and $\ln(p_A/p_B)$ with Price Framing. This figure  contains four subplots, and each represents a different preference domain: risk, time, social or food preference. The dark dashed lines represent human subjects in the human experiment, and the solid lines represent GPT observations.}
    \label{figs17}
\end{figure}

\begin{figure}[H]
    \centering
       \includegraphics[width=\linewidth]{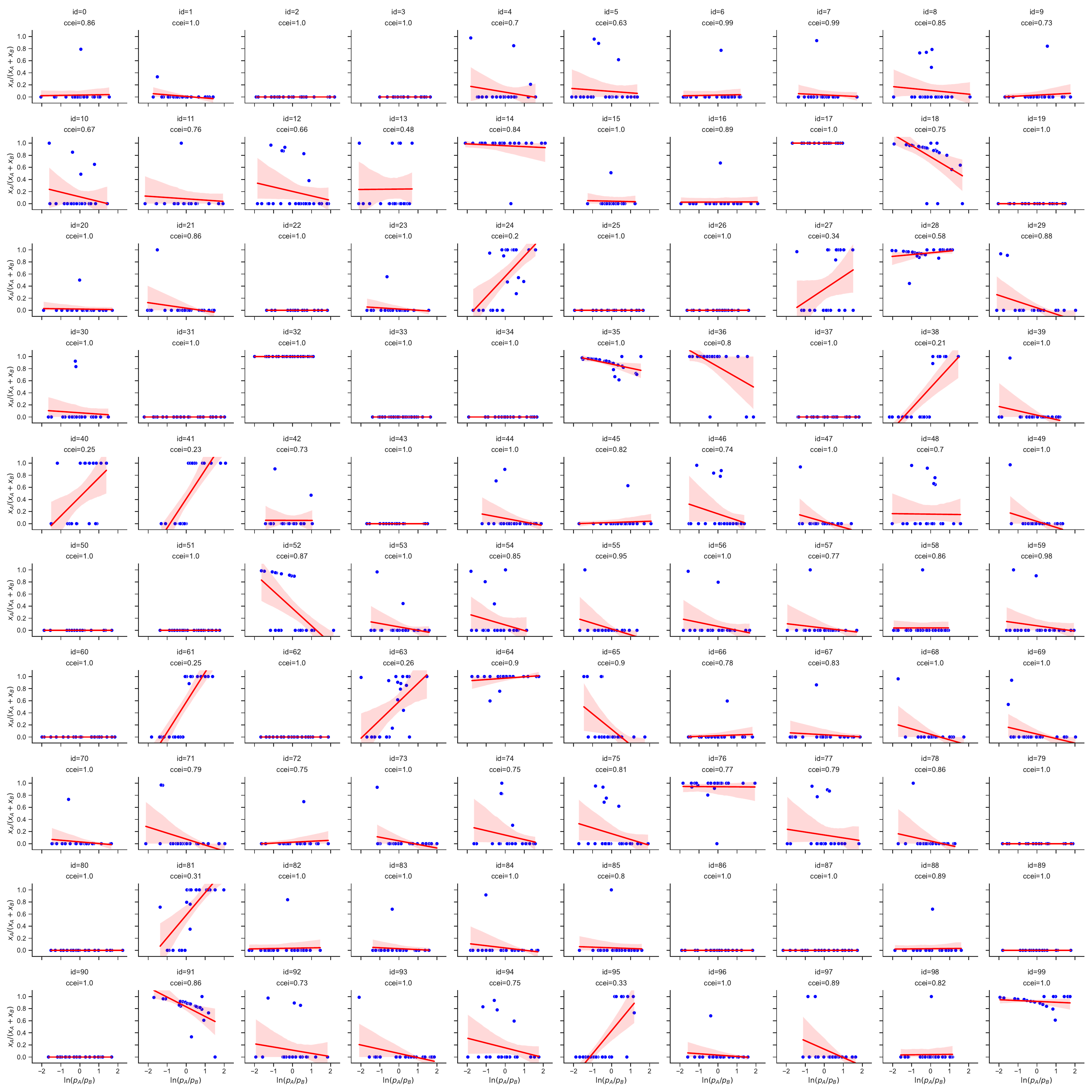}
    \caption{The relationship of the quantities share $x_A/(x_A+x_B)$ and the log-price ratio $\ln(p_A/p_B)$ for GPT observations in risk preference with discrete choice. This figure is composed of 100 subplots, and each represents one of the 100 experimental trials conducted on GPT in risk preference. The x-axis denotes the log-price ratio $\ln(p_A/p_B)$, and the y-axis represents the quantities share $x_A/(x_A+x_B)$. Each subplot contains 25 scatter points, which represent observed decisions, and a corresponding fitted line.}
    \label{figs18}

\end{figure}

\begin{figure}[H]
    \centering
       \includegraphics[width=\linewidth]{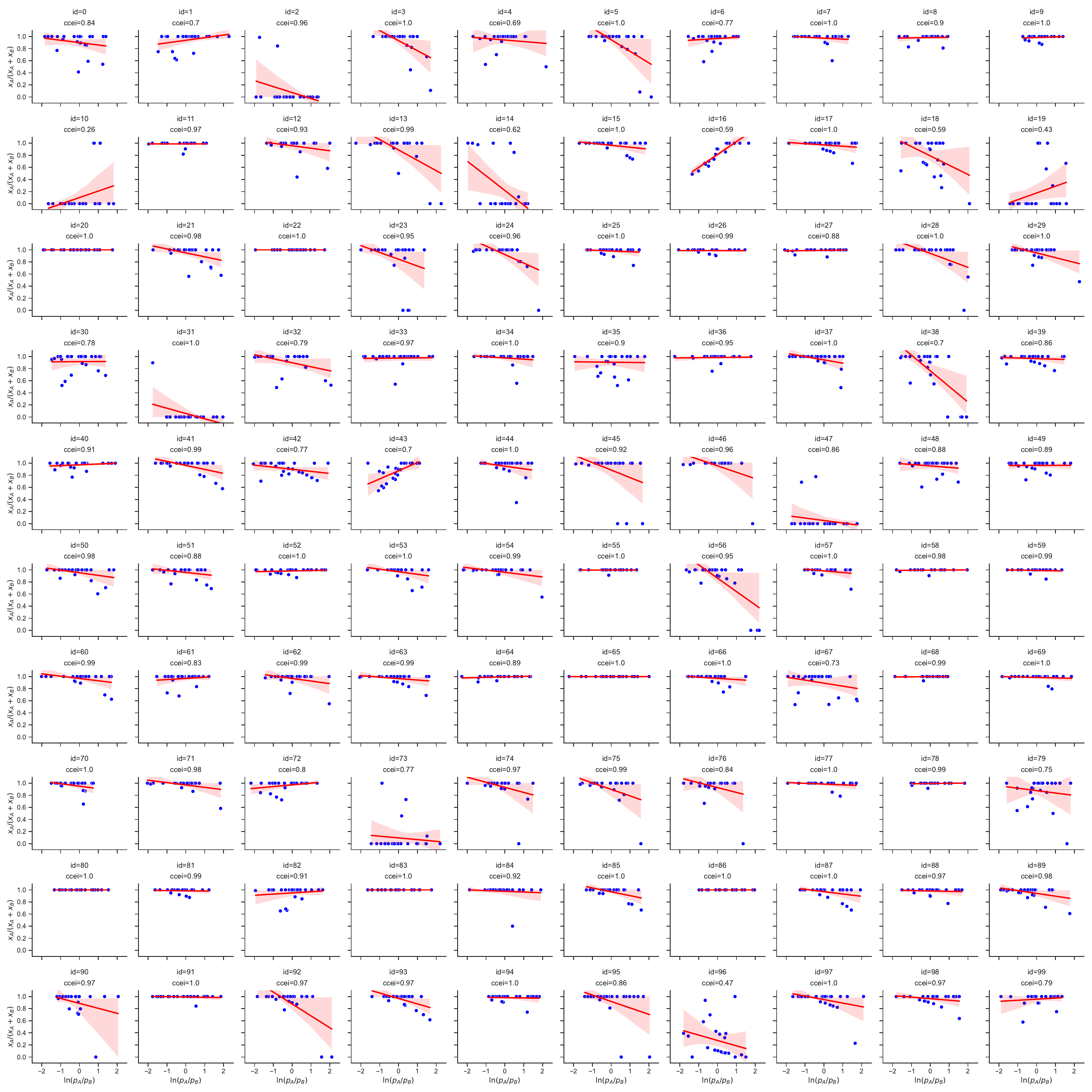}
    \caption{The relationship of the quantities share $x_A/(x_A+x_B)$ and the log-price ratio $\ln(p_A/p_B)$ for GPT observations in time preference with discrete choice. This figure is composed of 100 subplots, and each represents one of the 100 experimental trials conducted on GPT in time preference. The x-axis denotes the log-price ratio $\ln(p_A/p_B)$, and the y-axis represents the quantities share $x_A/(x_A+x_B)$. Each subplot contains 25 scatter points, which represent observed decisions, and a corresponding fitted line.}
    \label{figs19}

\end{figure}

\begin{figure}[H]
    \centering
       \includegraphics[width=\linewidth]{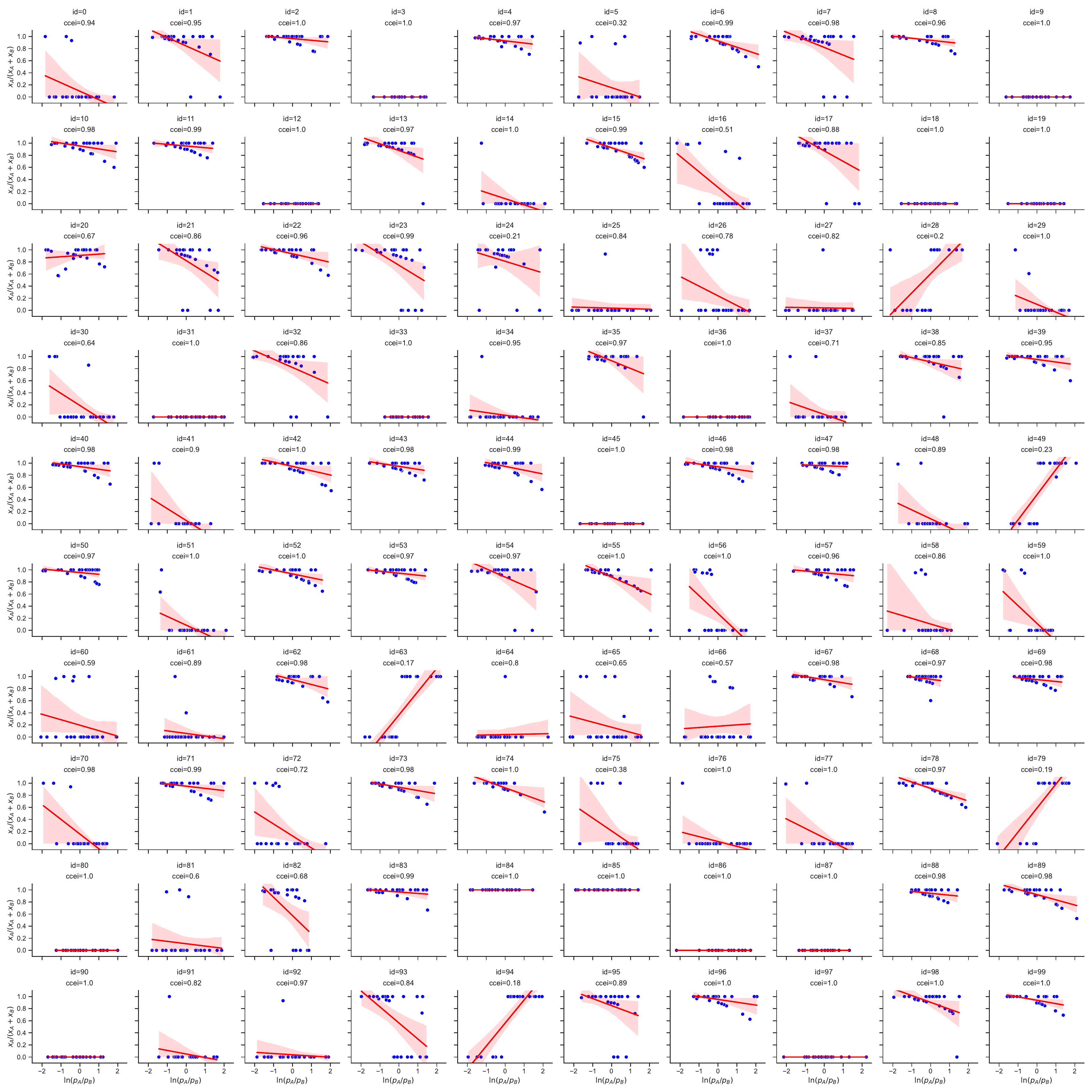}
    \caption{The relationship of the quantities share $x_A/(x_A+x_B)$ and the log-price ratio $\ln(p_A/p_B)$ for GPT observations in social preference with discrete choice. This figure is composed of 100 subplots, and each represents one of the 100 experimental trials conducted on GPT in social preference. The x-axis denotes the log-price ratio $\ln(p_A/p_B)$, and the y-axis represents the quantities share $x_A/(x_A+x_B)$. Each subplot contains 25 scatter points, which represent observed decisions, and a corresponding fitted line.}
    \label{figs20}

\end{figure}

\begin{figure}[H]
    \centering
       \includegraphics[width=\linewidth]{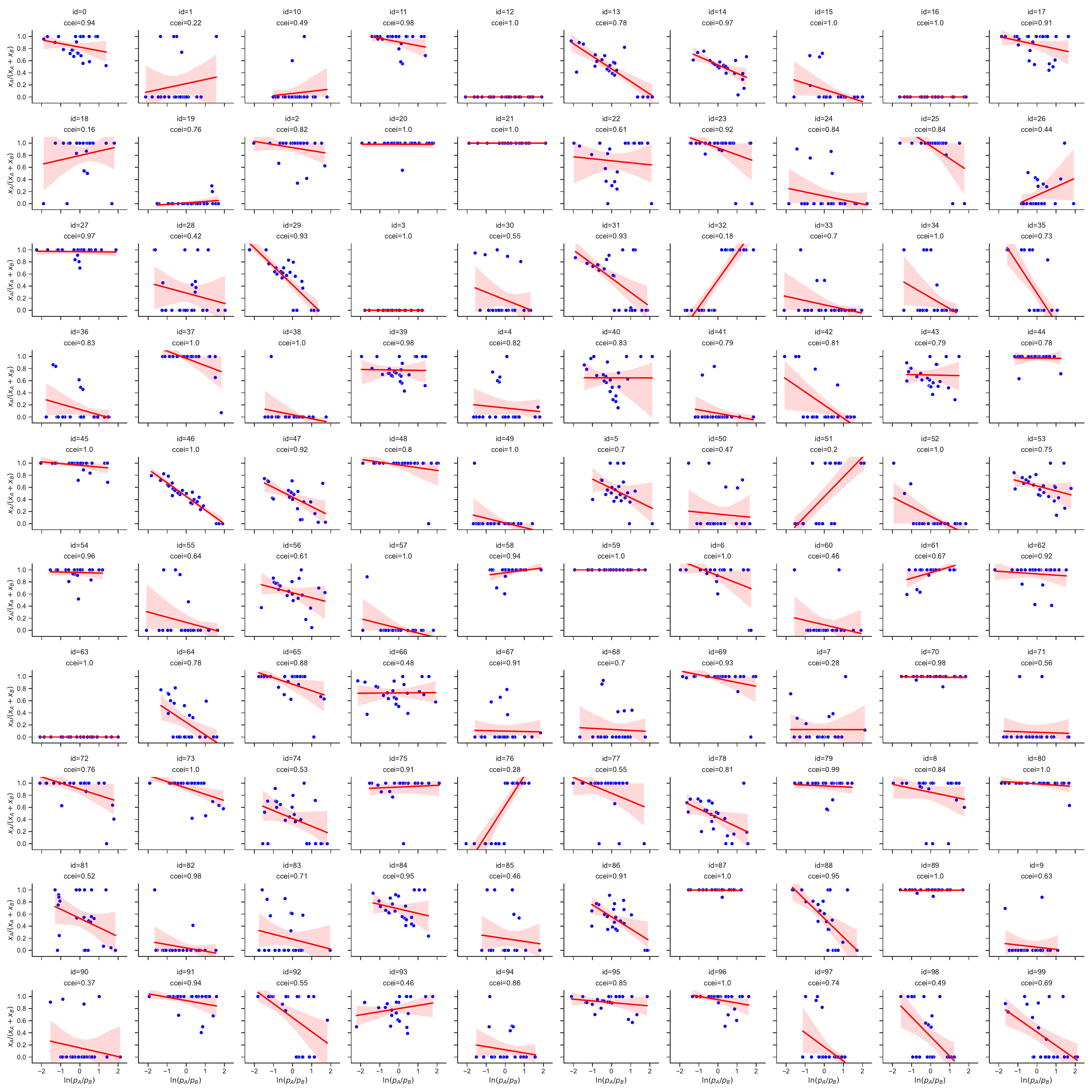}
    \caption{The relationship of the quantities share $x_A/(x_A+x_B)$ and the log-price ratio $\ln(p_A/p_B)$ for GPT observations in food preference with discrete choice. This figure is composed of 100 subplots, and each represents one of the 100 experimental trials conducted on GPT in risk preference. The x-axis denotes the log-price ratio $\ln(p_A/p_B)$, and the y-axis represents the quantities share $x_A/(x_A+x_B)$. Each subplot contains 25 scatter points, which represent observed decisions, and a corresponding fitted line.}
    \label{figs21}
\end{figure}

\begin{figure}[H]
    \centering
     \includegraphics[width=\linewidth]{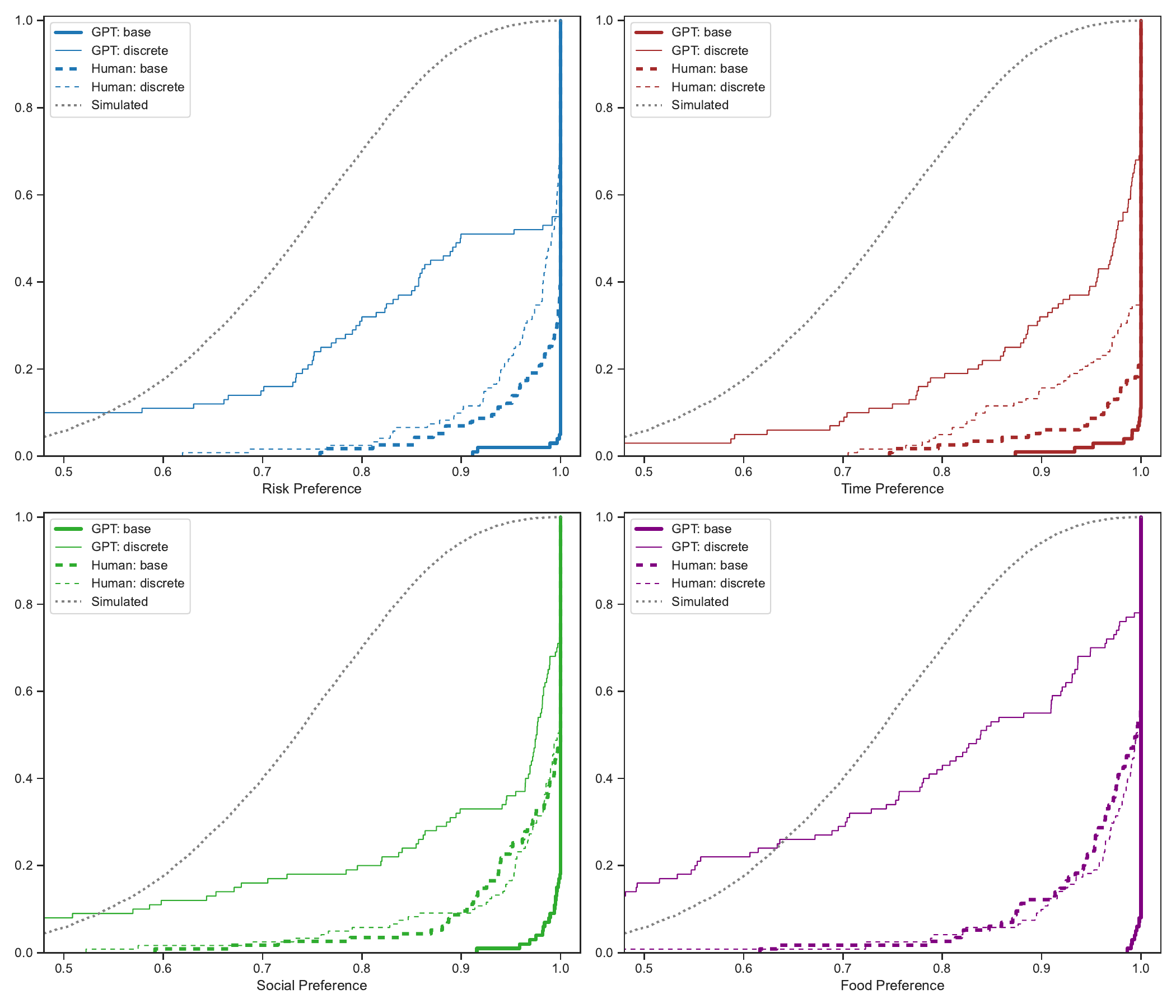}
    \caption{Cumulative Distributions of the CCEI Values with Discrete Choice. This figure contains four subplots, and each represent a different preference domain: risk, time, social or food preference. The light dotted lines represent simulated subjects, the dark dashed lines represent human subjects in the human experiment, and the solid lines represent GPT observations.}
    \label{figs22}
\end{figure}

\begin{figure}[H]
    \centering
     \includegraphics[width=\linewidth]{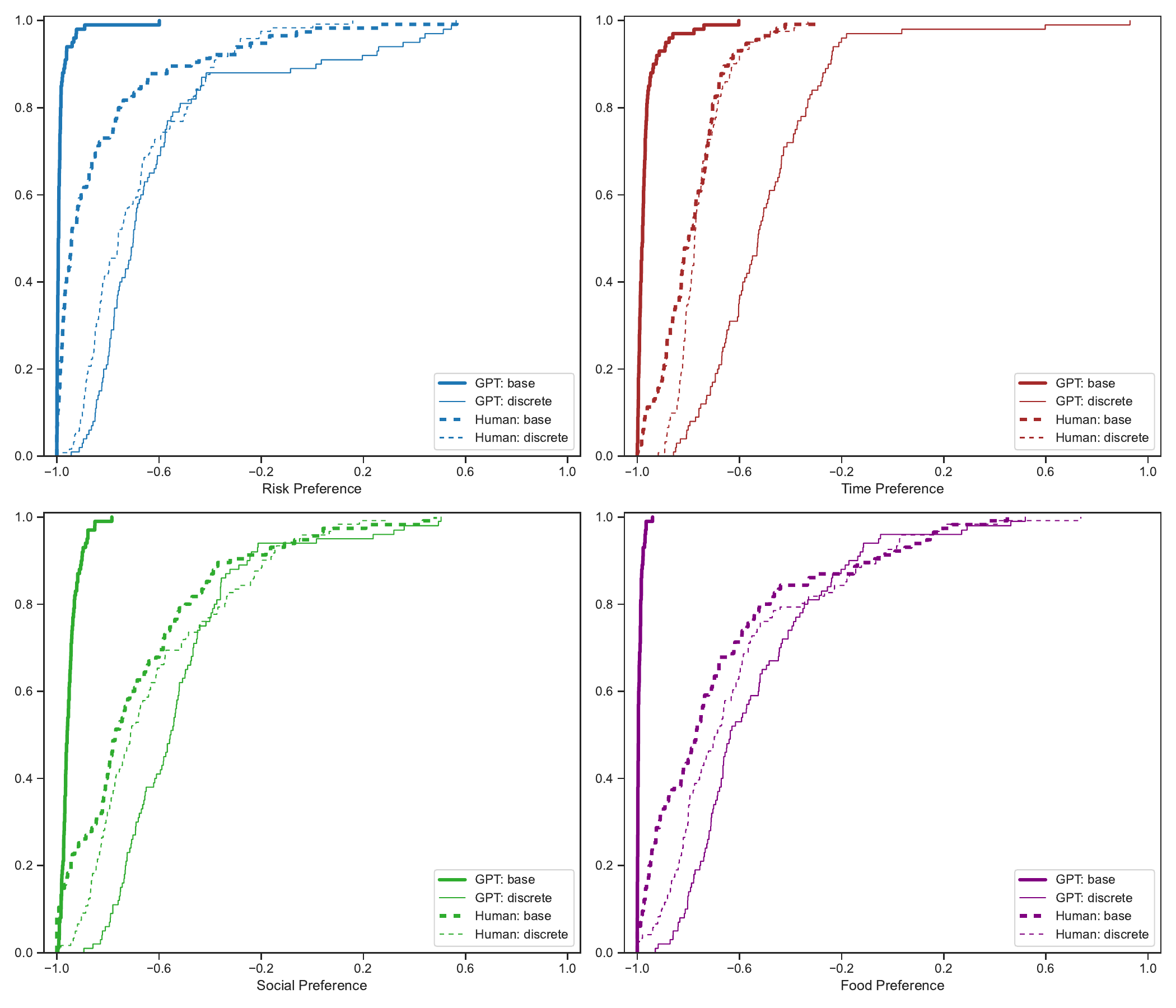}

    \caption{Cumulative Distributions of Spearman's Correlation Coefficients of $\ln(x_A/x_B)$ and $\ln(p_A/p_B)$ with Discrete Choice. This figure  contains four subplots, and each represents a different preference domain: risk, time, social or food preference. The dark dashed lines represent human subjects in the human experiment, and the solid lines represent GPT observations.}
    \label{figs23}

\end{figure}

\begin{figure}[H]
    \centering
    \includegraphics[width=0.99\linewidth]{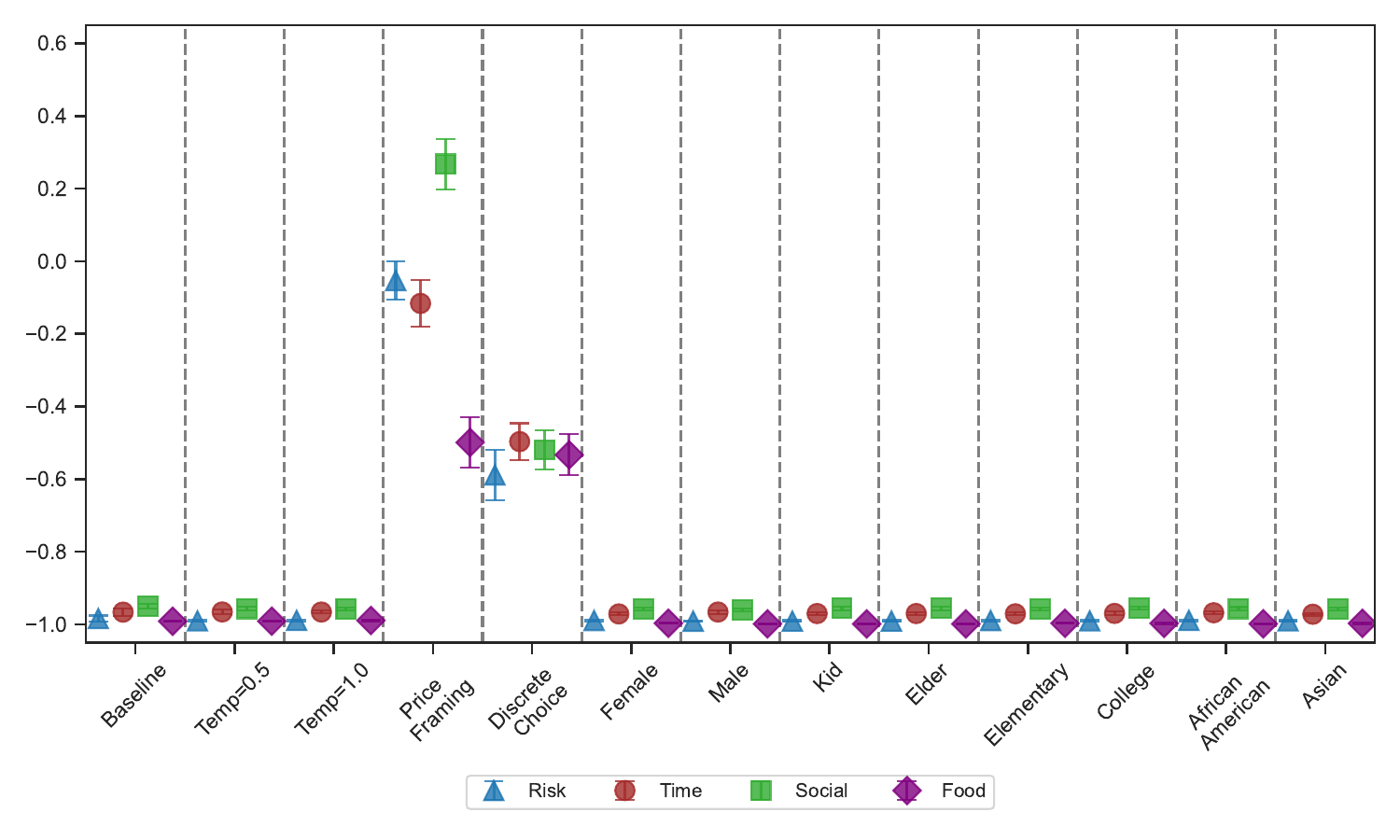}
    \caption{Mean Spearman's Correlation Coefficients of $\ln(x_A/x_B)$ and $\ln(p_A/p_B)$ of GPT across Different Variations. This figure displays the average of Spearman's correlation coefficient between $\ln(x_A/x_B)$ and $\ln(p_A/p_B)$, and 95\% confidence intervals for GPT under different conditions: baseline, temperature of 0.5, temperature of 1, price framing and discrete choices, and various demographic settings.}
    \label{figs24}    
\end{figure}

\clearpage

\begin{table}[H]
\captionsetup{font=footnotesize}
\footnotesize
\centering
\caption {\centering Demographics of  Subjects in the Human Experiment (obs=347) }\label{tables1}
 \makebox[\linewidth][c]{
\begin{threeparttable}  
   \begin{tabular}{R{2cm}L{8cm}C{2cm}}
    \toprule
    \multicolumn{1}{l}{Variable} & Definition & Mean \\   
  \midrule
    \multicolumn{1}{l}{$1_{\text{Female}}$} & =1 if Gender = Female; 0 otherwise & 51.3\% \\
    \multicolumn{1}{l}{$1_{\text{Age}<25}$} & =1 if Age $<$ 25; 0 otherwise & 10.7\% \\
    \multicolumn{1}{l}{$1_{\text{Age}>54}$} &  =1 if Age $>$ 54; 0 otherwise & 35.2\% \\
    \multicolumn{1}{l}{$1_{\text{Minority}}$} &  =1 if Race $\neq$ White; 0 otherwise & 22.8\% \\
    \multicolumn{1}{l}{$1_{ \leq \text{High School}}$} &  =1 if Education = Elementary school or High school; 0 otherwise & 31.4\% \\
    \multicolumn{1}{l}{$1_{> \text{Bachelor}}$} &  =1 if Education = Master's or Ph.D.; 0 otherwise & 14.4\% \\
    \bottomrule
    \end{tabular}%
    \begin{tablenotes}[flushleft]
    \item
\footnotesize \textit{Notes:} This table displays the demographic characteristics of the sample in our Human Experiment.
\end{tablenotes}
\end{threeparttable}
}
\end{table}
\clearpage

\begin{table}[H]
  \centering
  \footnotesize
  \caption{Aggregate Preference Parameters (Baseline)}\label{tables2}
      \makebox[\linewidth]{
\begin{threeparttable}  

    \begin{tabular}{lcccccccc}
    \toprule
          & \multicolumn{2}{c}{Risk Preference} & \multicolumn{2}{c}{Time Preference} & \multicolumn{2}{c}{Social Preference} & \multicolumn{2}{c}{Food Preference} \\
          & $\alpha_r$&$\rho_r$& $\alpha_t$ & $\rho_t$  & $\alpha_s$ & $\rho_s$  & $\alpha_f$ & $\rho_f$ \\
\cmidrule(l){2-3}  \cmidrule(l){4-5}   \cmidrule(l){6-7}  \cmidrule(l){8-9}      &       &       &       &       &       &       &       &  \\

 \textit{Human subjects} & 0.618 &    0.335	 &   0.513    &  0.981	   &0.735	 & 0.330	               &  0.583    &    0.386    \\
          & (0.010)  &  (0.018)  &     (2.849)  &    (0.004)   &   (0.012)    & (0.028)     &  (0.012)    &    (0.018)         \\

   \textit{GPT} &   0.508 &    0.488 &    0.504   &   0.466    &    0.512   &0.520  & 0.501    &    0.491              \\
          &(0.003) &  (0.003)  &    (0.003)   &   (0.003)    &   (0.004)     & (0.004)    &   (0.002)      &    (0.002)          \\
    
 \textit{GPT (Temp=0.5)} & 0.512 &   0.483	&  0.504	     &   0.473    &      0.516      &    0.518	  &    0.501   &    0.492   \\
         &   (0.004)    & (0.004)  &   (0.003)     &   (0.004)     &   (0.004)    &      (0.004)& (0.003)      &     (0.003)     \\
 \textit{GPT (Temp=1.0)} & 0.514&   0.482	 &    0.503   &  0.461	     &      0.513        & 0.522    &  0.502	    &    0.487   \\
          & (0.004) & (0.005)   &    (0.003)     &     (0.004)  &  (0.004)     &      (0.004)	   & (0.002)     &    (0.002)     \\    
  \textit{GPT (Female)} &  0.510 &   0.488&      0.504 &   0.470    &    0.512        &   0.521   &  0.501	    &  0.493     \\
          & (0.004) &   (0.004)  &  (0.003)     &     (0.003)  &   (0.004)         &   (0.003) &   (0.001)    &    (0.001)    \\     

  \textit{GPT (Male)} & 0.509  &  0.483& 0.503	           &  0.466     &   0.511	    &     0.525  &  0.501	    &   	0.492    \\
          &   (0.003)  &    (0.004)    &  (0.003)      &   (0.003)         &     (0.004)   &   (0.003)  &   (0.001)      &   (0.001)     \\  

  \textit{GPT (Kid)} & 0.510&   0.495    &  0.505     &   0.472   &   0.510	         &0.520  &    0.501    &  0.493     \\
          &   (0.004) &   (0.004)  &  (0.003)     &        (0.003)      &   (0.004)    & (0.004)   &     (0.001)    &    (0.001)      \\     

  \textit{GPT (Elder)} & 0.510  &  0.493&    0.504  &    0.465	   &    0.514        &     0.523 &  0.501    &   0.489     \\
          &  (0.004)&   (0.004)    & (0.003)      &    (0.003)         &   (0.004)    & (0.004)  & (0.001)     &     (0.001)   \\  

  \textit{GPT (Elementary School)} & 0.513	&  0.483     &        0.504     &    0.469   &   0.513	    &  0.518  &  0.502     &   0.493  \\
          & (0.004)  &    (0.004)   &  (0.003)     &  (0.003)     &        (0.004)     &   (0.003) & (0.001)    &   (0.001)    \\     

  \textit{GPT (College)} & 0.514  &    0.485	 &  0.505	     &      0.469     &    0.512   &  0.520 &   0.502    &  0.494      \\
          &  (0.004) &   (0.004)    &  (0.003)     &   (0.003)    &    (0.004)          & (0.003)   &  (0.001)     &   (0.001)    \\  
  \textit{GPT (African American)} & 0.511 &  0.489     &    0.505   &     0.467  &  0.512     &  0.522 &  0.501	    &   0.492         \\
          &(0.004)  &   (0.004)   &   (0.003)    &    (0.003)    &        (0.004)       & (0.003)    &  (0.001)      &   (0.001)      \\  
     \textit{GPT (Asian)} &0.514 &   0.491   & 0.504    &     0.472         & 0.512      & 0.517 & 0.501     &  0.494     \\
          & (0.004)&    (0.004)   &  (0.003) &   (0.003)     &  (0.004)     &    (0.003)         &   (0.001)   &   (0.001)    \\         
    \bottomrule
    \end{tabular}%
          \begin{tablenotes}[flushleft]
          \footnotesize
    \item \textit{Notes:} The table presents aggregate parameters estimation in risk, time, social and time preference of GPT and human subjects, respectively. Standard errors in parentheses are wicalculated via the delta method. For GPT and human subjects, we compare the estimated parameters by performing two-sample two-sided t-tests. All comparisons are statistically significant at 1\% level except $\alpha_t$.
    For each variation of GPT, we compare the estimated parameters with those estimated from the baseline (GPT) by conducting two-sample two-sided t-tests. None of the comparisons are statistically significant  at 10\% level.

\end{tablenotes}
    \end{threeparttable}}

\label{result_agg}
\end{table}%

\clearpage
\begin{table}[H]
  \centering
  \footnotesize
  \caption{Summary Statistics of Preference Parameters at the Individual Level (Baseline)}\label{tables3}
      \makebox[\linewidth]{
\begin{threeparttable}  

    \begin{tabular}{lcccccccc}
    \toprule
          & \multicolumn{2}{c}{Risk Preference} & \multicolumn{2}{c}{Time Preference} & \multicolumn{2}{c}{Social Preference} & \multicolumn{2}{c}{Food Preference} \\
          & $\alpha_r$&$\rho_r$& $\alpha_t$ & $\rho_t$  & $\alpha_s$ & $\rho_s$  & $\alpha_f$ & $\rho_f$ \\
\cmidrule(l){2-3}  \cmidrule(l){4-5}   \cmidrule(l){6-7}  \cmidrule(l){8-9}      &       &       &       &       &       &       &       &  \\

 \textit{Human subjects} & 0.561 &    -0.712	 &   0.547    &  0.825   &0.751	 & -0.557	               &  0.595    &   -0.890    \\
          & (0.120)  &  (4.384)  &     (0.103)  &    (0.295)   &   (0.203)    & (3.439)     &  (0.236)    &    (4.154)         \\

   \textit{GPT} &   0.509 &    0.486 &    0.504   &   0.468    &    0.513   &0.525  & 0.501    &    0.491              \\
          &(0.024) &  (0.072)  &    (0.020)   &   (0.072)    &   (0.019)     & (0.047)    &   (0.005)      &    (0.024)          \\
    
 \textit{GPT (Temp=0.5)} & 0.513 &   0.477	&  0.504	     &   0.463    &      0.517      &    0.519	  &    0.501   &    0.491   \\
         &   (0.037)    & (0.128)  &   (0.032)     &   (0.102)     &   (0.032)    &      (0.072)& (0.011)      &     (0.038)     \\
 \textit{GPT (Temp=1.0)} & 0.513&   0.479	 &    0.503   &  0.460	     &      0.513       & 0.521   &  0.502	    &    0.488   \\
          & (0.029) & (0.112)   &    (0.033)     &     (0.093)  &  (0.025)     &      (0.069)	   & (0.009)     &    (0.044)     \\    
  \textit{GPT (Female)} &  0.509 &   0.488&      0.505 &   0.468   &    0.513        &   0.521	   &  0.501	    &  0.493     \\
          & (0.017) &   (0.070)  &  (0.015)     &     (0.073)  &   (0.023)         &   (0.049) &   (0.005)    &    (0.019)    \\     

  \textit{GPT (Male)} & 0.509  &  0.482& 0.504	           &  0.464     &   0.511	    &     0.526  &  0.502	    &   	0.492    \\
          &   (0.019)  &    (0.078)    &  (0.015)      &   (0.072)         &     (0.022)   &   (0.045)  &   (0.004)      &   (0.020)     \\  

  \textit{GPT (Kid)} & 0.509&   0.494    &  0.506    &   0.469   &   0.511	         &0.519  &    0.502   &  0.493     \\
          &   (0.016) &   (0.072)  &  (0.017)     &        (0.076)      &   (0.023)    & (0.051)   &     (0.004)    &    (0.020)      \\     

  \textit{GPT (Elder)} & 0.511  &  0.490&    0.504   &    0.465   &    0.513        &     0.522 &  0.501    &   0.489     \\
          &  (0.020)&   (0.082)    & (0.018)      &    (0.068)         &   (0.021)    & (0.048)  & (0.005)     &     (0.024)   \\  

  \textit{GPT (Elementary School)} & 0.511	&  0.484     &        0.504     &    0.469   &   0.513	    &  0.519  &  0.502     &   0.492 \\
          & (0.020)  &    (0.080)   &  (0.015)     &  (0.074)     &        (0.022)     &   (0.048) & (0.004)    &   (0.021)    \\     

  \textit{GPT (College)} & 0.512  &    0.488	 &  0.504	     &      0.468     &    0.513   &  0.522 &   0.502    &  0.493      \\
          &  (0.020) &   (0.074)&  (0.016)     &   (0.076)    &    (0.021)          & (0.046)   &  (0.005)     &   (0.024)    \\  
  \textit{GPT (African American)} & 0.510 &  0.490     &    0.505   &     0.460  &  0.512     &  0.519 &  0.501	    &   0.491         \\
          &(0.017)  &   (0.077)   &   (0.017)    &    (0.075)    &        (0.021)       & (0.052)    &  (0.004)      &   (0.021)      \\  
     \textit{GPT (Asian)} &0.513 &   0.491    & 0.504     &     0.469       & 0.512      & 0.521 & 0.501     &  0.494     \\
          & (0.019)&    (0.076)   &  (0.017) &   (0.078)     &  (0.020)     &    (0.049)         &   (0.004)   &   (0.023)    \\         
    \bottomrule
    \end{tabular}%
          \begin{tablenotes}[flushleft]
          \footnotesize
    \item \textit{Notes:} The table presents mean values of preference parameters in risk, time, social and time preference of GPT observations and human subjects at the individual level, respectively. Standard deviations are in parentheses.  For GPT observations and human subjects, we compare the estimated parameters by performing two-sample two-sided t-tests. All comparisons are statistically significant at 1\% level. For each variation of GPT, we compare the estimated parameters with those estimated from the baseline (GPT) by conducting two-sample two-sided t-tests. None of the comparisons are statistically significant at 10\% level.

\end{tablenotes}
    \end{threeparttable}}

\label{result_ind}
\end{table}%

\clearpage

\begin{table}[H]
\footnotesize
  \centering
  \caption{OLS Regression Analyses}\label{tables4}
        \makebox[\linewidth]{
\begin{threeparttable}  
   \begin{tabular}{lcccc}
\toprule
 & \multicolumn{2}{c}{Dep.Var: CCEI} & \multicolumn{2}{c}{Dep.Var: Spearman Correlation Coefficient} \\
 & (1) & (2) & (3) & (4) \\
 & Baseline$+$Price Framing & Baseline$+$Discrete Choice & Baseline$+$Price Framing & Baseline$+$Discrete Choice \\ \cmidrule(l){2-3}  \cmidrule(l){4-5}   
\\
$1_{\text{GPT}}$ & 0.024*** & 0.024*** & -0.231*** & -0.231*** \\
 & (0.007) & (0.008) & (0.021) & (0.017) \\
$1_{\text{Price Framing}}$ & -0.027*** &  & 0.045** & \multicolumn{1}{l}{} \\
 & (0.007) & \multicolumn{1}{l}{} & (0.020) & \multicolumn{1}{l}{} \\
$1_{\text{GPT}}\times 1_{\text{Price Framing}}$ & -0.127*** &  & 0.828*** & \multicolumn{1}{l}{} \\
 & (0.010) &  & (0.030) & \multicolumn{1}{l}{} \\
$1_{\text{Discrete Choice}}$ &  & -0.009 & \multicolumn{1}{l}{} & 0.079*** \\
 &  & (0.007) & \multicolumn{1}{l}{} & (0.016) \\
$1_{\text{GPT}}\times 1_{\text{Discrete Choice}}$ &  & -0.138*** & \multicolumn{1}{l}{} & 0.360*** \\
 &  & (0.011) & \multicolumn{1}{l}{} & (0.023) \\ \midrule
Preference Fixed Effects & Yes & Yes & Yes & Yes \\
N & 1704 & 1744 & 1704 & 1744 \\
R$^{2}$ & 0.268 & 0.206 & 0.528 & 0.299 \\ \bottomrule
\end{tabular}
    \begin{tablenotes}[flushleft]
    \item

\footnotesize \textit{Notes:} In this table, we pool the data of the GPT experiment and the human experiment. Columns 1 and 3 use data from baseline and price framing conditions, while columns 2 and 4 use data from baseline and discrete choice conditions. In columns 1 and 2, the dependent variable is the CCEI values, while in columns 3 and 4, the dependent variable is the Spearman's correlation coefficients of $\ln(x_A/x_B)$ and $\ln(p_A/p_B)$. $1_{\text{GPT}}$ is a binary variable that equals 1 if the data is from the GPT experiment, and 0 otherwise. $1_{\text{Price Framing}}$ is a binary variable that equals 1 if the data is from the price framing condition, and 0 otherwise. $1_{\text{Discrete Choice}}$ is a binary variable that equals 1 if the data is from the discrete choice condition, and 0 otherwise. Standard errors are in parentheses. {*} \(p<0.10\), {**} \(p<0.05\), {***} \(p<0.01\).
\end{tablenotes}
\end{threeparttable}}
\label{reg_ccei}
\end{table}%

\begin{table}[H]
\footnotesize
  \centering
  \caption{OLS Regressions Analyses on Demographics}\label{tables5}
    \makebox[\linewidth]{
\begin{threeparttable}  
\begin{tabular}{L{4cm}C{4cm}C{4cm}}
\toprule
 & Dep.Var: CCEI & Dep.Var: Spearman Correlation Coefficient \\
 & (1) & (2) \\
\cmidrule(l){2-2}     \cmidrule(l){3-3}
\\
$1_{\text{Female}}$ & -0.005 & -0.010 \\
 & (0.005) & (0.016) \\
$1_{\text{Age}<25}$ & -0.014* & -0.000 \\
 & (0.008) & (0.027) \\
$1_{\text{Age}>54}$ & -0.009* & 0.014 \\
 & (0.005) & (0.018) \\
$1_{\text{Minority}}$ & -0.031*** & 0.058*** \\
 & (0.006) & (0.019) \\
$1_{ \leq \text{High School}}$ & -0.003 & 0.003 \\
 & (0.006) & (0.018) \\
$1_{> \text{Bachelor}}$ & 0.008 & -0.019 \\
 & (0.007) & (0.024) \\
\midrule
Preference Fixed Effects & Yes & Yes \\
Condition Fixed Effects & Yes & Yes \\
N & 1388 & 1388 \\
R$^{2}$ & 0.039 & 0.038\\
\bottomrule
\end{tabular}
        \begin{tablenotes}[flushleft]
    \item
\footnotesize \textit{Notes:} In this table, we use the data of the human experiment. In column 1, the dependent variable is the CCEI values, while in column 2, the dependent variable is the Spearman's correlation coefficients of $\ln(x_A/x_B)$ and $\ln(p_A/p_B)$. $1_{\text{Female}}$ is a binary variable that equals 1 if the subject is female, and 0 otherwise. $1_{\text{Age}<25}$ and $1_{\text{Age} >54}$ are the two dummies indexing age, with the group of age between 25 and 55 being the reference group. $1_{\text{Minority}}$ is a binary variable that equals 1 if the subject's race is not white, and 0 otherwise. $1_{\leq \text{High School}}$ and $1_{>\text{Bachelor}}$ are the two dummies indexing education, with the group that receives the medium level of education (Associate's, or Bachelor's) being the reference group. Standard errors are in parentheses. {*} \(p<0.10\), {**} \(p<0.05\), {***} \(p<0.01\).
\end{tablenotes}
    \end{threeparttable}}
\end{table}%

\begin{table}[H]
\footnotesize
  \centering
  \caption{Tobit Regressions Analyses on Demographics}\label{tables6}
    \makebox[\linewidth]{
\begin{threeparttable}  
    \begin{tabular}{lcccccccc}
    \toprule
          & Dep.Var: $\alpha_r$ & Dep.Var: $\rho_r$   & Dep.Var: $\alpha_t$  &Dep.Var: $\rho_t$   & Dep.Var: $\alpha_s$ & Dep.Var: $\rho_s$   & Dep.Var: $\alpha_f$ & Dep.Var: $\rho_f$ \\
                    & (1)   & (2)   & (3)   & (4)   & (5)   & (6)   & (7)   & (8) \\

                      & \multicolumn{2}{c}{Risk Preference} & \multicolumn{2}{c}{Time Preference} & \multicolumn{2}{c}{Social Preference} & \multicolumn{2}{c}{Food Preference} \\
\cmidrule(l){2-3}  \cmidrule(l){4-5}   \cmidrule(l){6-7}  \cmidrule(l){8-9}           & \multicolumn{1}{l}{} & \multicolumn{1}{l}{} & \multicolumn{1}{l}{} & \multicolumn{1}{l}{} & \multicolumn{1}{l}{} & \multicolumn{1}{l}{} & \multicolumn{1}{l}{} & \multicolumn{1}{l}{} \\

  $1_{\text{Female}}$& 0.019 & -0.769 & 0.002 & 0.016 & -0.026 & -0.952* & -0.029 & 0.109 \\
          & (0.019) & (0.625) & (0.018) & (0.166) & (0.029) & (0.558) & (0.030) & (0.499) \\
$1_{\text{Age}<25}$& -0.033 & -0.083 & -0.054* & 0.212 & -0.023 & 0.703 & 0.026 & -0.553 \\
          & (0.030) & (1.005) & (0.031) & (0.277) & (0.048) & (0.945) & (0.053) & (0.867) \\
$1_{\text{Age}>54}$ & 0.025 & -0.194 & 0.019 & -0.053 & 0.016 & -0.610 & 0.025 & 0.036 \\
          & (0.021) & (0.704) & (0.020) & (0.184) & (0.031) & (0.612) & (0.033) & (0.546) \\
 $1_{\text{Minority}}$ & 0.059** & -0.625 & 0.064*** & -0.052 & -0.002 & -0.114 & -0.007 & -0.270 \\
          & (0.023) & (0.768) & (0.024) & (0.212) & (0.035) & (0.696) & (0.039) & (0.636) \\
$1_{ \leq \text{High School}}$  & 0.050** & -1.577** & 0.034 & -0.185 & -0.001 & -0.184 & 0.097*** & 0.431 \\
          & (0.021) & (0.710) & (0.021) & (0.190) & (0.032) & (0.633) & (0.034) & (0.560) \\
$1_{> \text{Bachelor}}$ & -0.015 & 0.322 & -0.018 & -0.039 & -0.001 & 1.161 & 0.051 & -1.177 \\
          & (0.029) & (0.958) & (0.027) & (0.246) & (0.042) & (0.833) & (0.046) & (0.759) \\
    \midrule
    N     & 275   & 275   & 283   & 283   & 281   & 281   & 283   & 283 \\
    \bottomrule
    \end{tabular}%
        \begin{tablenotes}[flushleft]
    \item
\footnotesize \textit{Notes:} In this table, we use the data of the human experiment. Columns 1-2, 3-4, 5-6, 7-8 use data of risk preference, time preference, social preference, and food preference with CCEI above 0.95, respectively. In columns 1,3,5,7, the dependent variable is the CCEI values, while in columns 2,4,6,8, the dependent variable is the Spearman's correlation coefficients of $\ln(x_A/x_B)$ and $\ln(p_A/p_B)$. $1_{\text{Female}}$ is a binary variable that equals 1 if the subject is female, and 0 otherwise. $1_{\text{Age}<25}$ and $1_{\text{Age} >54}$ are the two dummies indexing age, with the group of age between 25 and 55 being the reference group. $1_{\text{Minority}}$ is a binary variable that equals 1 if the subject's race is not white, and 0 otherwise. $1_{\leq \text{High School}}$ and $1_{>\text{Bachelor}}$ are the two dummies indexing education, with the group that receives the medium level of education (Associate's, or Bachelor's) being the reference group. Standard errors are in parentheses. {*} \(p<0.10\), {**} \(p<0.05\), {***} \(p<0.01\).
\end{tablenotes}
    \end{threeparttable}}
\end{table}%

\clearpage

\newpage

\end{document}